\documentclass[12pt]{cernrep}

\usepackage{graphicx}

\begin{document}
\begin{titlepage}
\def\thefootnote{\fnsymbol{footnote}}       

\begin{center}
\mbox{ }

\end{center}
\begin{flushright}
\Large
\mbox{\hspace{10.4cm} physics/0604199} \\
\mbox{\hspace{9.0cm} D0 Note CONF-5085 (2006)} \\
\end{flushright}
\begin{center}
\vskip 2.0cm
{\boldmath \LARGE\bf
{Alignment of the Central D\O\ Detector\footnote{Presented at the 
workshop on Tracking In high Multiplicity Environments,
TIME'05, Zurich, October 3-7, 2005. Proceedings {\sl Nucl. Inst. Methods} {\bf A}.}}
}
\vskip 2cm
{\rm\Large 
        Andr\'e Sopczak \\

on behalf of the D\O\ Collaboration}
\bigskip
\large

Lancaster University   \\

\normalsize
\vskip 2.5cm
\centerline{\Large \bf Abstract}
\end{center}
\vspace*{5mm}
\begin{picture}(0.001,0.001)(0,0)
\put(,0){
\begin{minipage}{16cm}
\renewcommand{\baselinestretch} {1.2}
\normalsize
The alignment procedure of the Silicon Microstrip Tracker, SMT, 
and the Central Fiber Tracker, CFT, is described. Alignment
uncertainties and resulting systematic errors in physics 
analyses are addressed.

\renewcommand{\baselinestretch} {1.}

\end{minipage}
}
\end{picture}

\end{titlepage}


\newpage
\normalsize
\setcounter{page}{1}
\pagestyle{plain}

\section{Introduction}
\begin{figure}[hp]
\centering
\vspace*{2mm}
\includegraphics[width=\columnwidth]{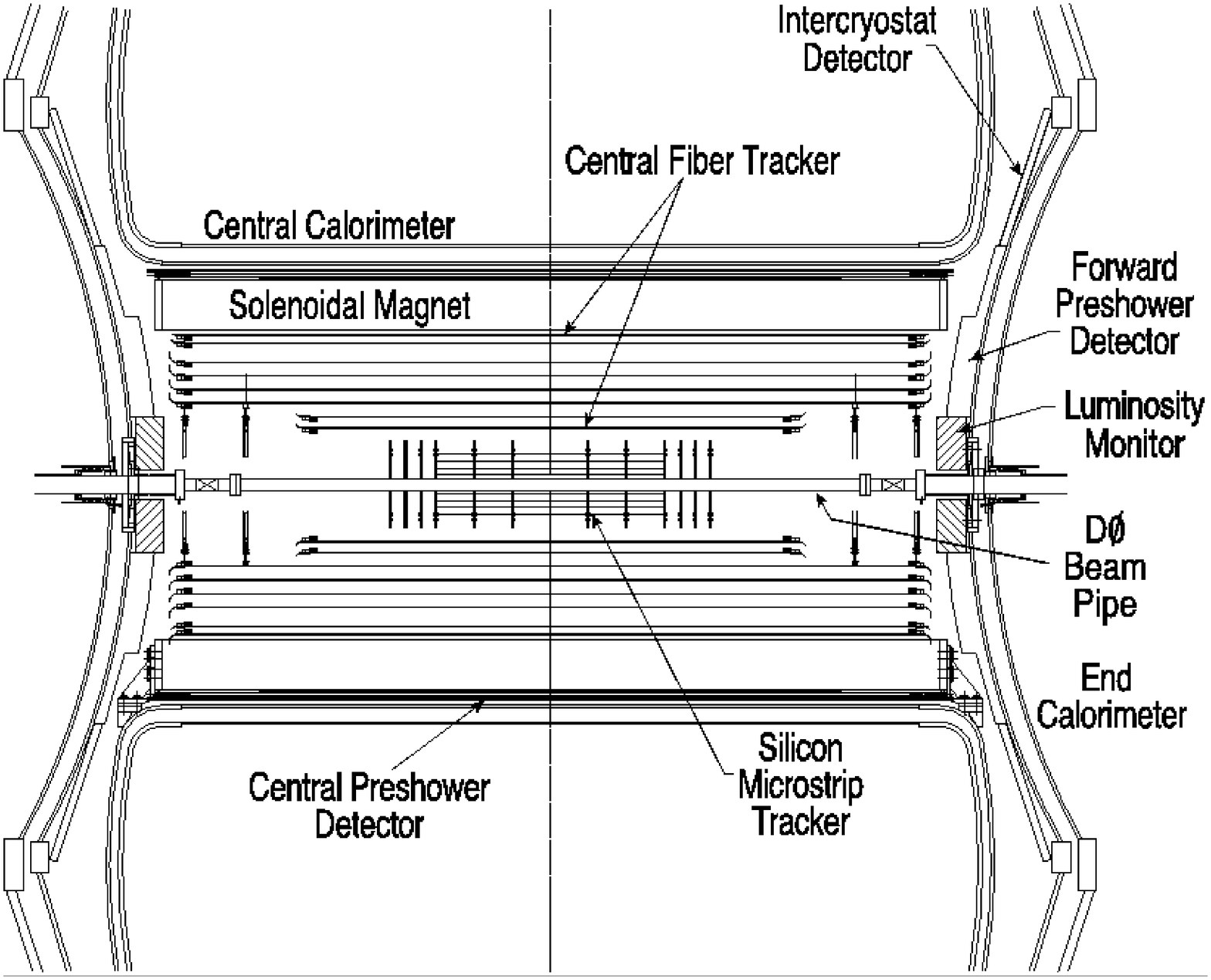}
\vspace*{-5mm}
\caption{\label{fig:geo}
Central D\O\ detector. The Silicon Microstrip Tracker (SMT) and the Central Fiber Tracker (CFT)
are used for the central tracking.
}
\end{figure}

\begin{figure}[h!]
\centering
\hspace*{20mm}
\includegraphics[width=0.4\columnwidth]{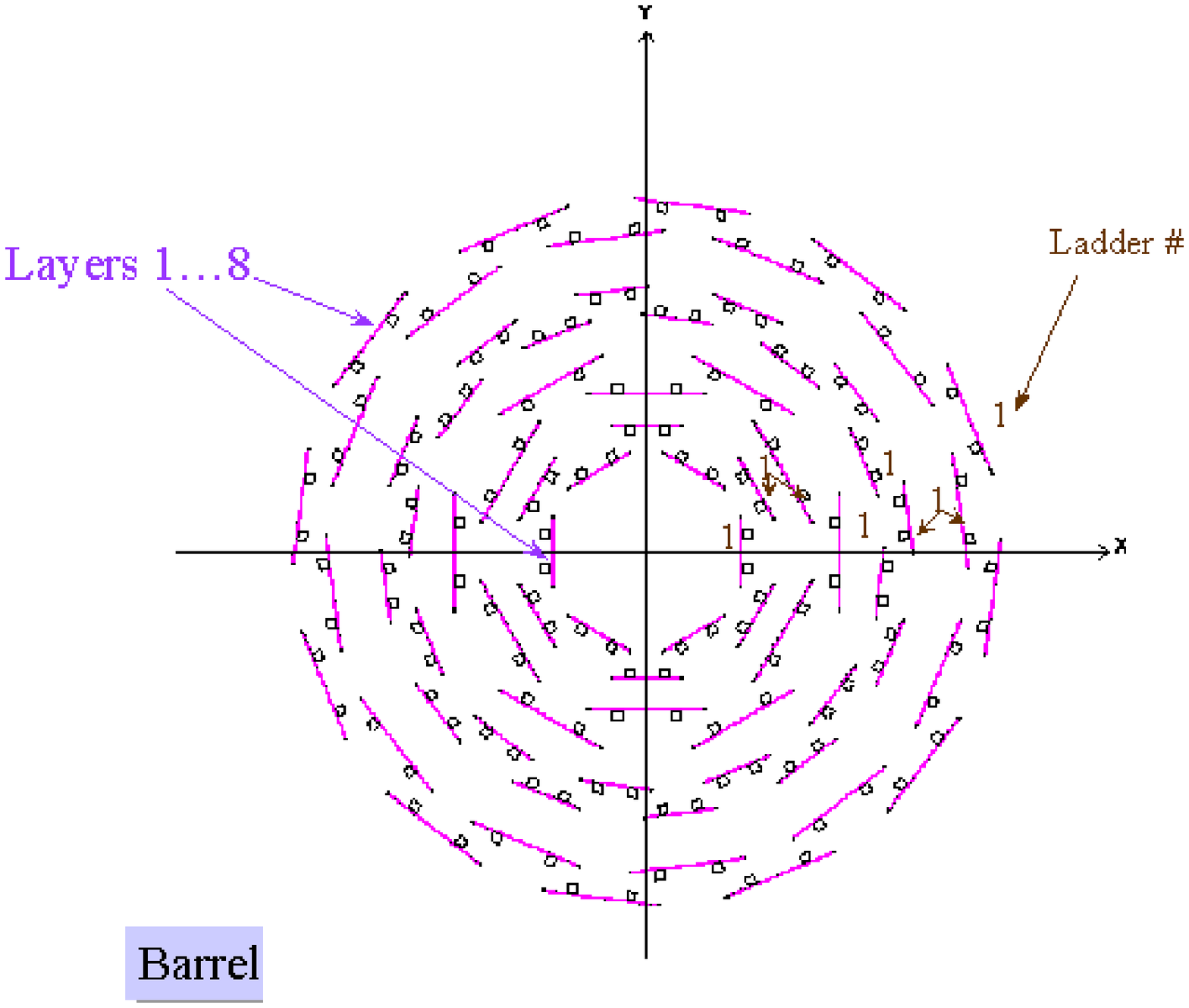} \hfill
\hspace*{-1cm}
\includegraphics[width=0.4\columnwidth]{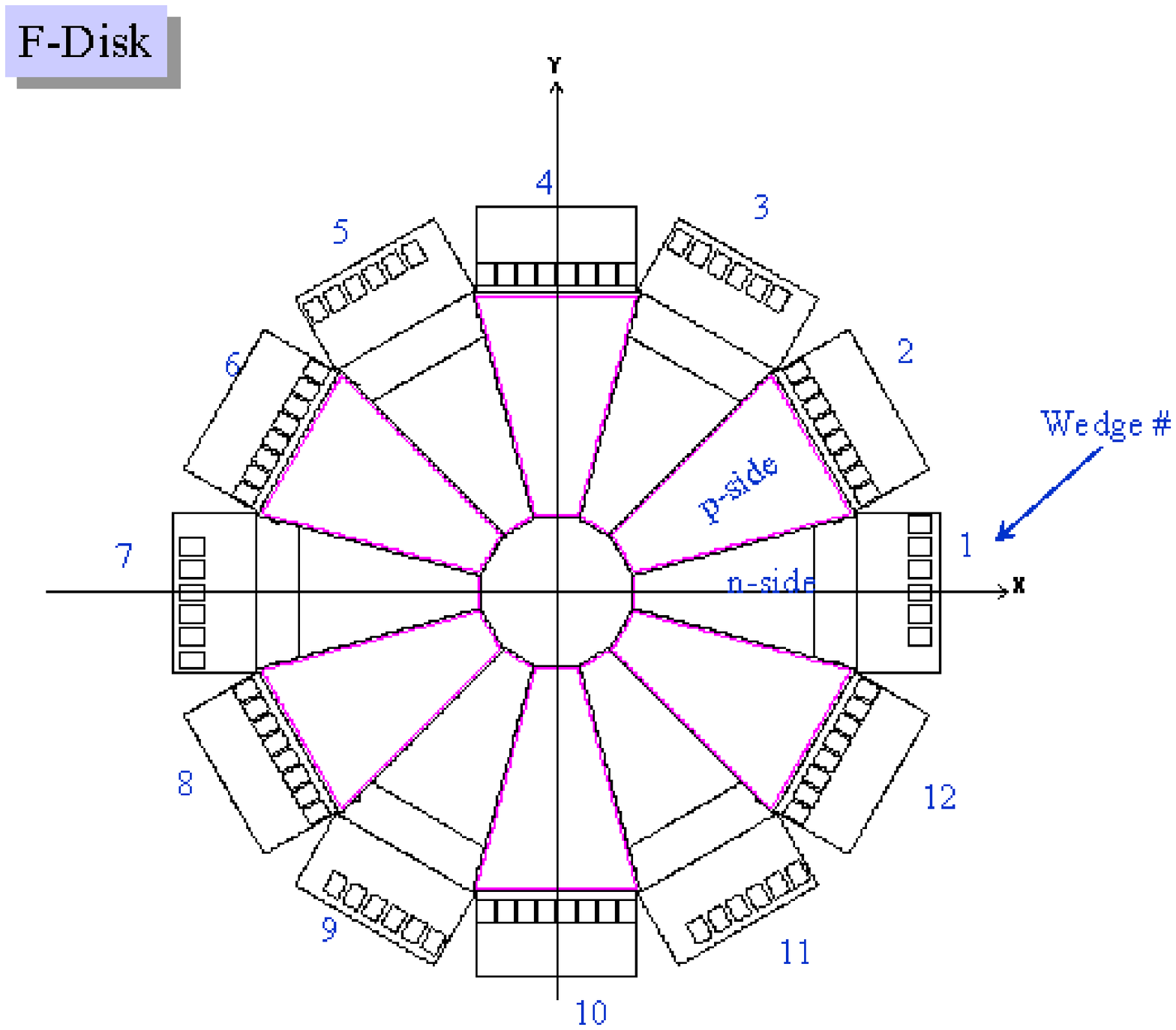}
\hspace*{2mm}
\caption{\label{fig:geo2}
Left: Barrel wafer positions.
Right: F-disk wedges. 
}
\end{figure}

\begin{figure}[h!]
\vspace*{3mm}
\centering
\includegraphics[width=0.8\columnwidth]{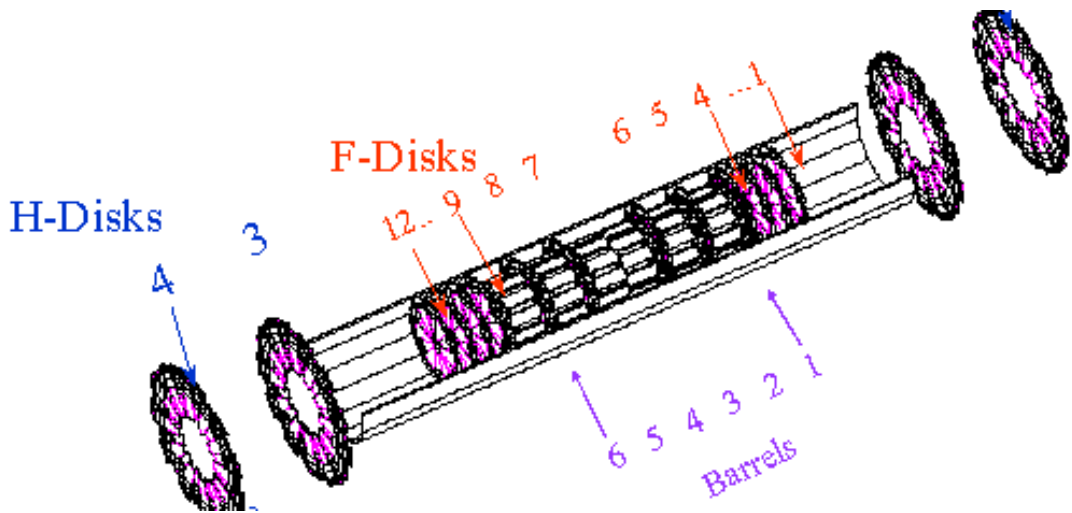}
\vspace*{-4mm}
\caption{\label{fig:geo3}
SMT: Side-view of barrels, F-disks and H-disks. 
}
\end{figure}

The alignment of the D\O\ tracking detectors is crucial for many physics analyses.
The precision determination of the detector element positions improves the 
track reconstruction and the precision measurements at the interaction point. This is
particularly important for Higgs, top and B-physics, and for an impact parameter trigger.
A general overview of the D\O\ detector~\cite{nim} and operation~\cite{ron} has recently 
been given.
Figure~\ref{fig:geo} shows the central D\O\ detector, Fig.~\ref{fig:geo2}
gives the positions of the barrel wafers and F-disk wedges, and Fig.~\ref{fig:geo3}
displays a side-view of the SMT with barrels, F-disks and H-disks.

\section{Method and alignment procedure}

\begin{figure}[h!]
\vspace*{3mm}
\centering
\includegraphics[width=0.73\columnwidth]{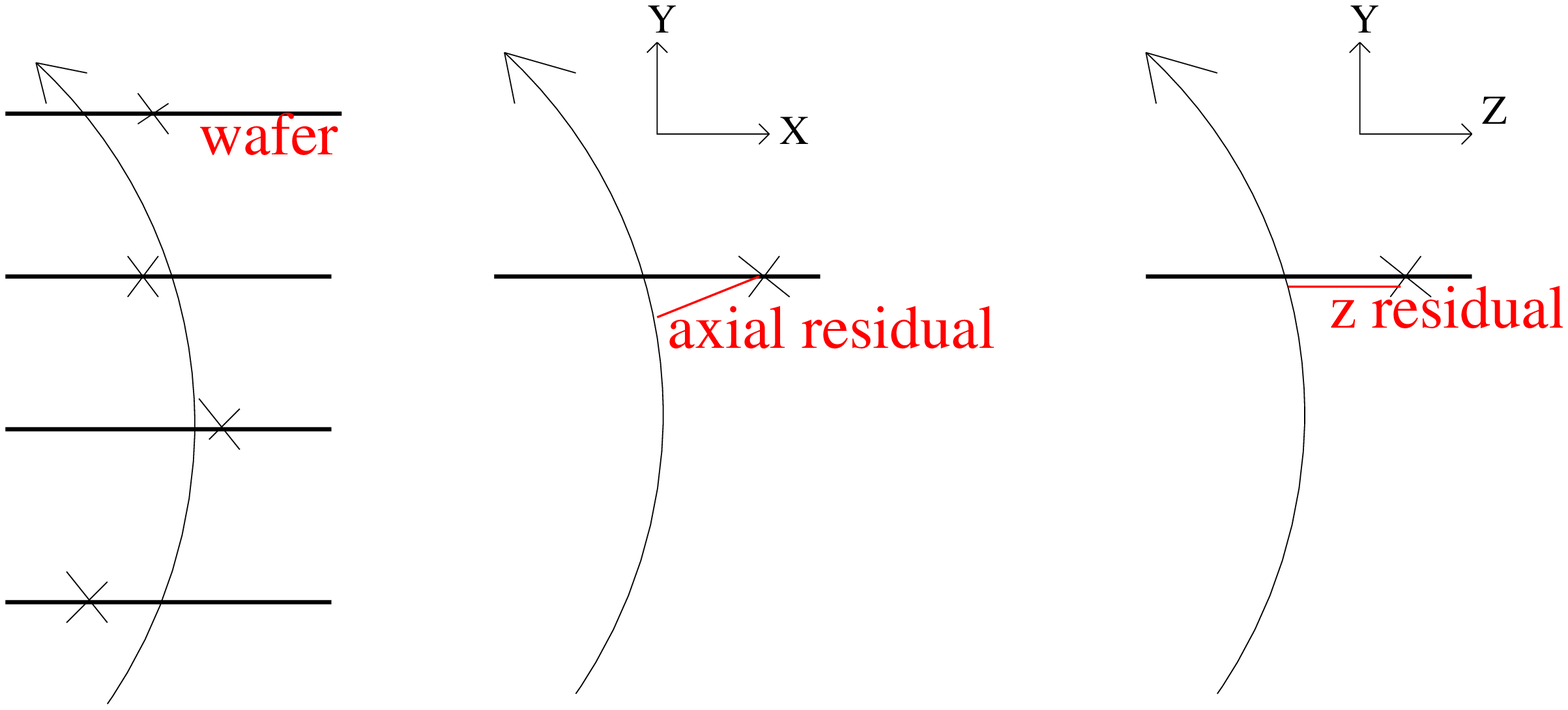}
\vspace*{-3mm}
\caption{\label{fig:method}
Definition of axial- and z-residuals for the SMT barrel wafers
as the deviations of the fitted track (curved line) from the true hit position (marked by $\times$).
}
\vspace*{3mm}
\end{figure}

The basic method to align the wafers is to minimize axial-residuals and z-residuals
(Fig.~\ref{fig:method}). In the SMT, 
there are 432 barrel wafers, 144 F-disk wedges, and 96 H-disk wedges. The CFT encompasses
304 ribbons 
(each with 256 parallel scintillating fibers, 1.6 or 2.5~m long).
In total 976 elements require alignment.
The initial position of the sensitive barrel elements were determined from metrology measurements. 

The alignment procedure is as follows: 
a track is fitted with all hits, except the hit from the sensitive element to be aligned.
Then, axial-residuals and z-residuals of the hit on the wafer to be aligned, 
are determined.
The pull (residual/error) is calculated and the corresponding $\chi^2$ as a sum of 
pulls from all tracks on the wafer are determined.
The $\chi^2$ is minimized as a function of the wafer position (three space coordinates and 
three angles). All wafer positions are determined and these positions serve as input geometry
for the next iterative step. 
The iterative process continues 
until a convergence criterion is reached. 
A wafer is considered aligned if the shift divided by its uncertainty of 
a sensitive element between two iterations is less than a certain value. This value is called 
the `shift limit'.

\clearpage

\begin{figure}[htbp]
\centering
\includegraphics[width=0.323\columnwidth]{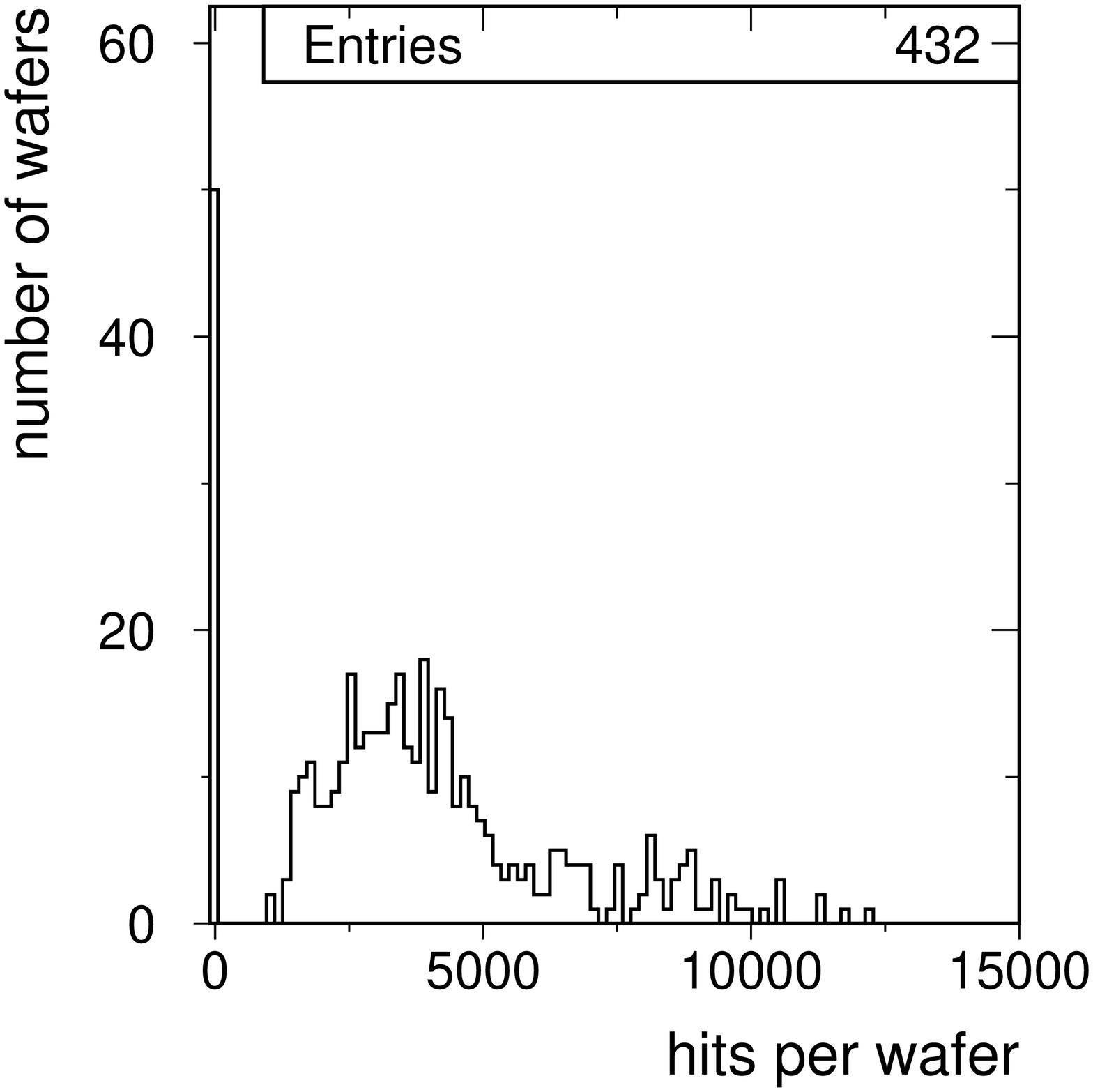} \hfill
\includegraphics[width=0.323\columnwidth]{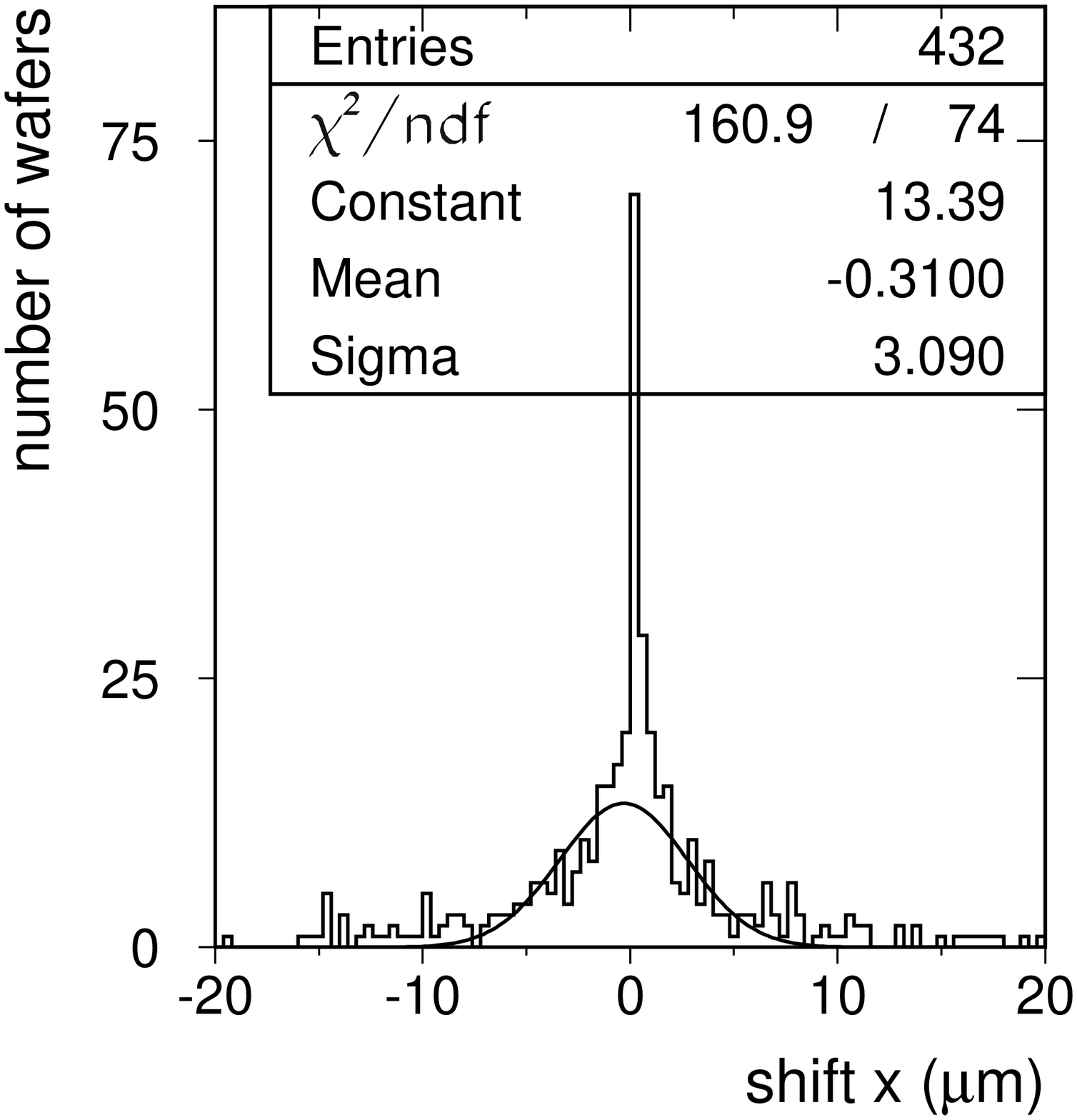} \hfill
\includegraphics[width=0.323\columnwidth]{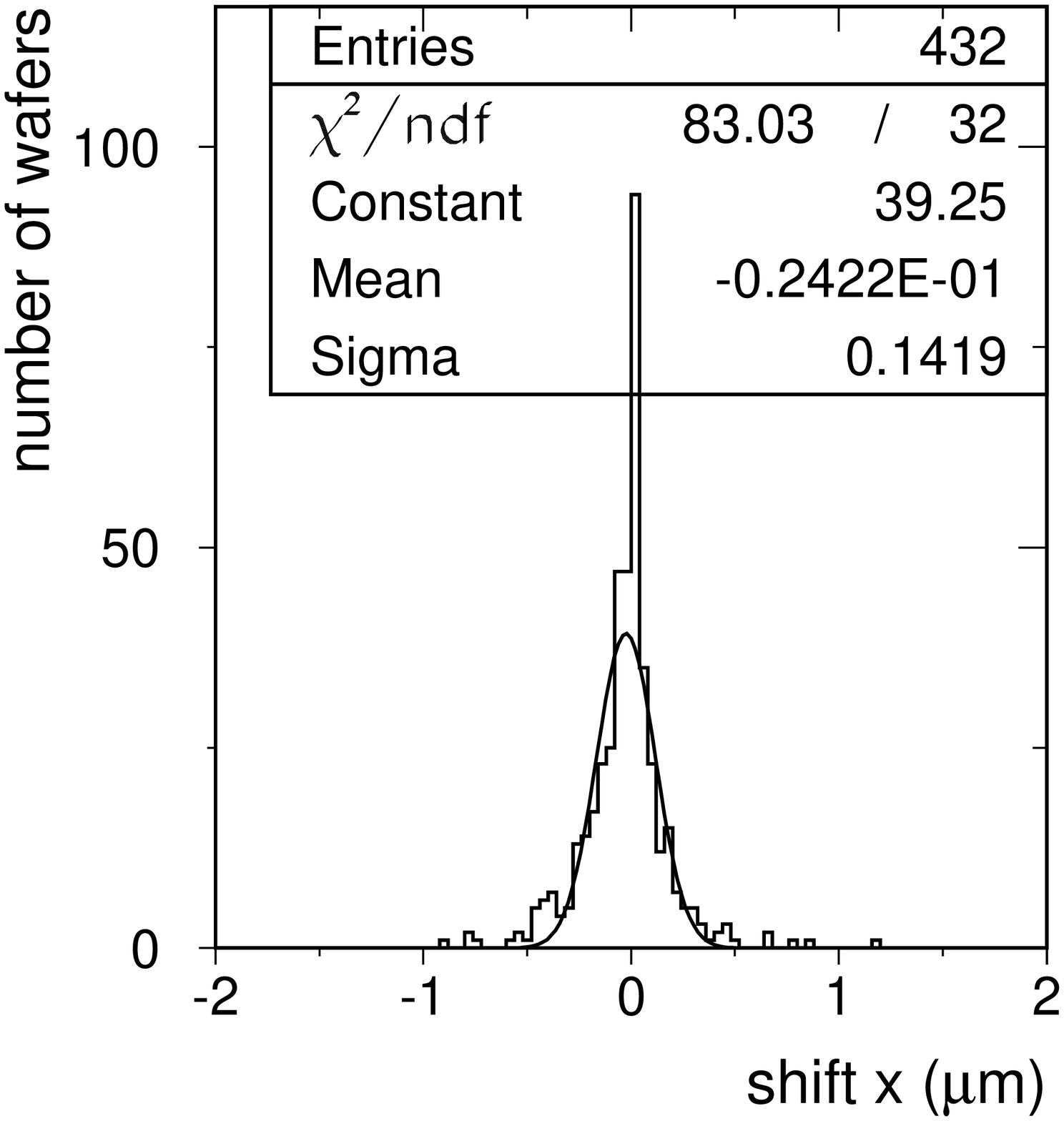}
\vspace*{-7mm}
\caption{\label{fig:barrel_first_last}
Left: number of hits per wafer for 50,000 data events in the SMT barrel. 
Center: shifts of the 432 wafers for the first iteration.
Right: shifts for the last iteration.
}
\end{figure}

\begin{figure}[htbp]
\centering
\includegraphics[width=0.323\columnwidth]{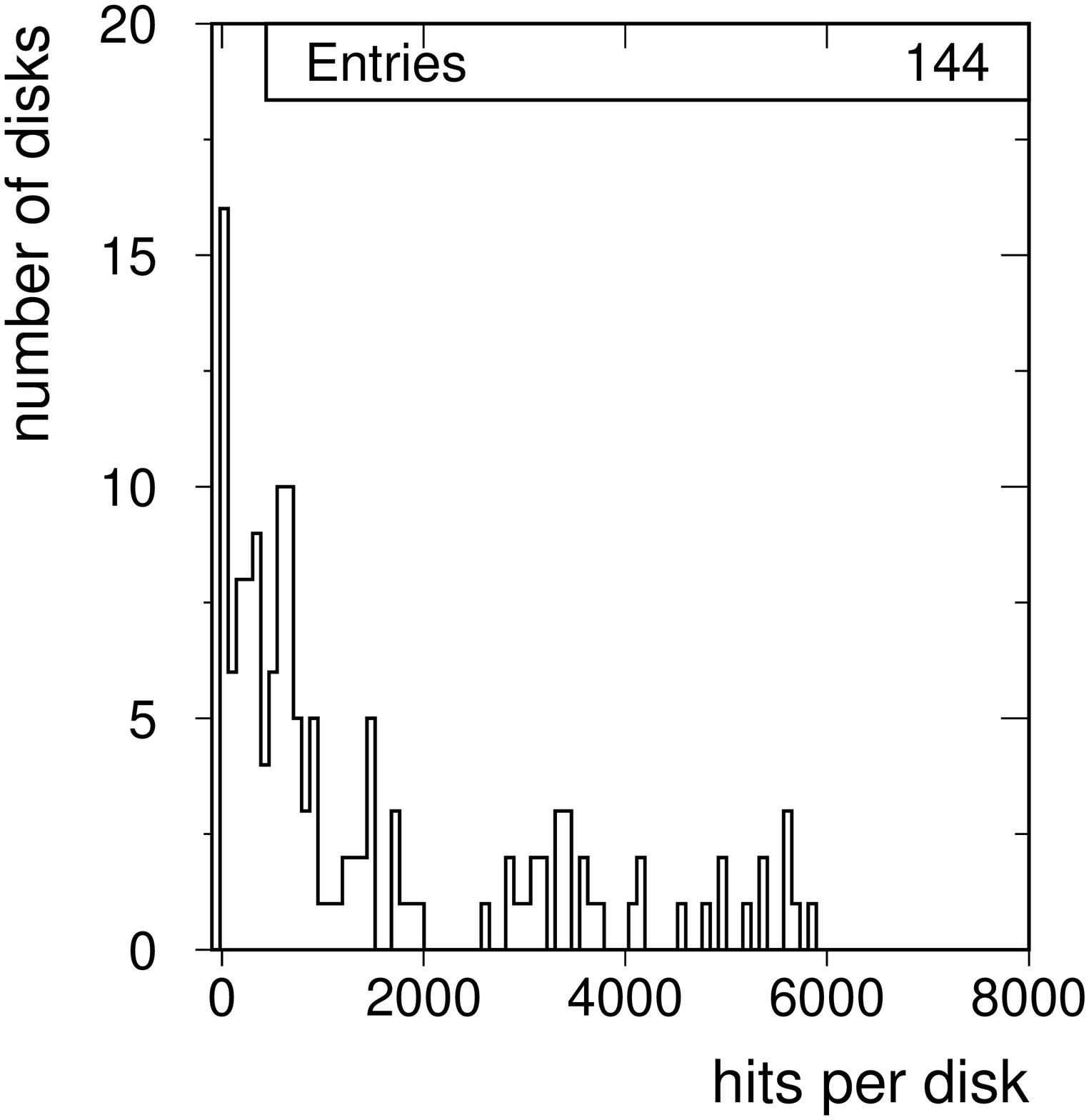} \hfill
\includegraphics[width=0.323\columnwidth]{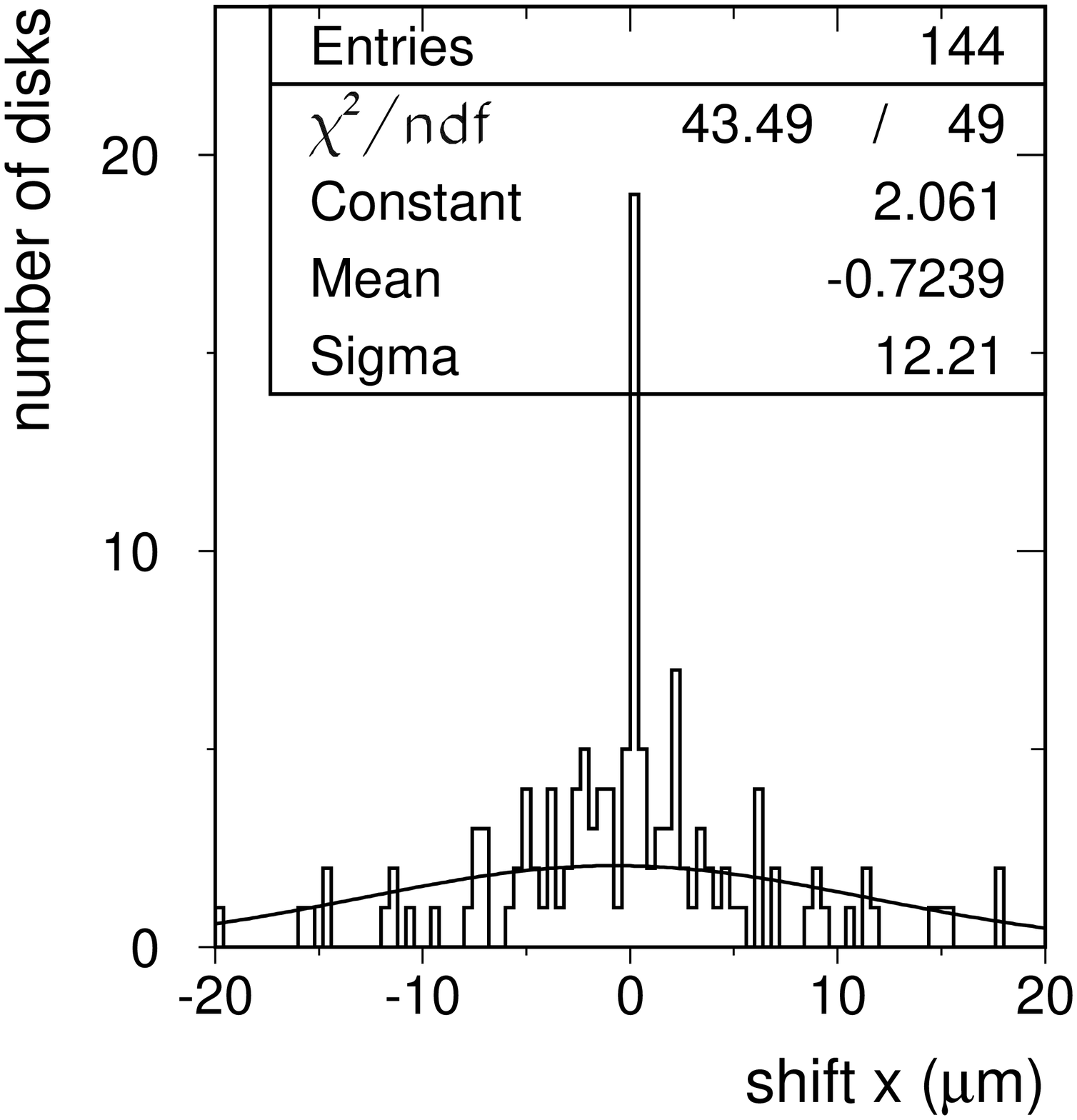} \hfill
\includegraphics[width=0.323\columnwidth]{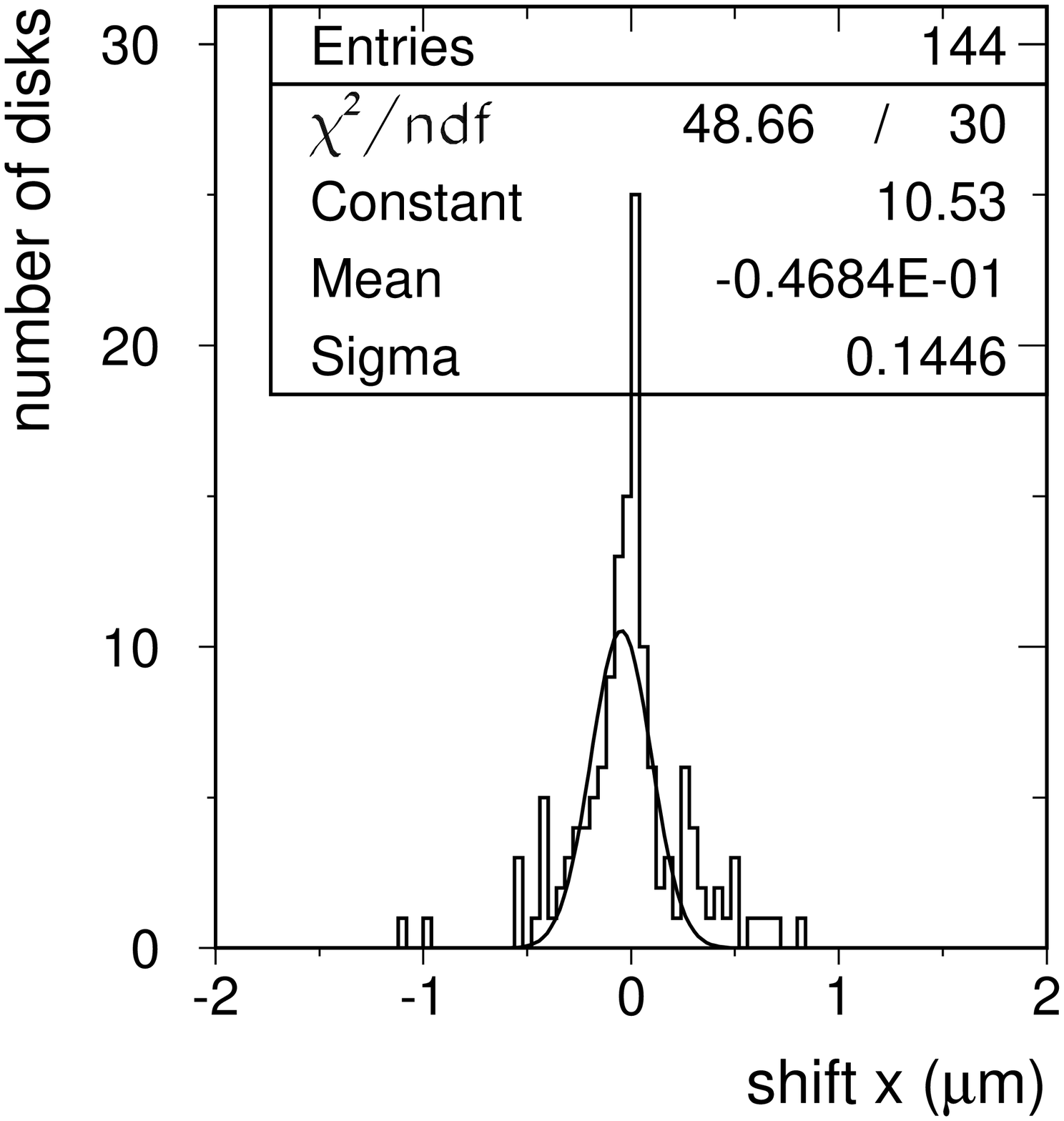}
\vspace*{-7mm}
\caption{\label{fig:fdisk_first_last}
Left: hits per disk for 50,000 data events for the F-disks. 
Center: shifts of the 144 wedges for the first iteration. 
Right: shifts for the last iteration.
}
\end{figure}

\begin{figure}[h!]
\centering
\includegraphics[width=0.323\columnwidth]{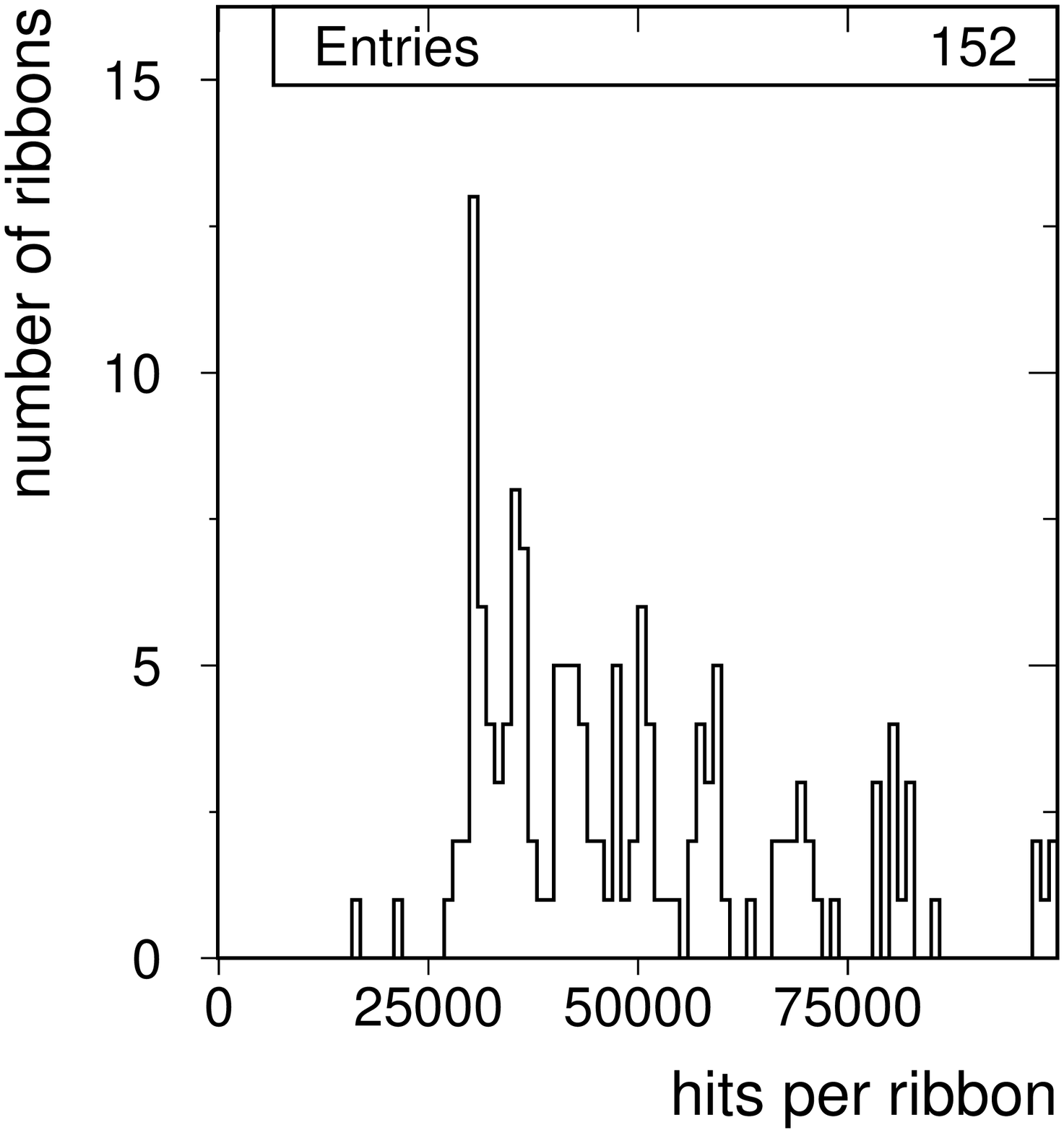} \hfill
\includegraphics[width=0.323\columnwidth]{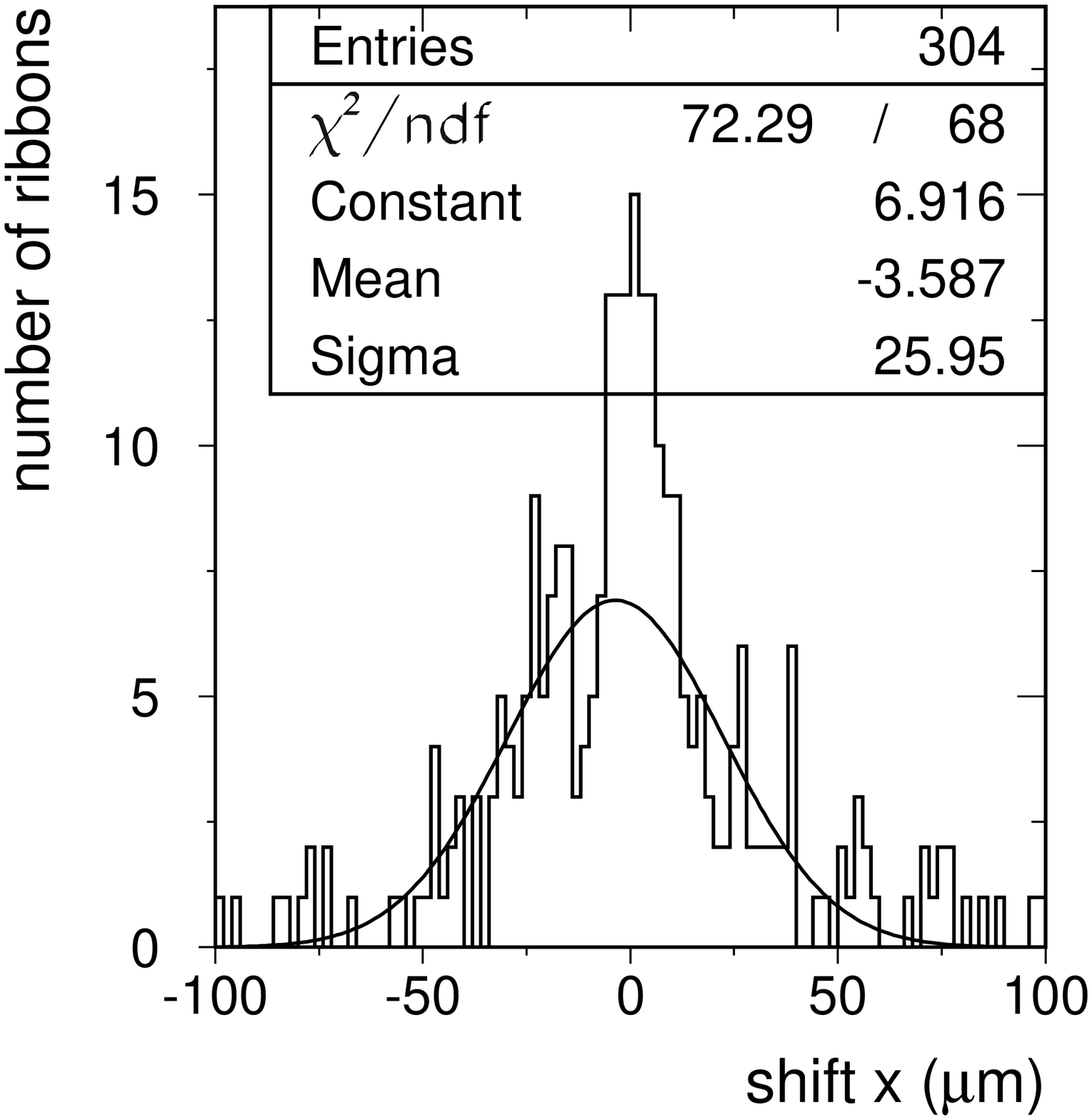} \hfill
\includegraphics[width=0.323\columnwidth]{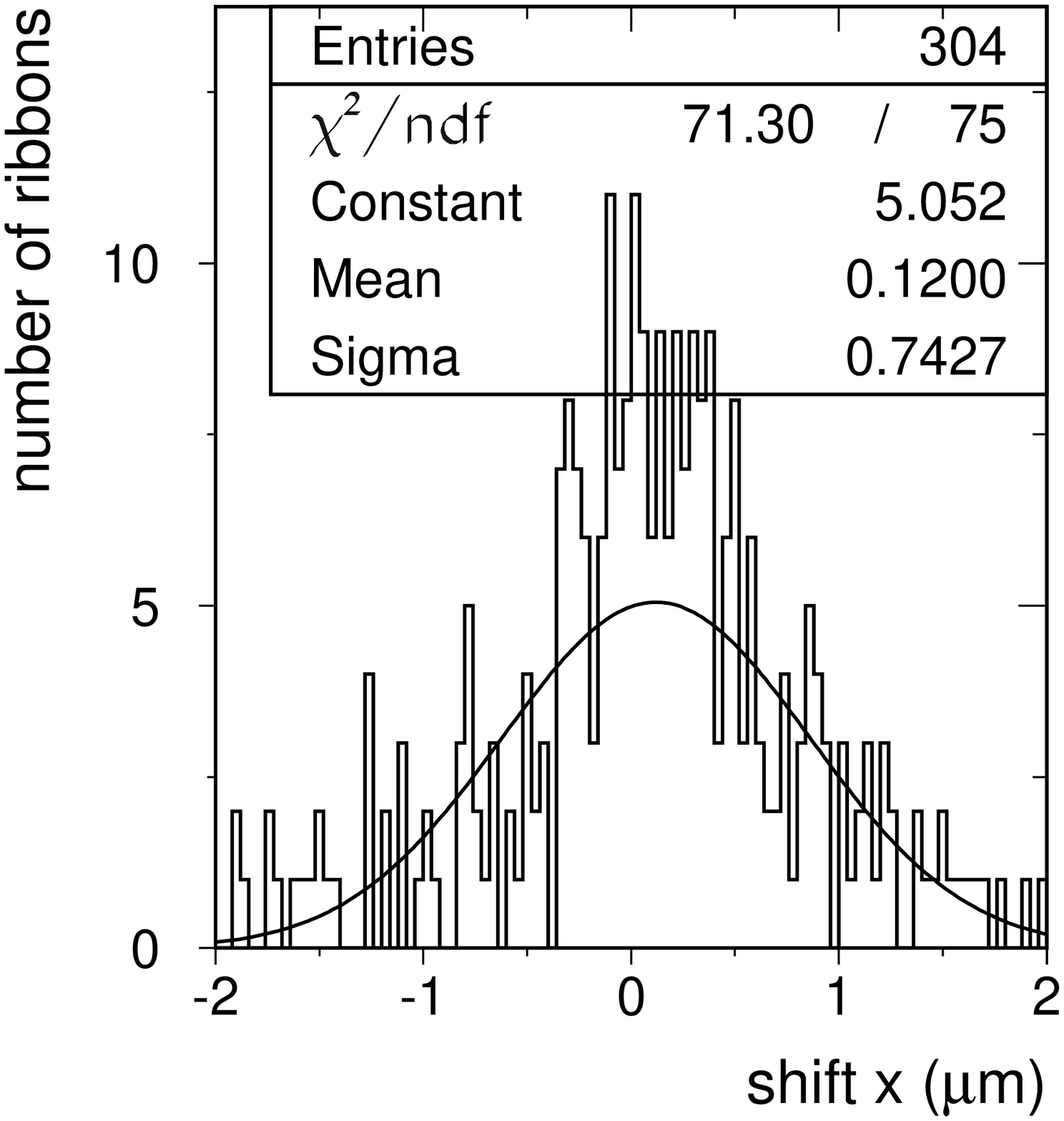}
\vspace*{-7mm}
\caption{\label{fig:cft_first_last}
Left: hits per ribbon for 50,000 data events for the CFT. 
Center: shifts of the 304 ribbons for the first iteration. 
Right: shifts for the last iteration.
}
\vspace*{-2mm}
\end{figure}

\begin{figure}[htbp]
\vspace*{1mm}
\centering
\includegraphics[width=0.45\columnwidth]{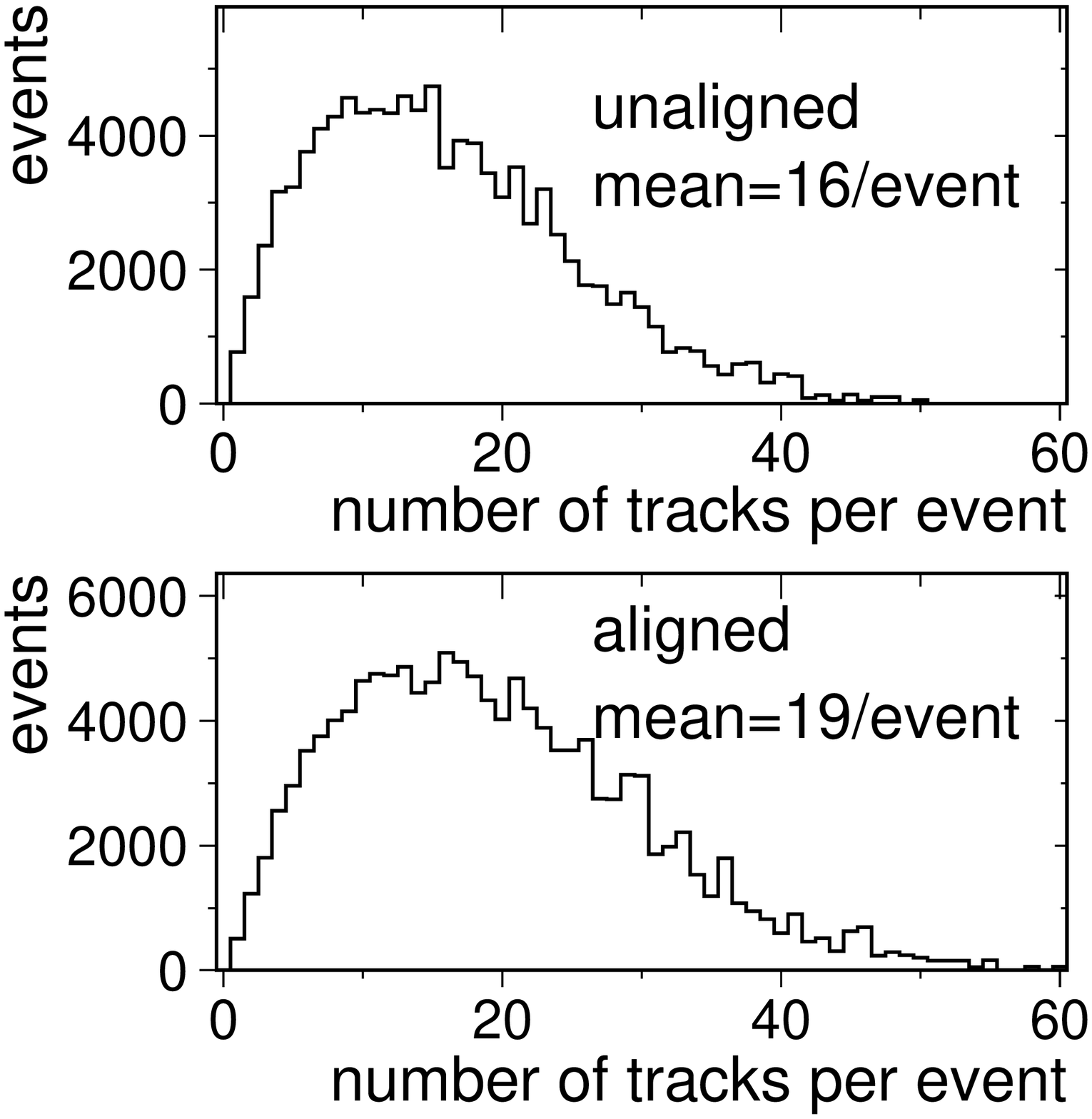} \hfill
\includegraphics[width=0.45\columnwidth]{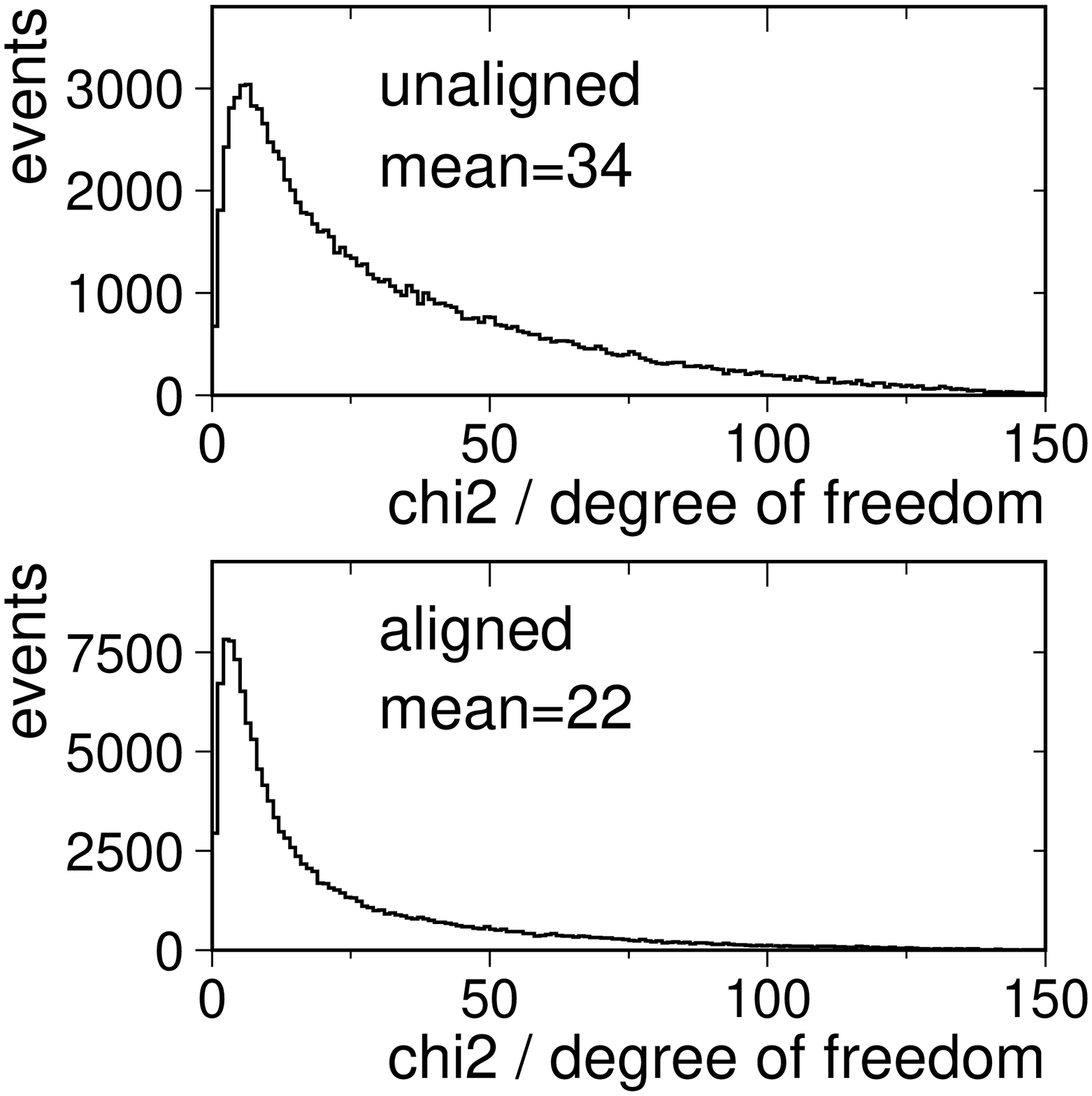}
\vspace*{-9mm}
\caption{\label{fig:ntr}
Left: number of tracks per event. Right: $\rm \chi^2$ per degree of freedom.
Both, before and after alignment.
}
\vspace*{1cm}
\end{figure}

For the barrel, Fig.~\ref{fig:barrel_first_last} shows a histogram  of the number of hits per wafer 
for 50,000 events, and shows the shifts of each wafer between two consecutive iterations for the 
first and last iteration, illustrating the convergence.
The corresponding plots for the F-disk and CFT are shown in 
Figs.~\ref{fig:fdisk_first_last} and~\ref{fig:cft_first_last}.
The number of tracks per event and the $\rm \chi^2$ per degree of freedom for the track
reconstruction show an improvement after alignment (Fig.~\ref{fig:ntr}).
A shift limit of 0.07 is applied.

\clearpage
\section{Residuals}
\begin{figure}[h!]
\centering
\includegraphics[width=0.49\columnwidth,height=7cm]{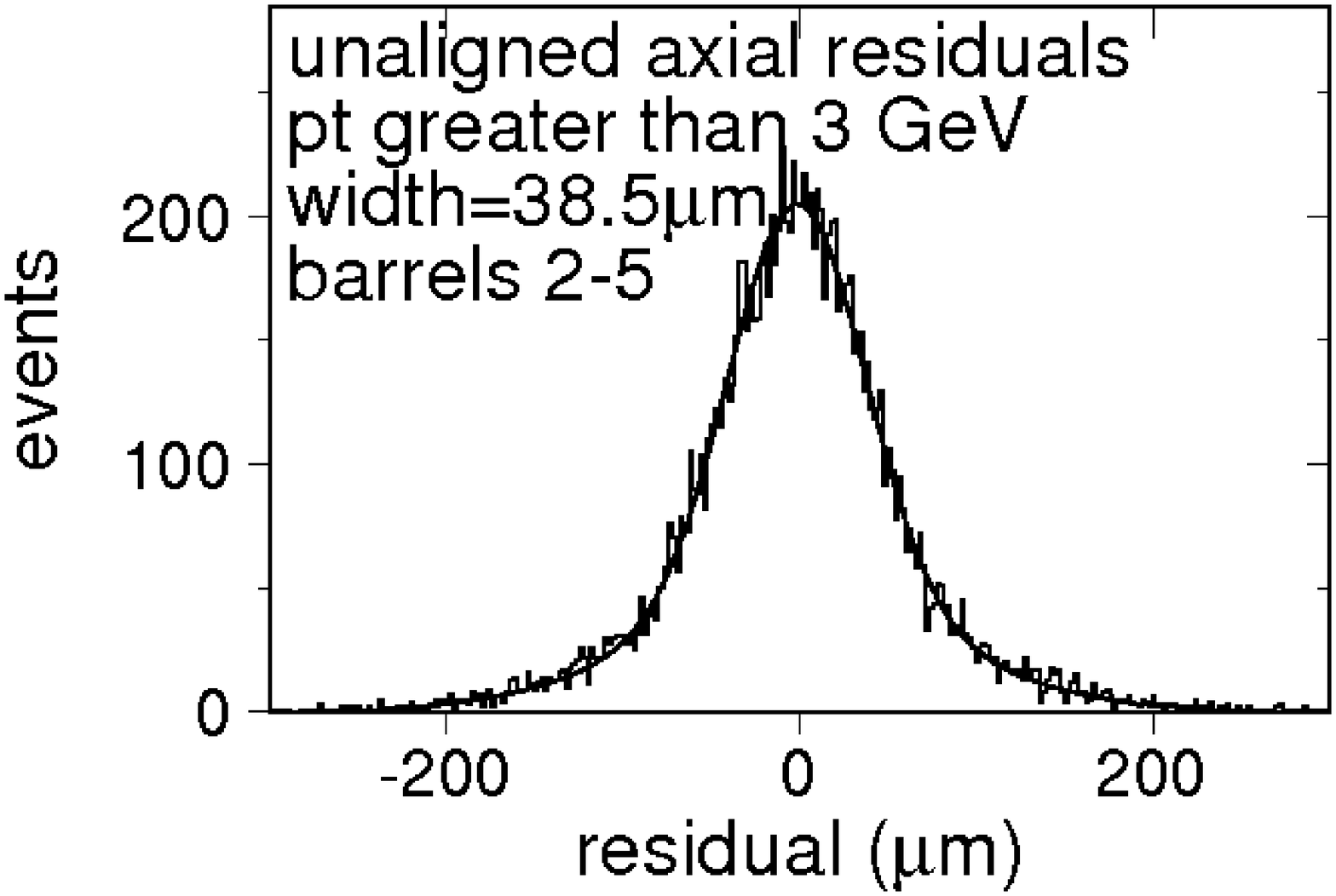} \hfill
\includegraphics[width=0.49\columnwidth,height=7cm]{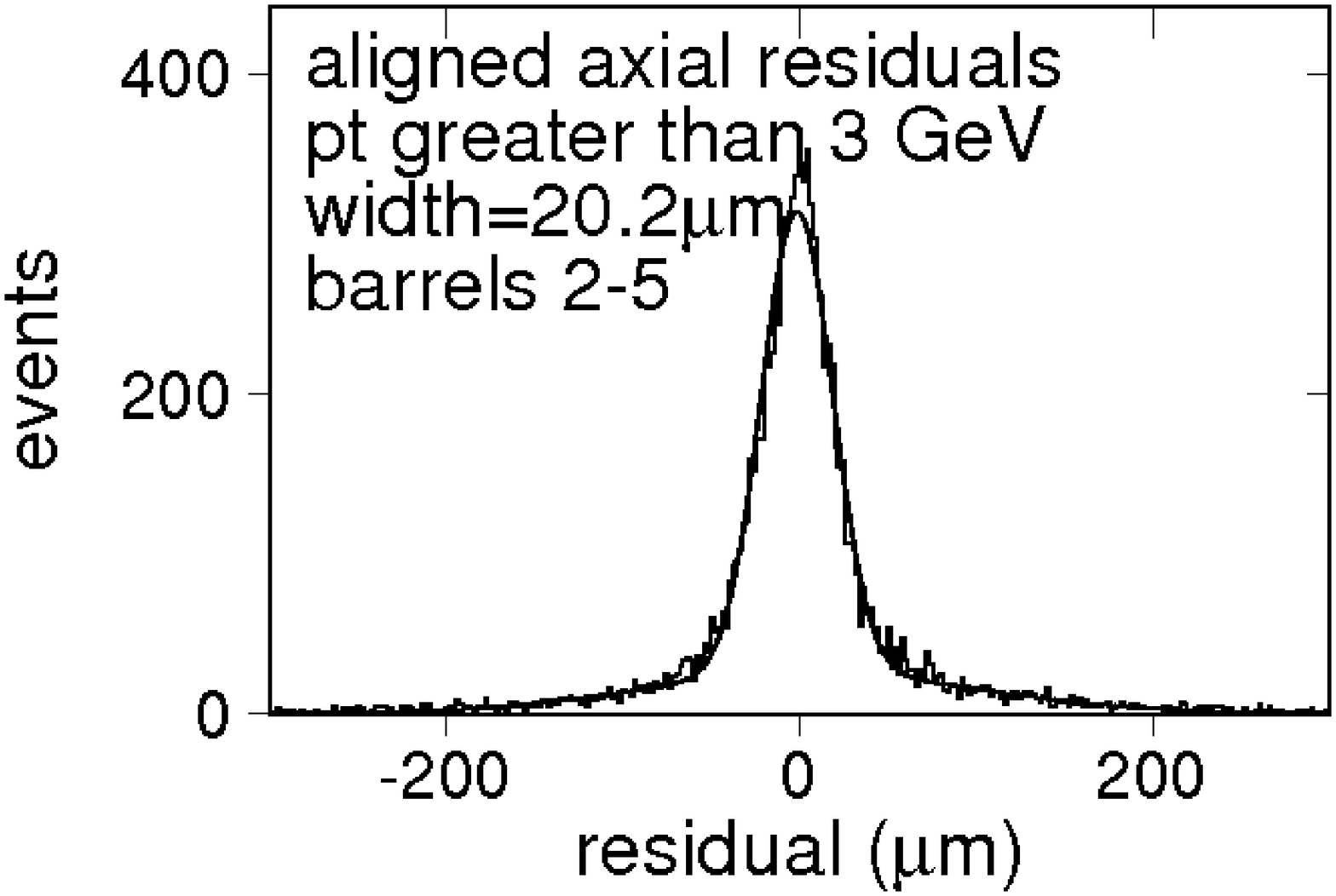}   \\
\includegraphics[width=0.49\columnwidth]{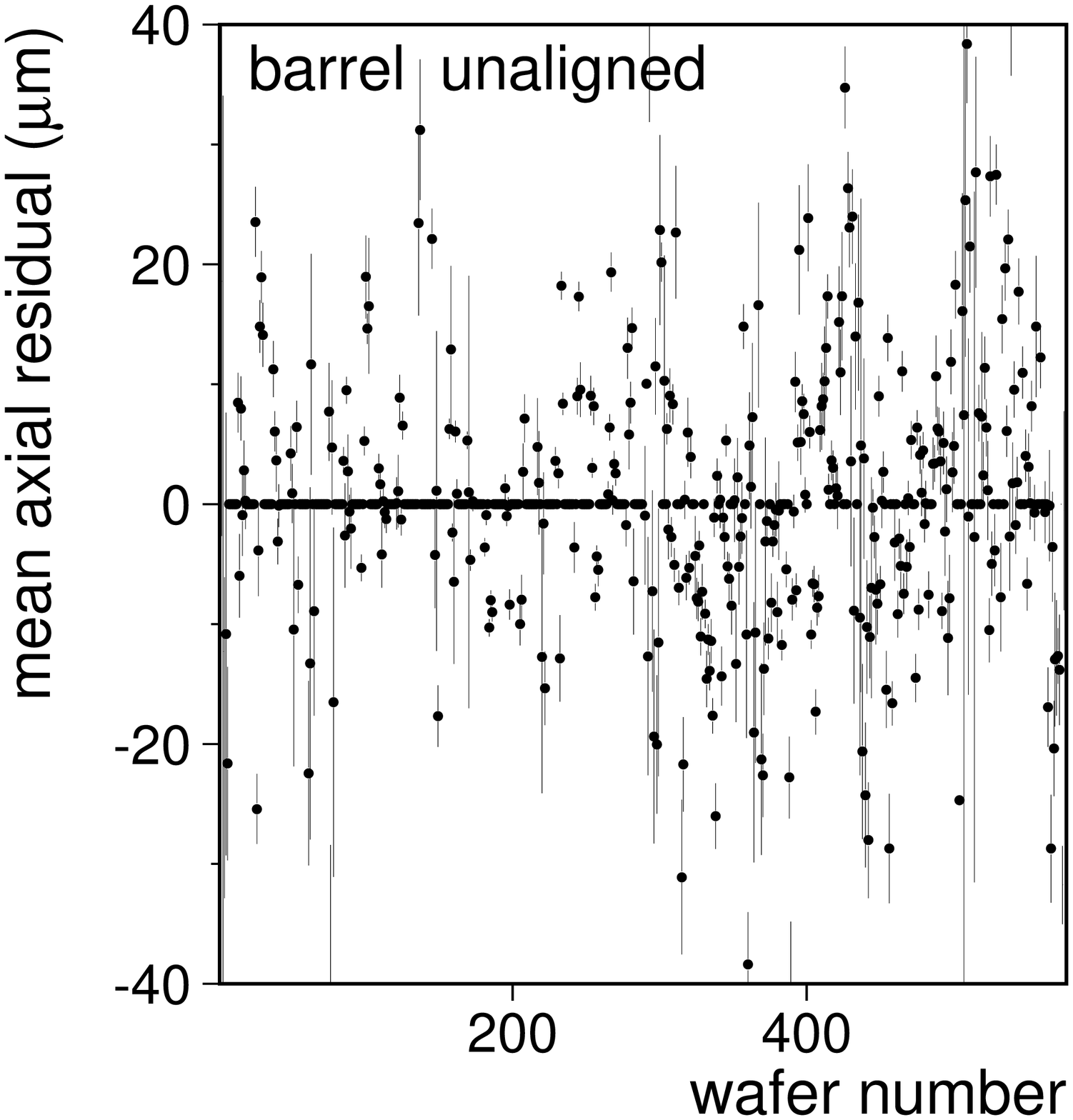}\hfill
\includegraphics[width=0.49\columnwidth]{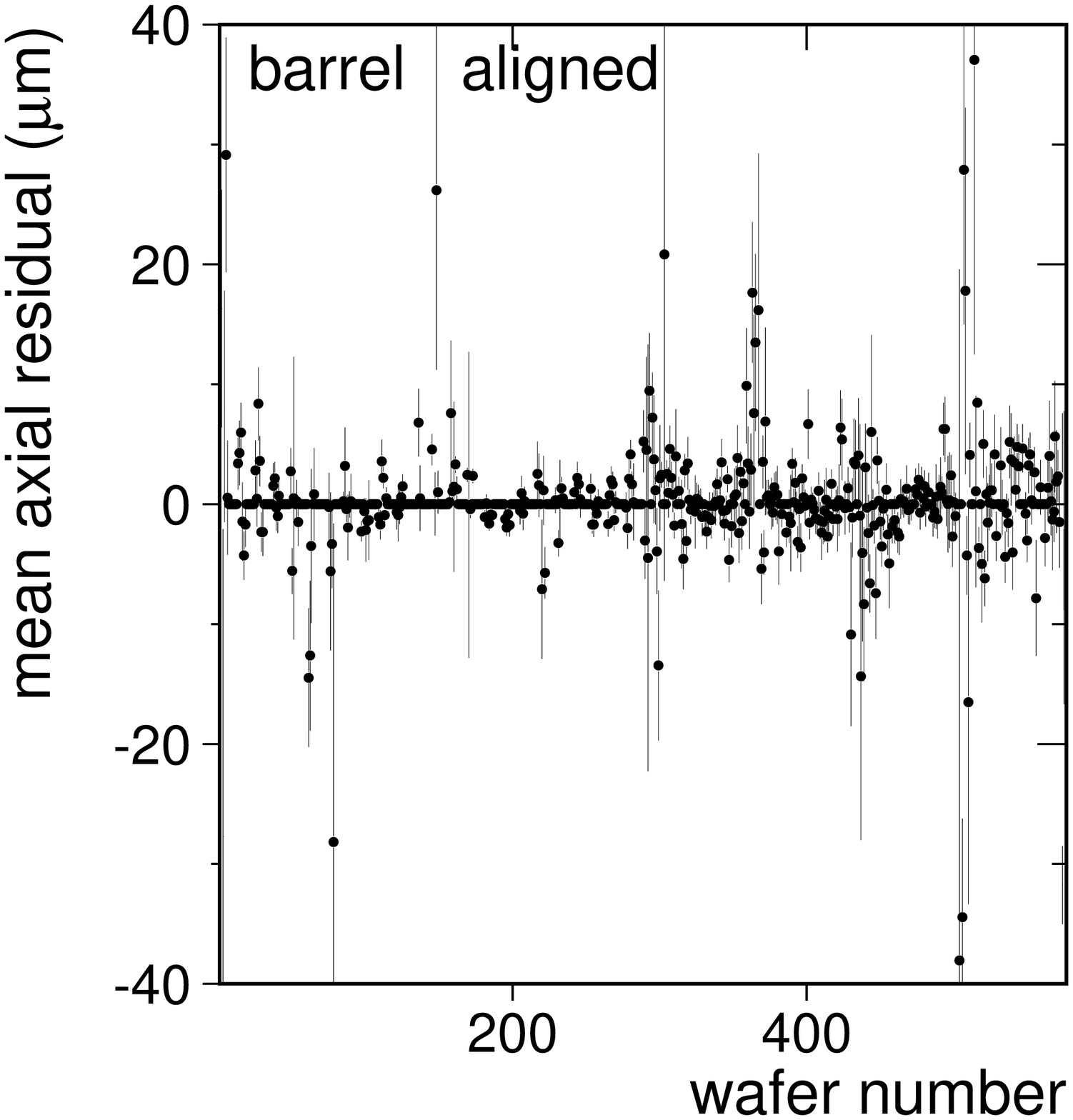}
\vspace*{-8mm}
\caption{\label{fig:barrel}
Barrel axial residuals. 
Upper left:  unaligned for all wafers. 
Upper right: aligned for all wafers.
Lower left: unaligned for individual wafers.
Lower right: aligned for individual wafers.
}
\end{figure}

\begin{figure}[cp]
\centering
\includegraphics[width=0.49\columnwidth,height=7cm]{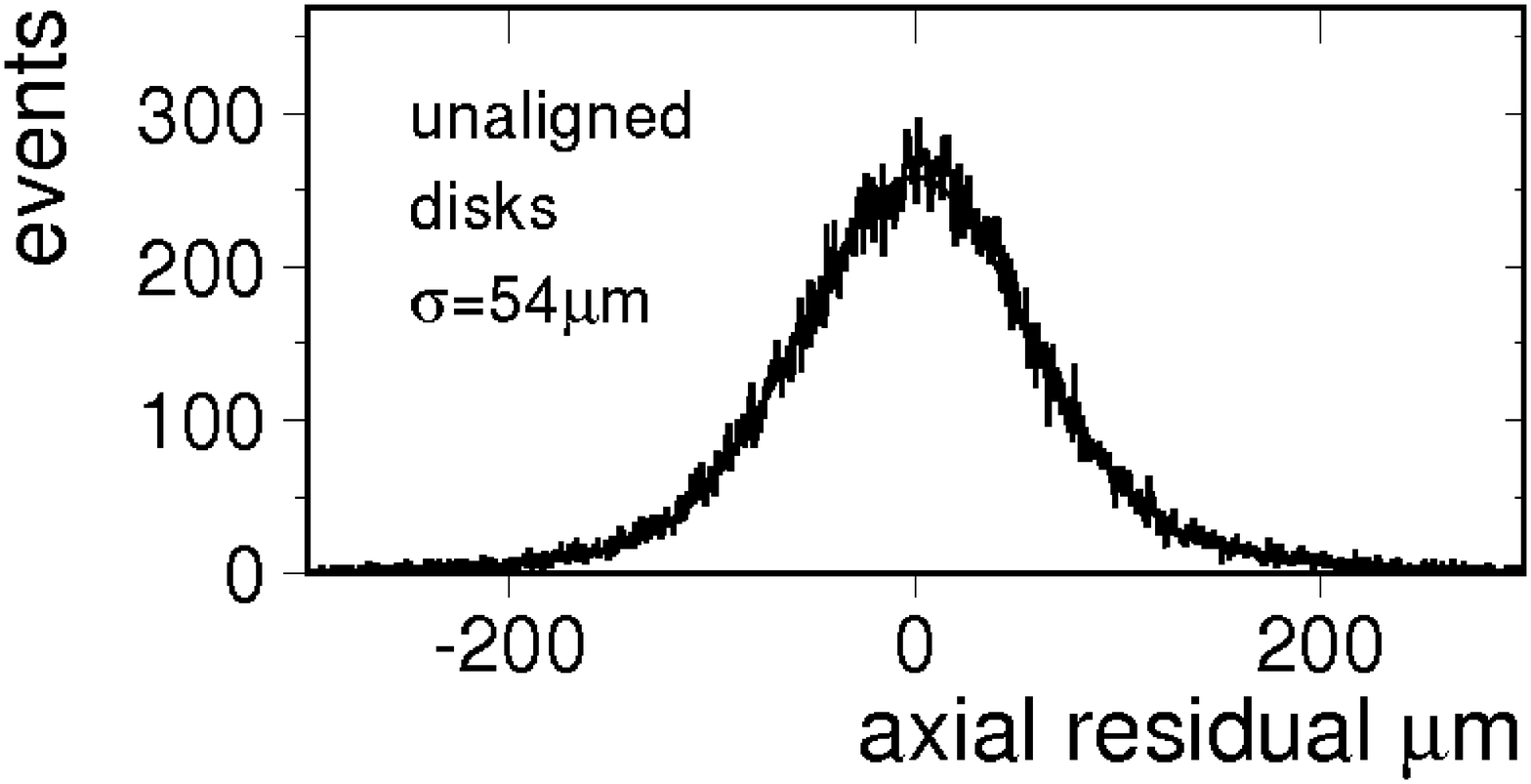} \hfill
\includegraphics[width=0.49\columnwidth,height=7cm]{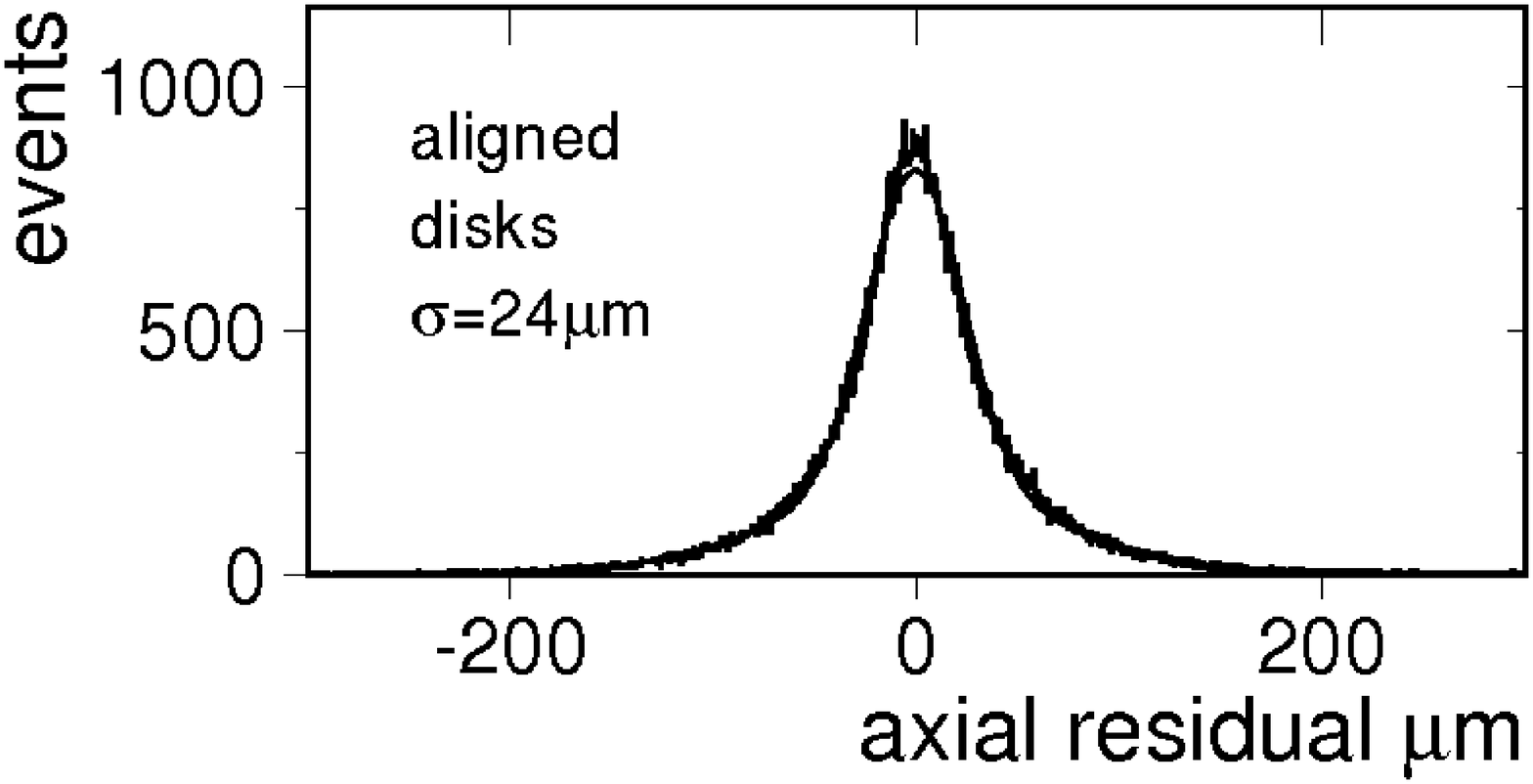}   \\
\includegraphics[width=0.49\columnwidth]{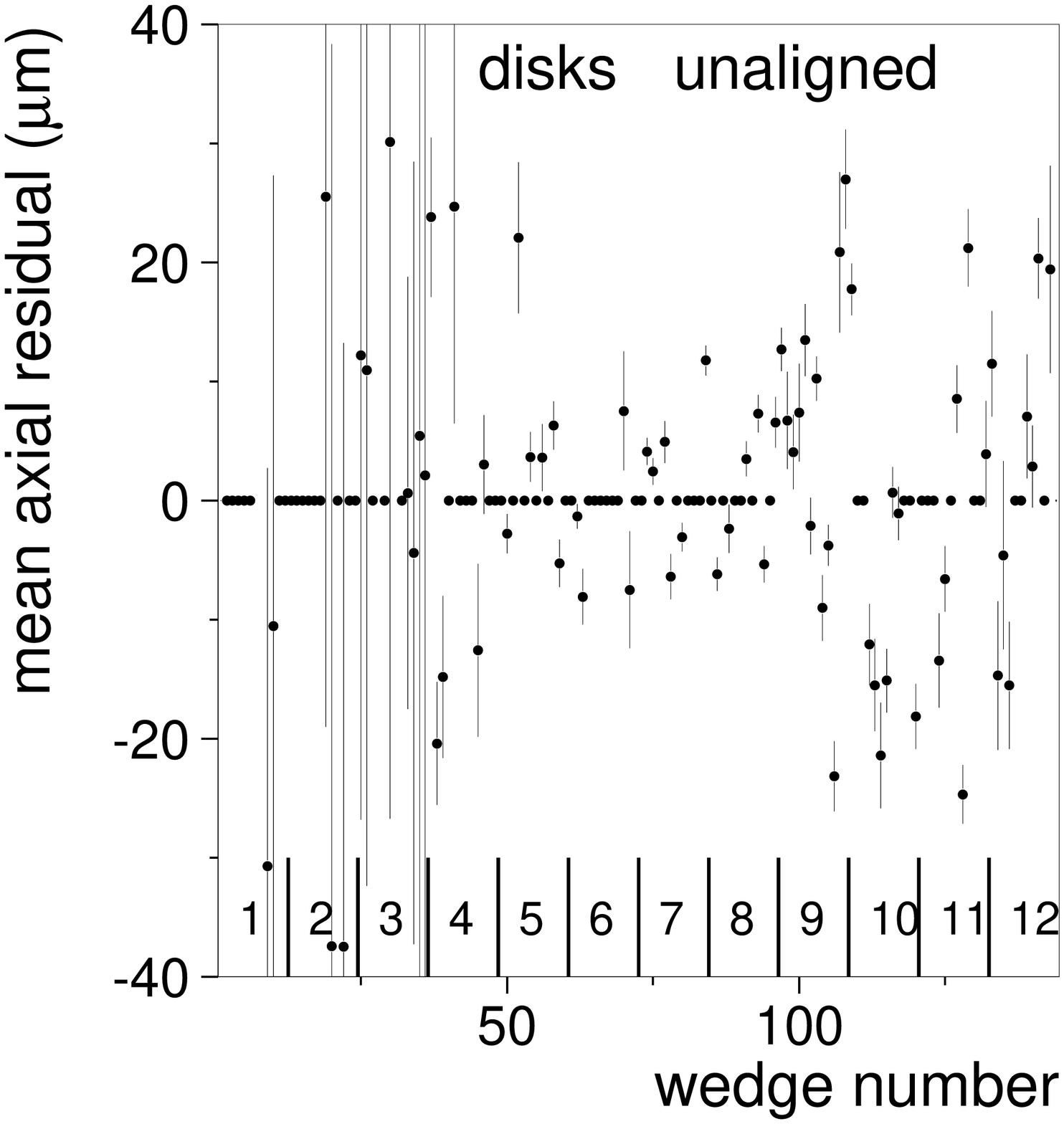}\hfill
\includegraphics[width=0.49\columnwidth]{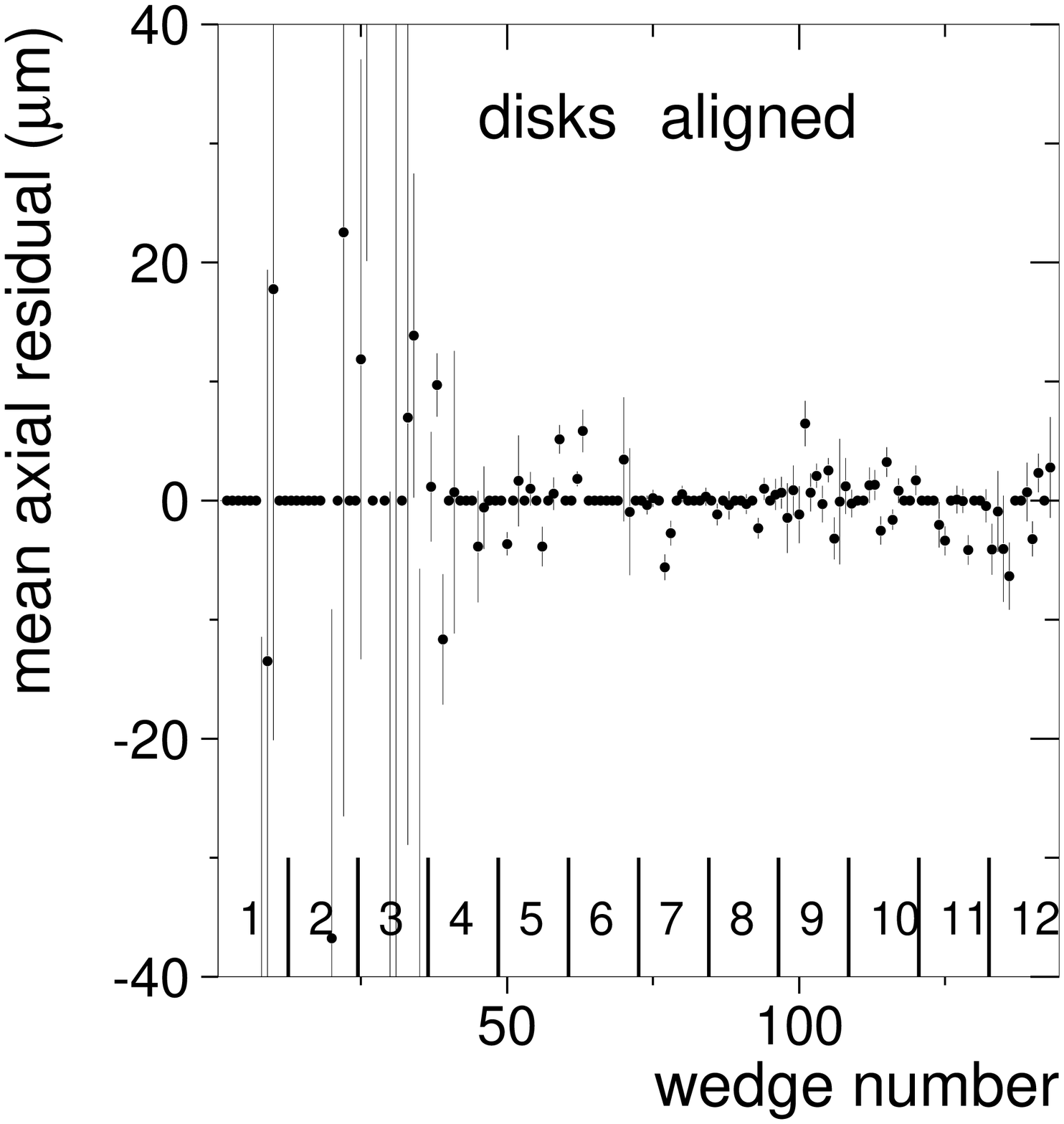}
\vspace*{-7mm}
\caption{\label{fig:fdisk}
F-disk axial residuals. 
Upper left: unaligned for all wedges. 
Upper right: aligned for all wedges.
Lower left: unaligned for individual wedges. 
Lower right: aligned for individual wedges.
Each of the 12 indicated disks contain 12 wedges.
}
\end{figure}

\begin{figure}[cp]
\centering
\includegraphics[width=0.49\columnwidth,height=7cm]{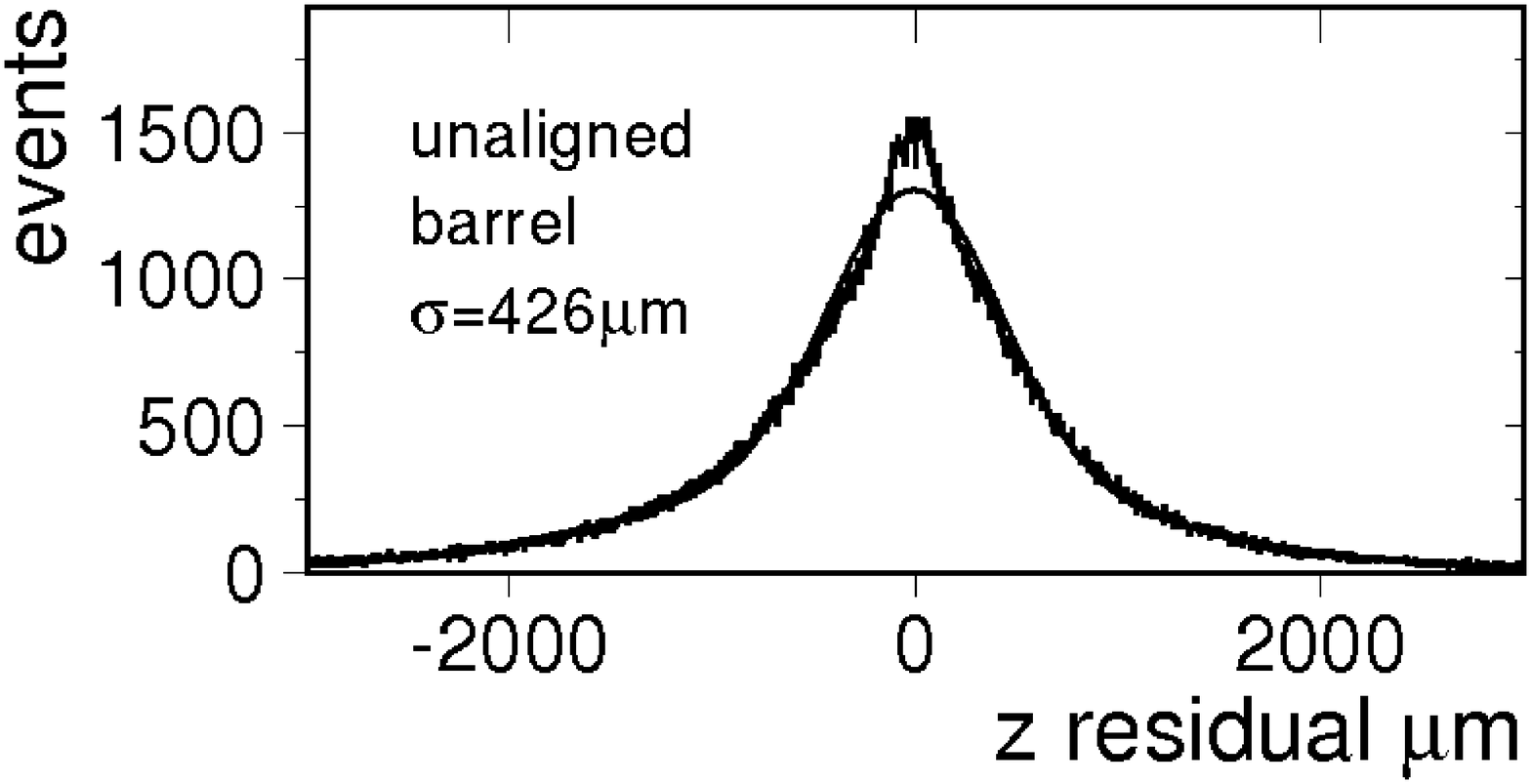} \hfill
\includegraphics[width=0.49\columnwidth,height=7cm]{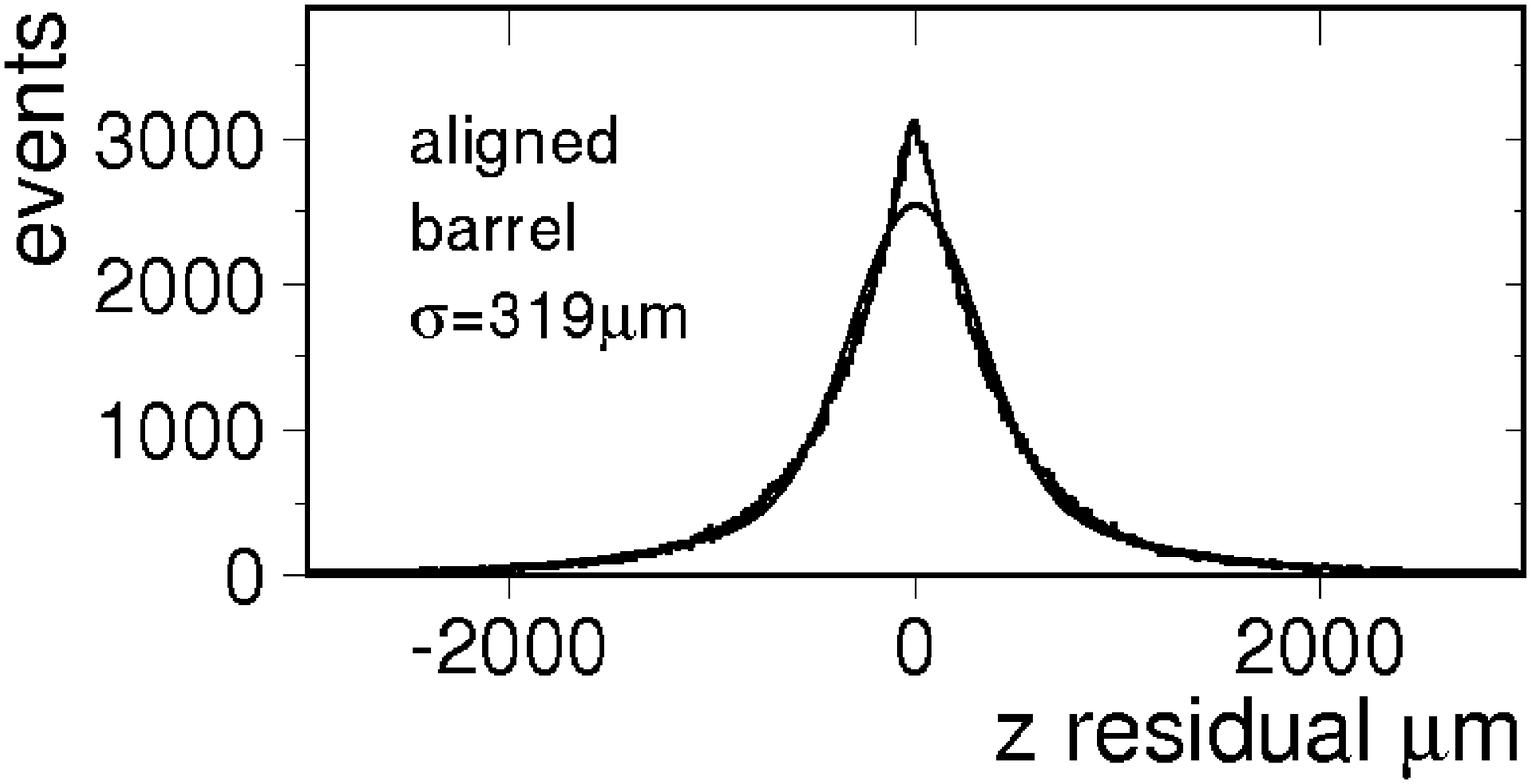}   \\
\includegraphics[width=0.49\columnwidth]{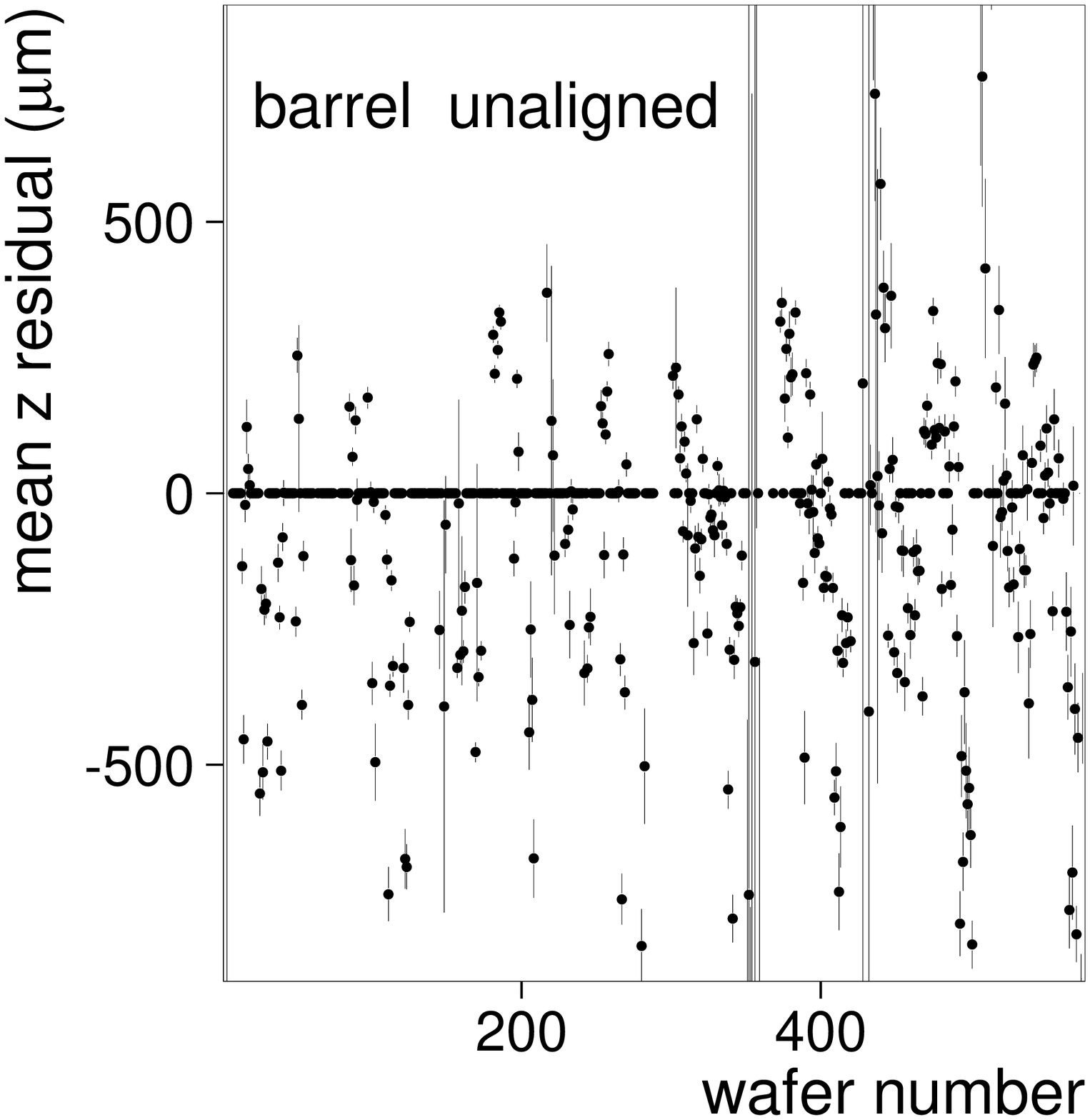}\hfill
\includegraphics[width=0.49\columnwidth]{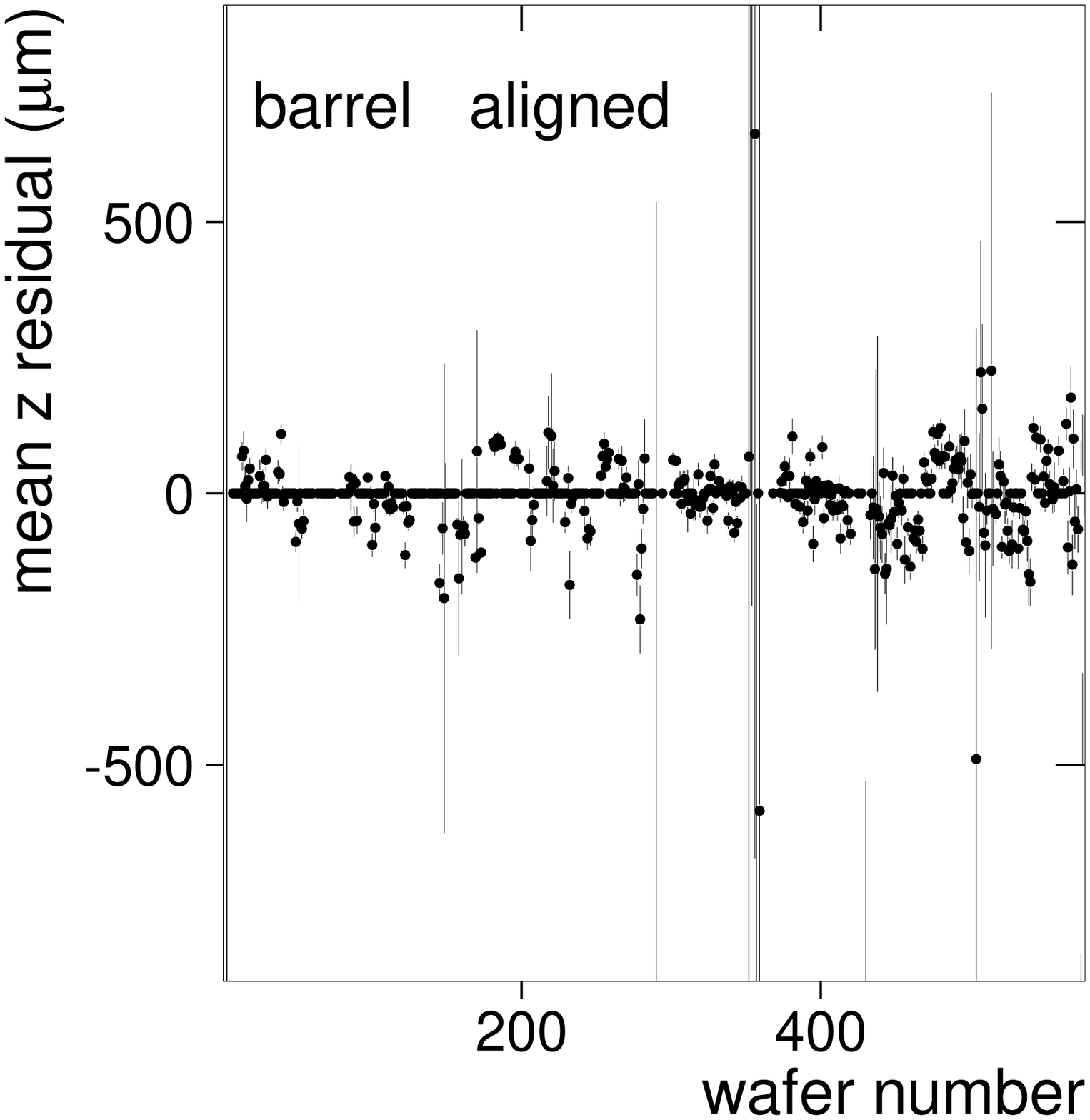}
\vspace*{-7mm}
\caption{\label{fig:barrel-z}
Barrel residuals in the z-direction. 
Upper left:  unaligned for all wafers. 
Upper right: aligned for all wafers.
Lower left: unaligned for individual wafers.
Lower right: aligned for individual wafers.
}
\end{figure}

\begin{figure}[h!]
\centering
\includegraphics[width=0.49\columnwidth,height=7cm]{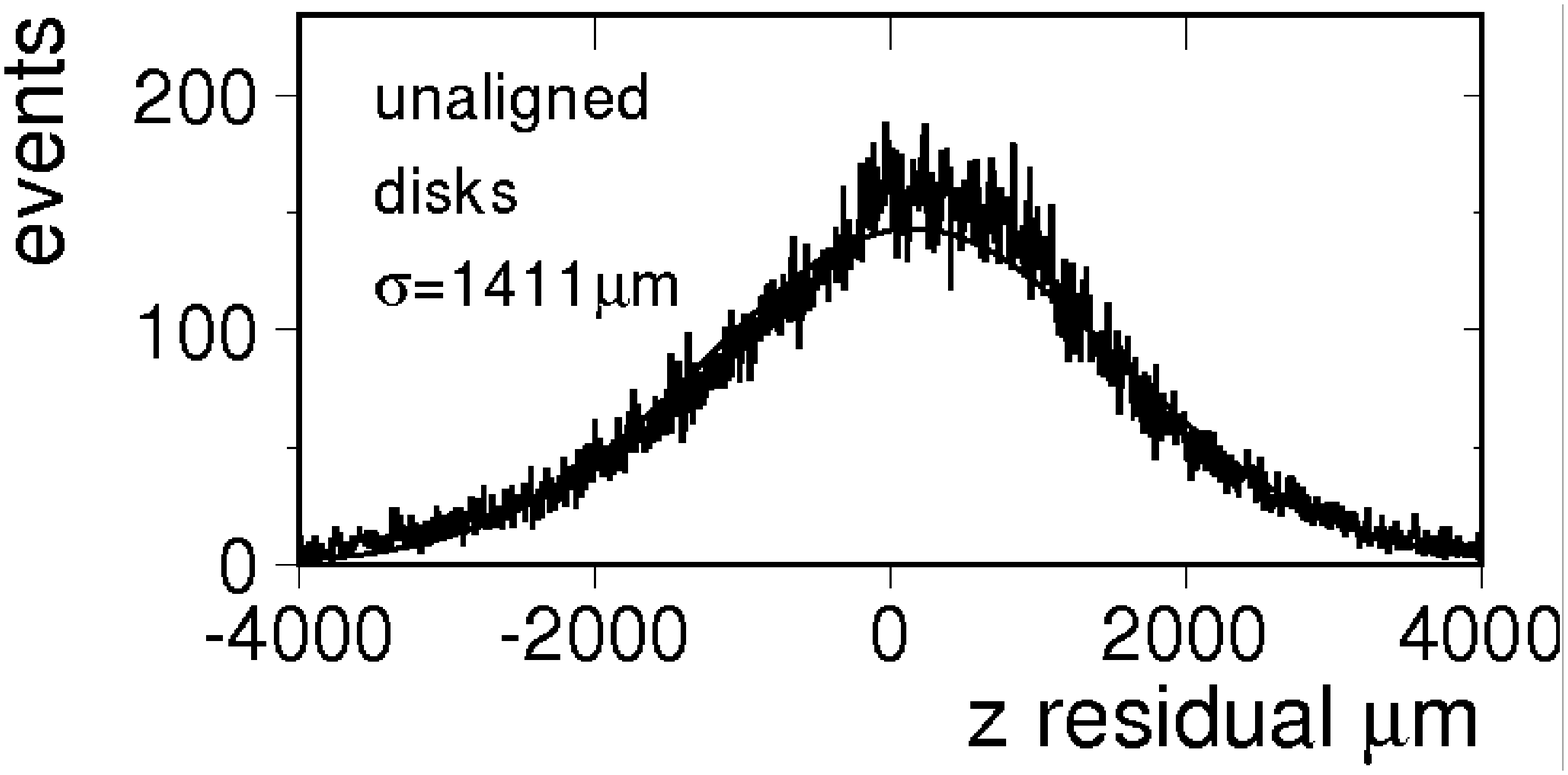}\hfill
\includegraphics[width=0.49\columnwidth,height=7cm]{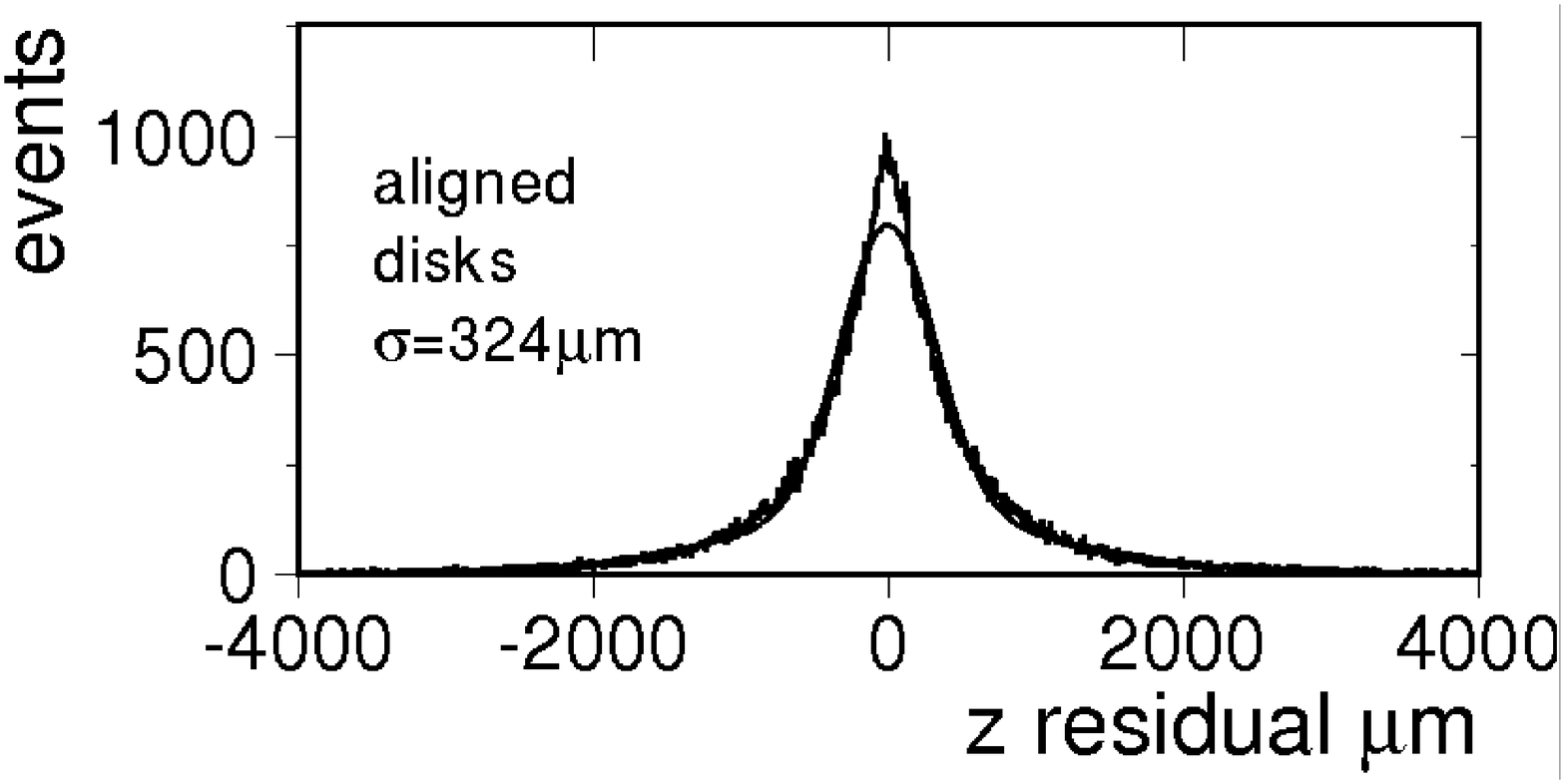} \\
\includegraphics[width=0.49\columnwidth]{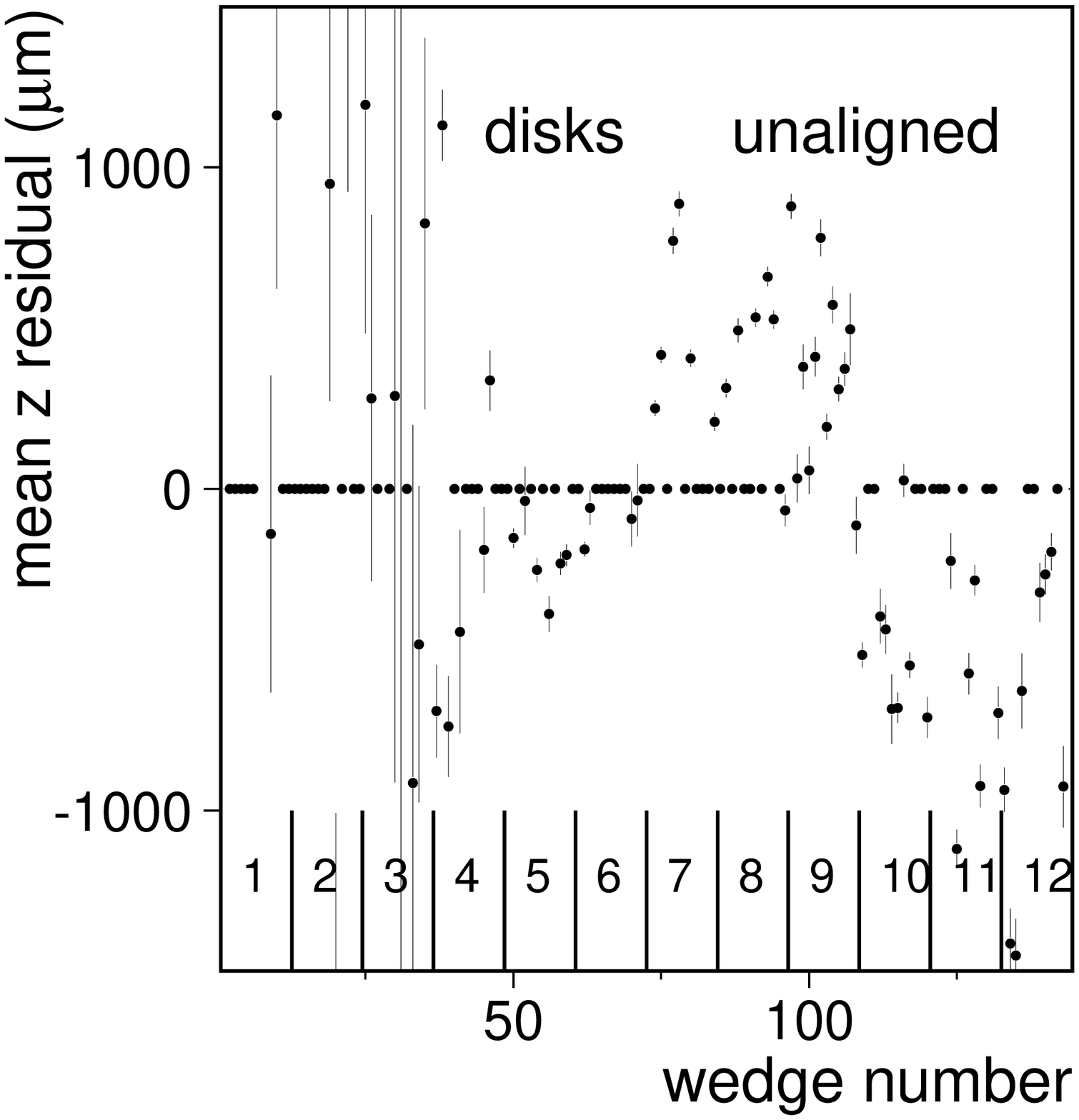} \hfill
\includegraphics[width=0.49\columnwidth]{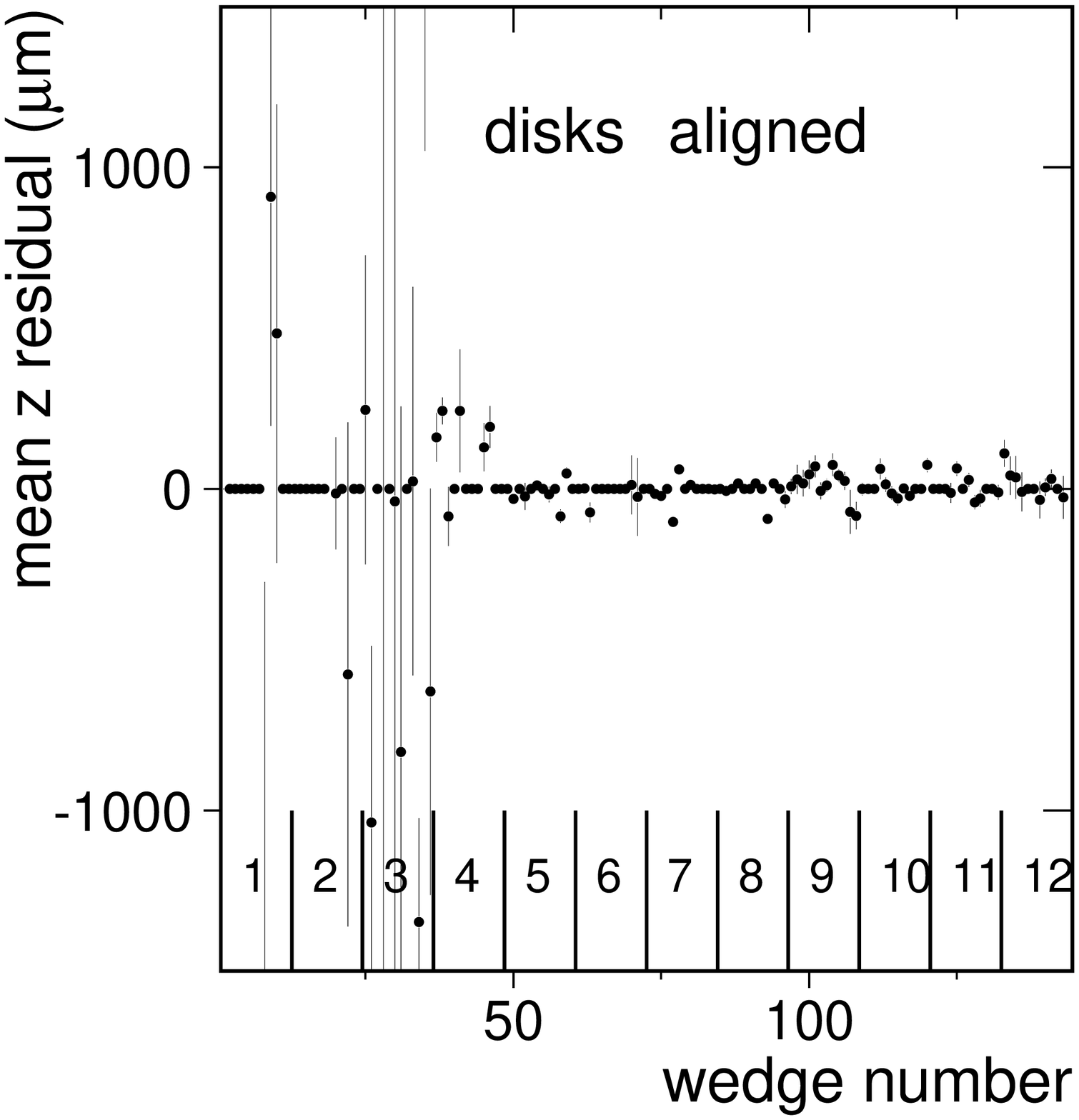}
\vspace*{-8mm}
\caption{\label{fig:fdisk-z}
F-disk residuals in z-direction. 
Upper left: unaligned for all wedges. 
Upper right: unaligned for individual wedges.
Lower left: unaligned for all wedges. 
Lower right: aligned for individual wedges.
Each of the 12 indicated disks contain 12 wedges.
}
\end{figure}

For the barrel, the axial residuals before and after alignment are shown for all wafers, and for
each individual wafer (Fig.~\ref{fig:barrel}). Figure~\ref{fig:fdisk} shows the corresponding
plots for the F-disks. The residuals in the z-direction are given in Figs.~\ref{fig:barrel-z}
and~\ref{fig:fdisk-z}.


\clearpage

\section{Optimization of convergence}
\begin{figure}[h!]
\centering
\includegraphics[width=0.45\columnwidth]{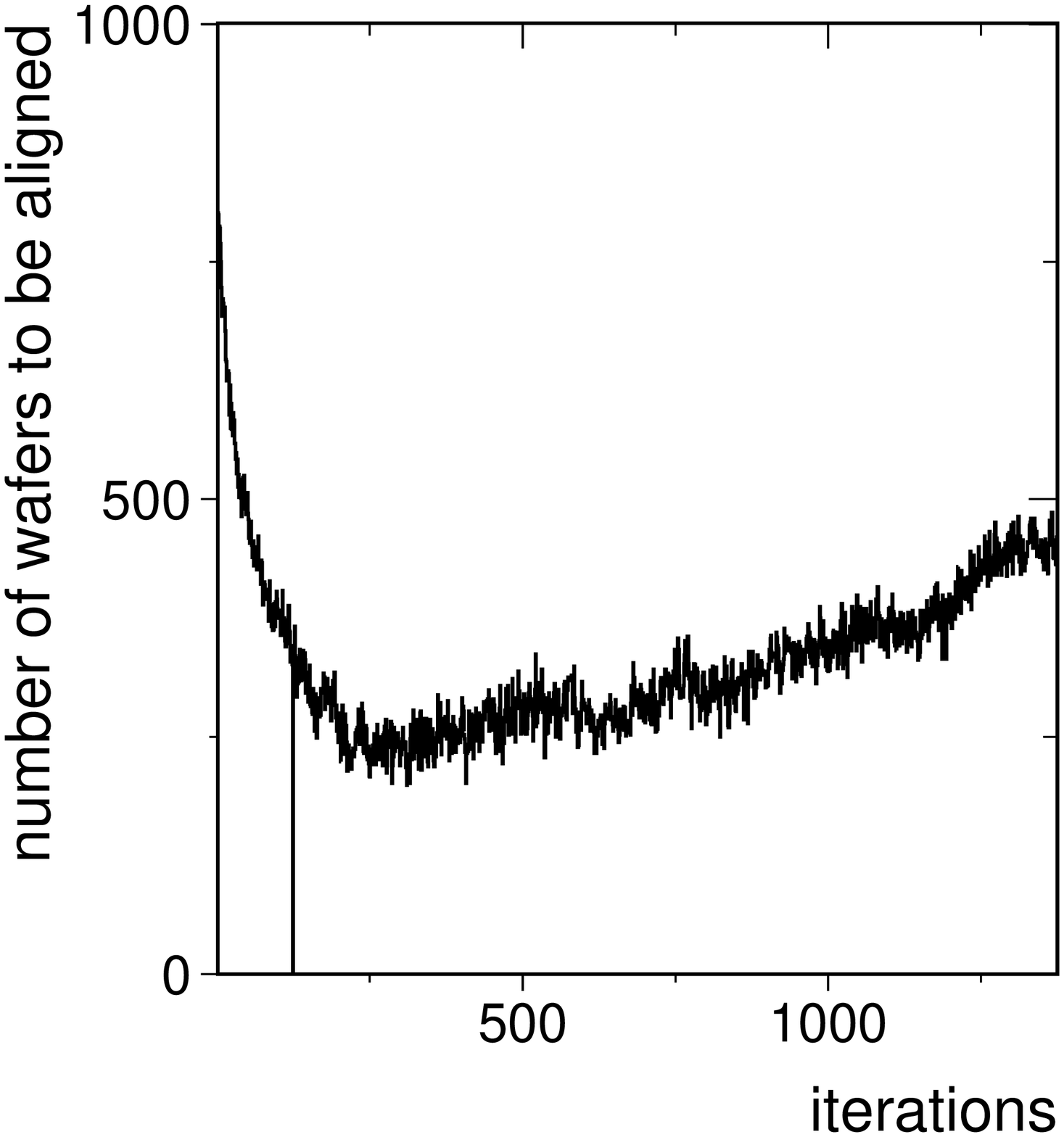} \hfill
\includegraphics[width=0.45\columnwidth]{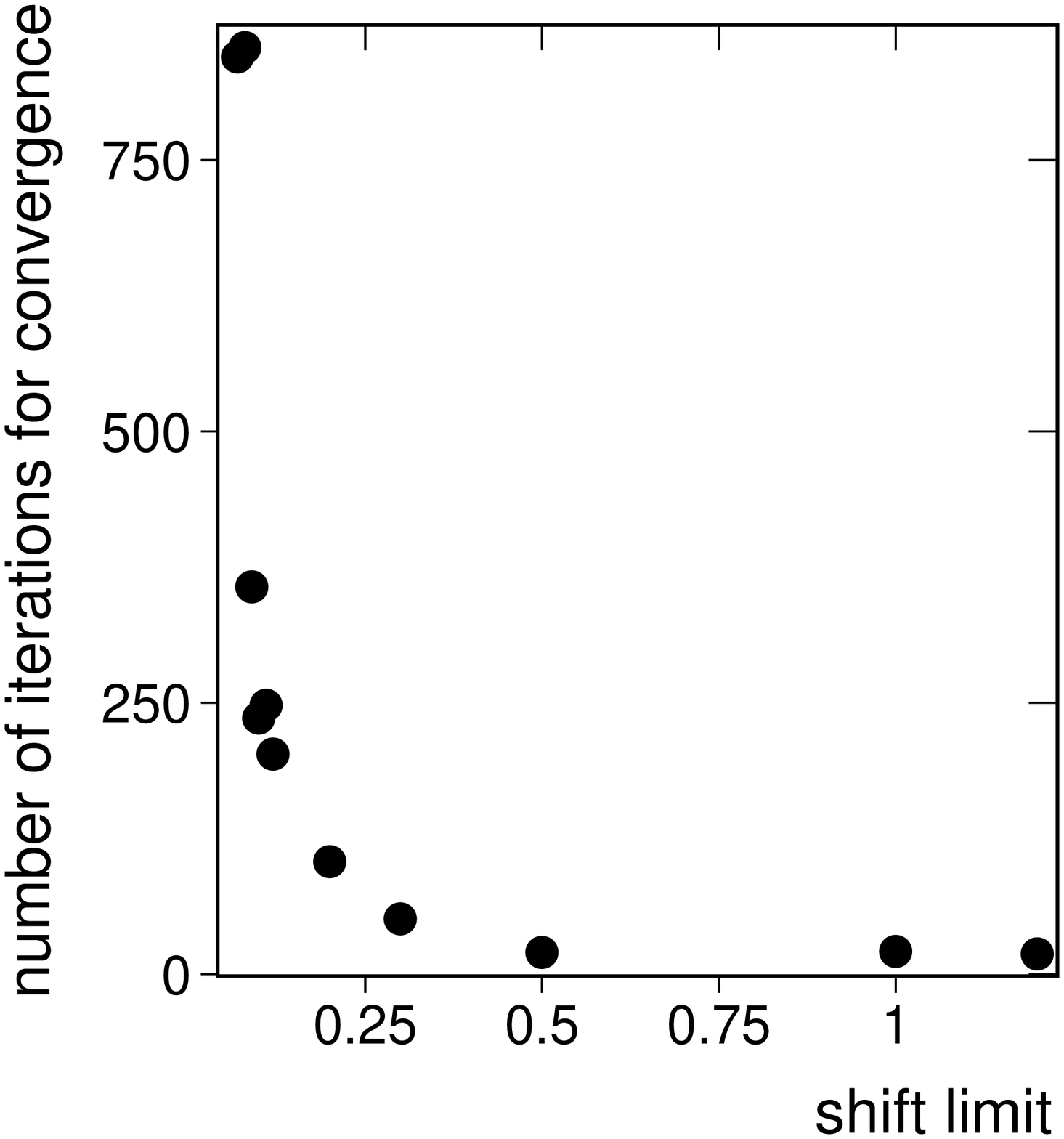}
\vspace*{-6mm}
\caption{\label{fig:conv}
Convergence of iterative process. 
Left: number of wafers to be aligned for shift limit 0.05 as a function of the number of iterations. 
The alignment process does not converge.
Right: number of iterations required for convergence as a function of the shift limit
in the range 0.07-1.2.
}
\end{figure}

In order to improve the alignment process, the convergence speed, the accuracy and 
the dependence on the number of input events have been studied. Figure~\ref{fig:conv}
shows the number of wafers to be aligned as a function of the iteration number for a 
shift limit of 0.05. For this shift limit value no convergence is obtained. The required 
numbers of iterations for convergence with larger shift limits is also shown.

\begin{figure}[h!]
\centering
\includegraphics[width=0.45\columnwidth]{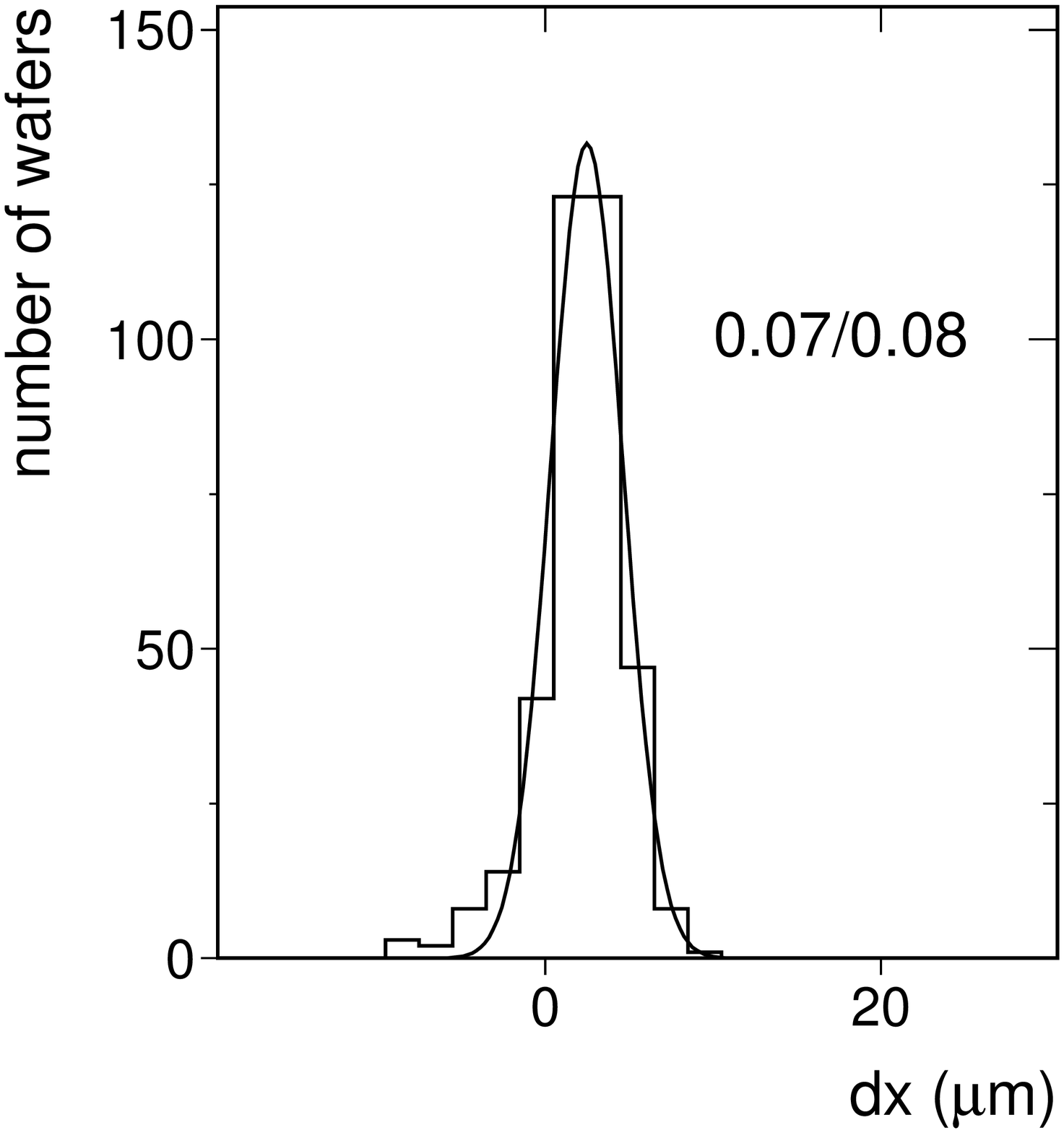} \hfill
\includegraphics[width=0.45\columnwidth]{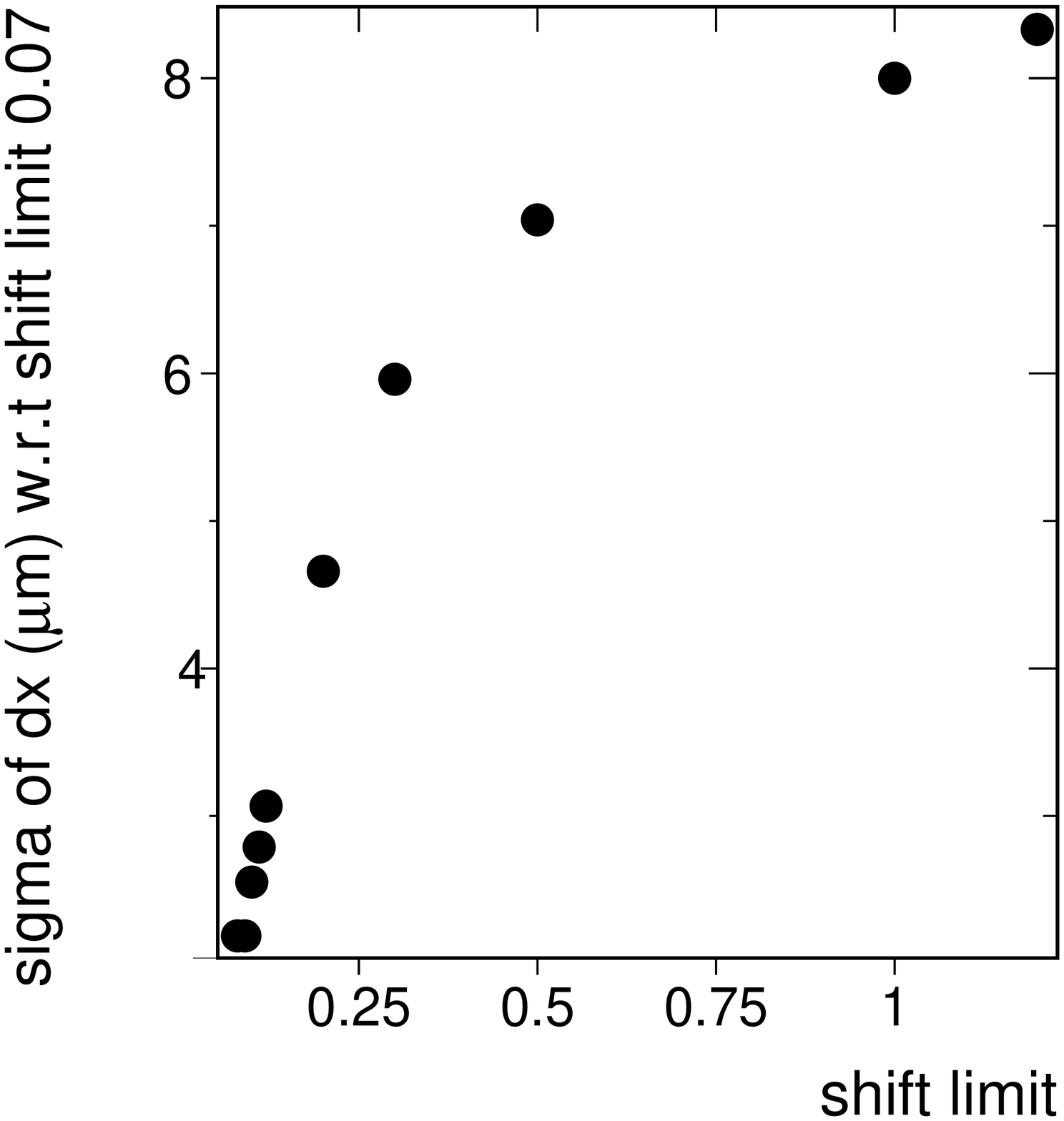}
\vspace*{-6mm}
\caption{\label{fig:shiftlimit}
Barrel difference in x-direction between aligned geometries produced with different 
shift-limits. 
Left: all wafers combined between shift limit 0.07 and 0.08. 
Right: as a function of the shift limit w.r.t. 0.07.
}
\end{figure}

\begin{figure}[tp]
\centering
\includegraphics[width=0.45\columnwidth]{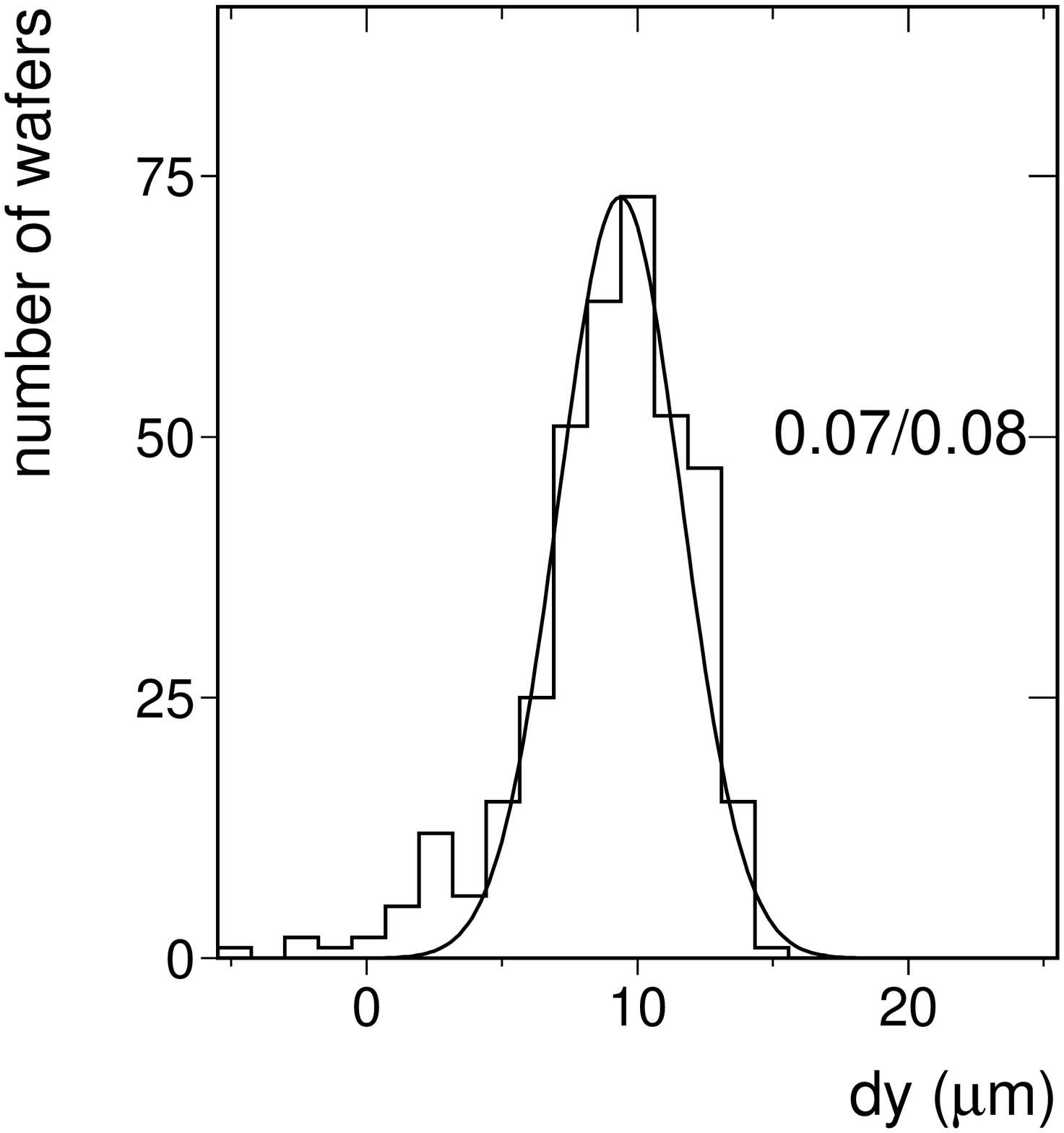}  \hfill
\includegraphics[width=0.45\columnwidth,height=8cm]{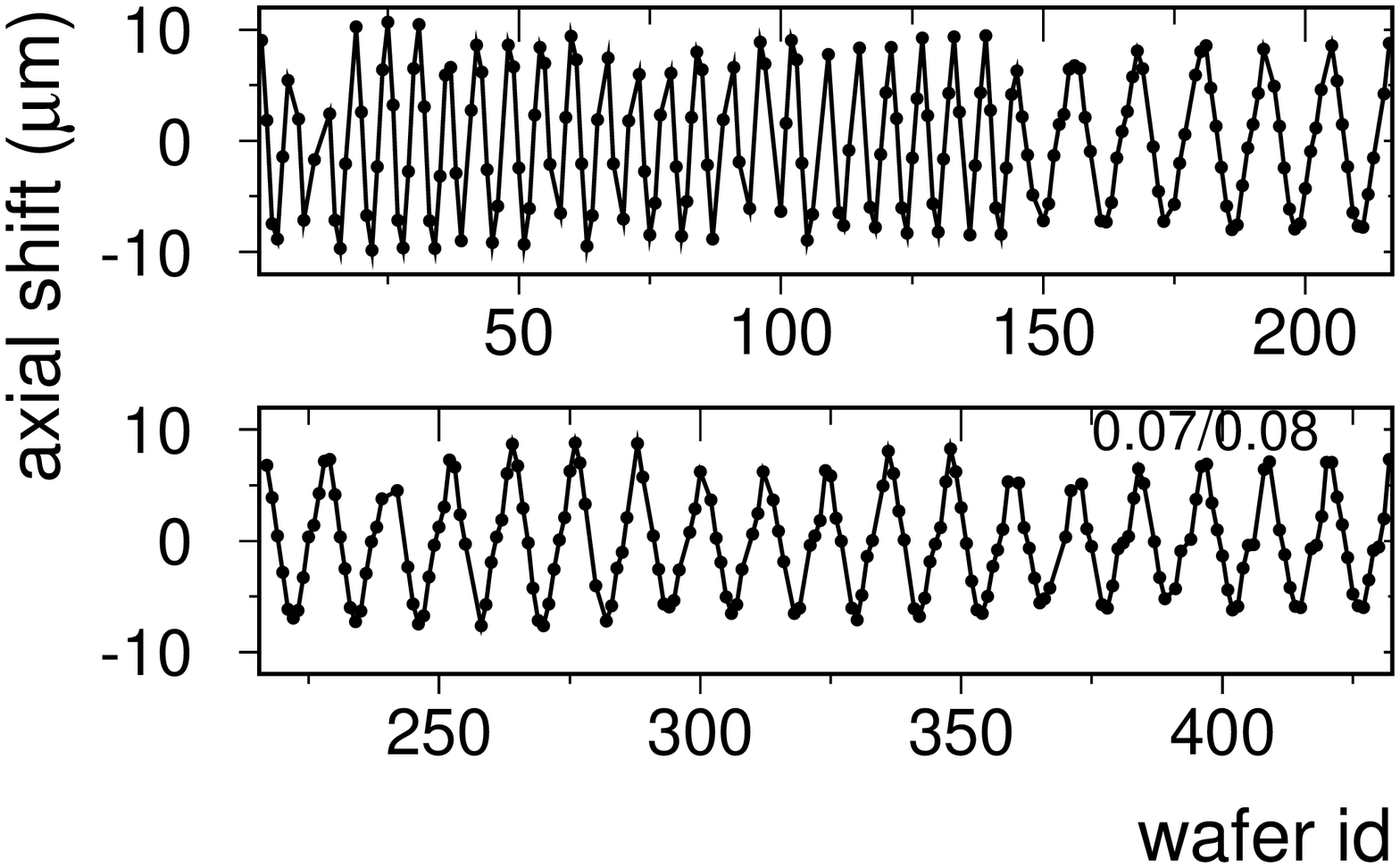}
\vspace*{-6mm}
\caption{\label{fig:shift_geo}
Left: barrel difference in y-direction between aligned geometries produced with different 
shift-limits. 
Right: corresponding axial shift. There are 6 wafers per inner layers, and 12 per 
outer layers beyond wafer id 144.
}
\end{figure}

The variation between two aligned geometries for the same data and two different shift limits
has been studied. 
The differences in x and y-directions between two barrel geometries for one geometry produced 
for shift limit 0.07, and the second one for 0.08 are shown in Figs.~\ref{fig:shiftlimit}
and~\ref{fig:shift_geo}.
The differences in x-direction between geometries produced with different shift limits 
are also shown in Fig.~\ref{fig:shiftlimit} as a function of the shift limit.
For small shift limits the variation is below $3\mu$m.
While a very good precision of the relative wafer positions is obtained,
a larger shift of the entire SMT position is possible even for a small variation of the shift 
limit parameter.
The example in Fig.~\ref{fig:shift_geo} shows a relative shift in y-direction of the SMT 
wafers with $\sigma = 2.3\mu$m, a very similar value as in the x-direction. However, the entire
SMT is shifted between the two geometries by about 9$\mu$m.
This corresponds to an oscillating behaviour of the axial shift as a function of the wafer id.


\clearpage

\begin{figure}[h!]
\centering
\includegraphics[width=0.45\columnwidth]{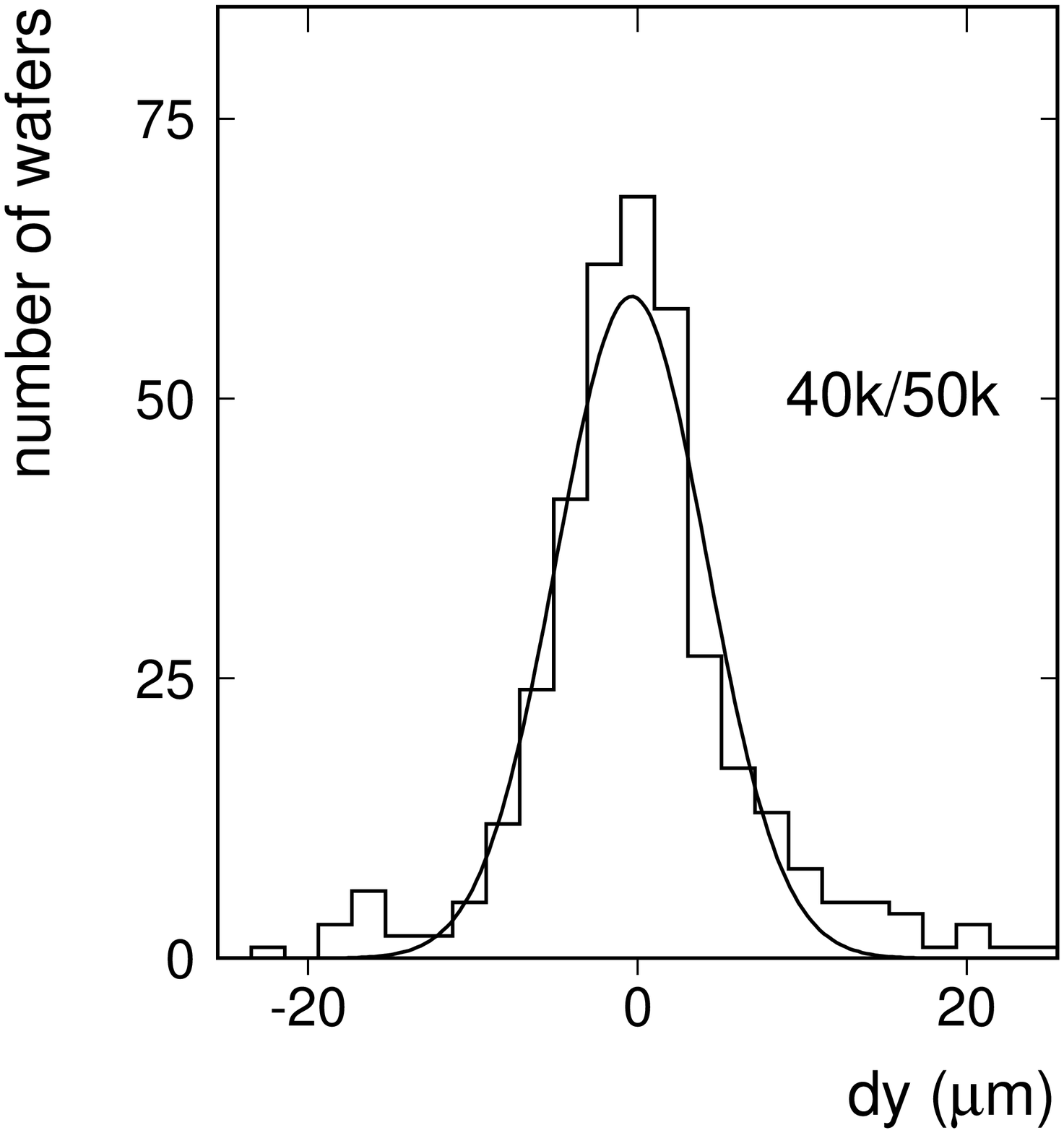} \hfill
\includegraphics[width=0.45\columnwidth]{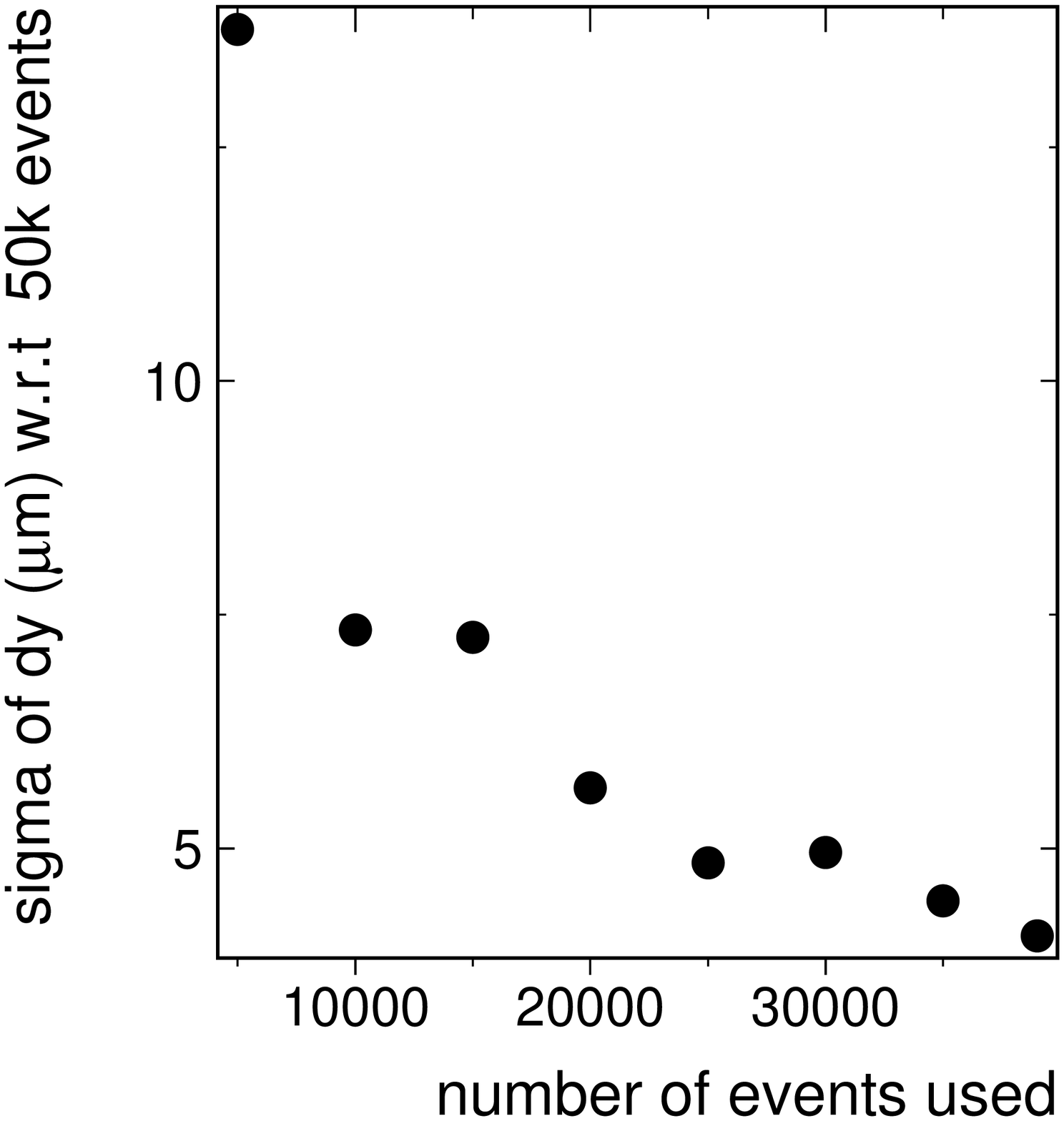}
\vspace*{-6mm}
\caption{\label{fig:number}
Barrel difference in y-direction between aligned geometries produced with different 
shift-limits. 
Left: all wafers combined between 40,000 and 50,000 data input events. 
Right: as a function of the number of input events w.r.t. 50,000 events.
}
\end{figure}

\begin{figure}[h!]
\centering
\includegraphics[width=0.45\columnwidth]{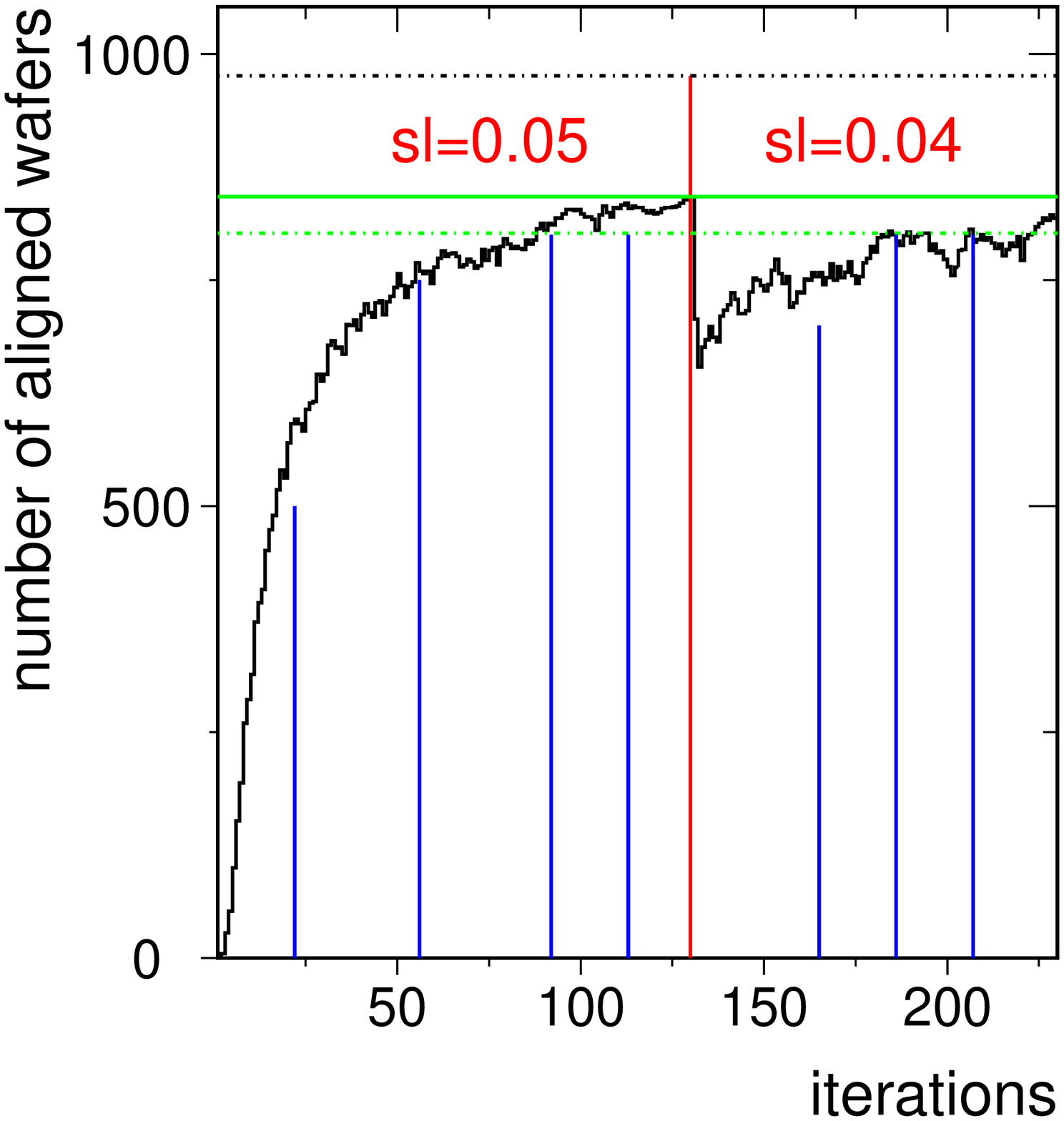} \hfill
\includegraphics[width=0.45\columnwidth]{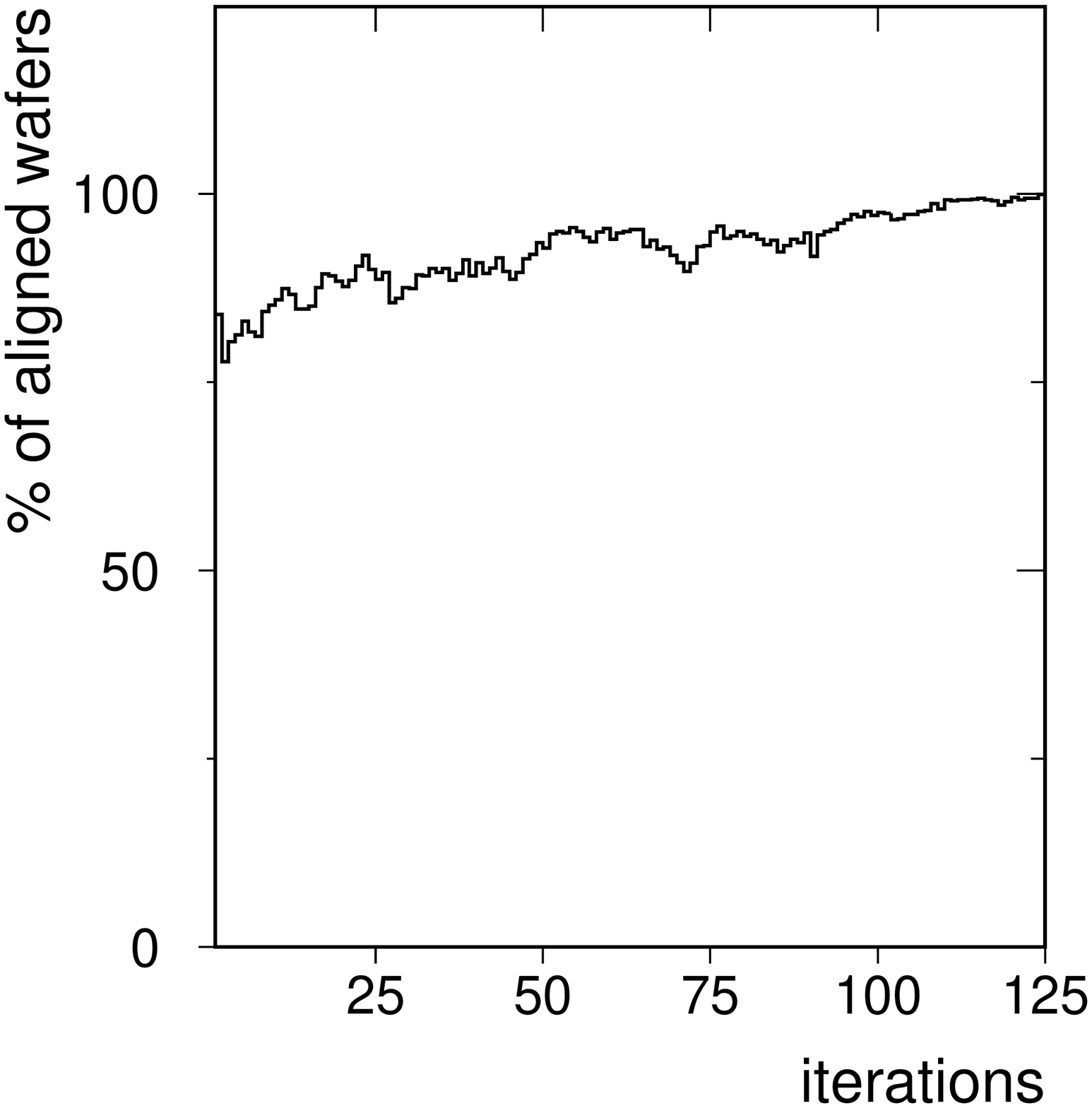}
\vspace*{-6mm}
\caption{\label{fig:second_step_conv}
Convergence of iterative process for two additional alignment steps,
starting with the aligned geometry for shift limit 0.07.
Left: number of aligned wafers for shift limit 0.05 and 0.04.
Vertical lines correspond to 72h CPU time.
Right: number of iterations required for shift limit 0.04 (in \%) w.r.t. the converged 0.05 alignment.
}
\end{figure}

In addition, the dependence of the wafer positions on the number of input 
events has been studied (Fig.~\ref{fig:number}). 
Variations of the wafer positions in the aligned geometries below 
$5\mu$m are expected for more than 30,000 data input events.
In a first step the shift limit has been reduced from 0.07 to 0.05, and
in a further step it has been reduced from 0.05 to 0.04. Convergence
was achieved by using the aligned 0.05 geometry as starting geometry for the 0.04 run.
The convergence is illustrated in Fig.~\ref{fig:second_step_conv}.

\clearpage

\section{Uncertainties from procedure variations}
\begin{figure}[h!]
\begin{center}
\vspace*{-0.1cm}
\includegraphics[width=0.45\columnwidth]{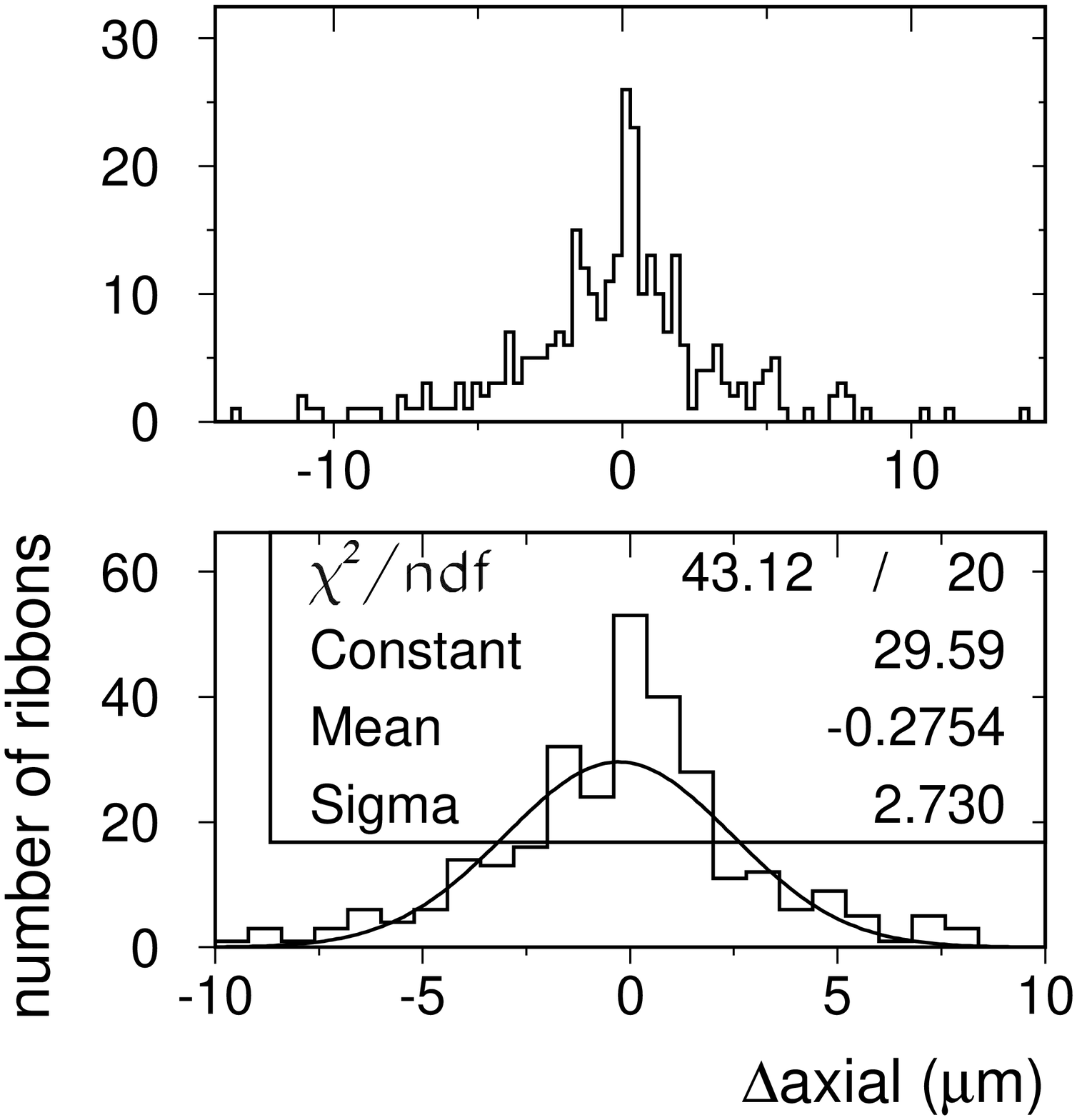} \hfill
\includegraphics[width=0.45\columnwidth]{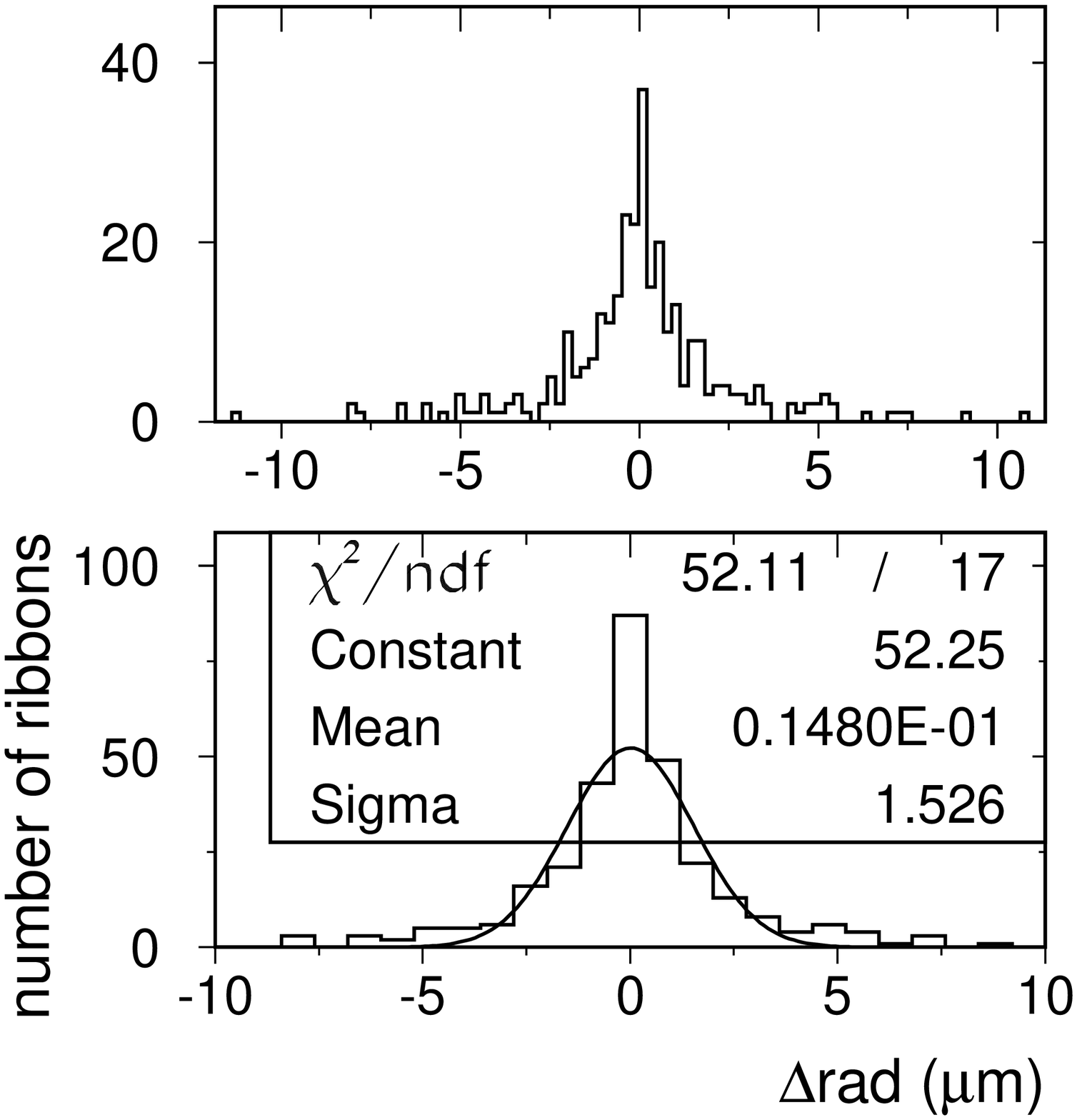}
\end{center}
\vspace*{-10mm}
\caption{\label{fig:procedure}
Differences between geometries as explained in the text.
}
\vspace*{2mm}
\end{figure}

In order to determine the uncertainty in the alignment procedure
the CFT geometries are compared for two cases 
a) when SMT and CFT were aligned simultaneously, and 
b) when the SMT was aligned first, and then the CFT was aligned.
No significant effect on the alignment was observed (Fig.~\ref{fig:procedure}).

\section{Single wafer re-alignment precision}
\begin{figure}[h!]
\begin{center}
\vspace*{-0.2cm}
\includegraphics[width=0.5\columnwidth,height=8cm]{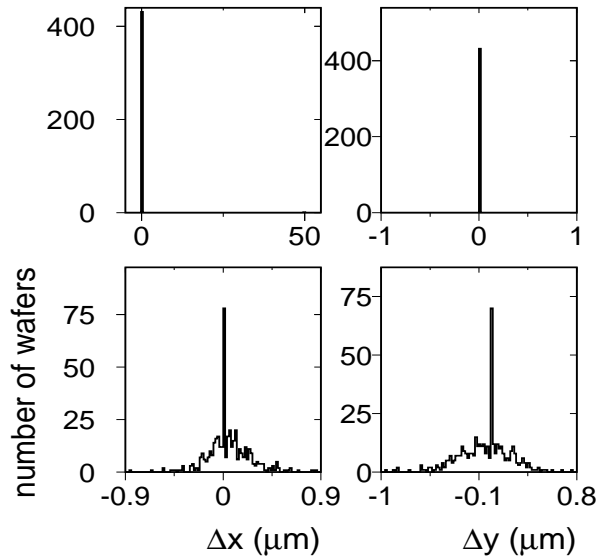}
\end{center}
\vspace*{-13mm}
\caption{\label{fig:single_ladder}
Upper: one wafer is shifted 50$\mu$m in x-direction. The other 431 wafers remain at $\Delta x=\Delta y =0$.
Lower: all wafers are re-aligned within 1$\mu$m.
}
\vspace*{-0.3cm}
\end{figure}

Furthermore, in order to test the re-alignability and the corresponding~sys\-tematic uncertainty,
one wafer was misaligned by 50$\mu$m w.r.t. the original aligned geometry,\,and 
subsequently re-aligned. Remarkably,
in the first iteration of the re-alignment 432 elements were shifted. After re-alignment
all elements were within $1\mu$m of the original position.
The geometries before and after re-alignment are compared to the original
geometry (Fig.~\ref{fig:single_ladder}).

\section{Longevity / variation of active elements}
\begin{figure}[h!]
\begin{center}
\includegraphics[width=0.9\columnwidth]{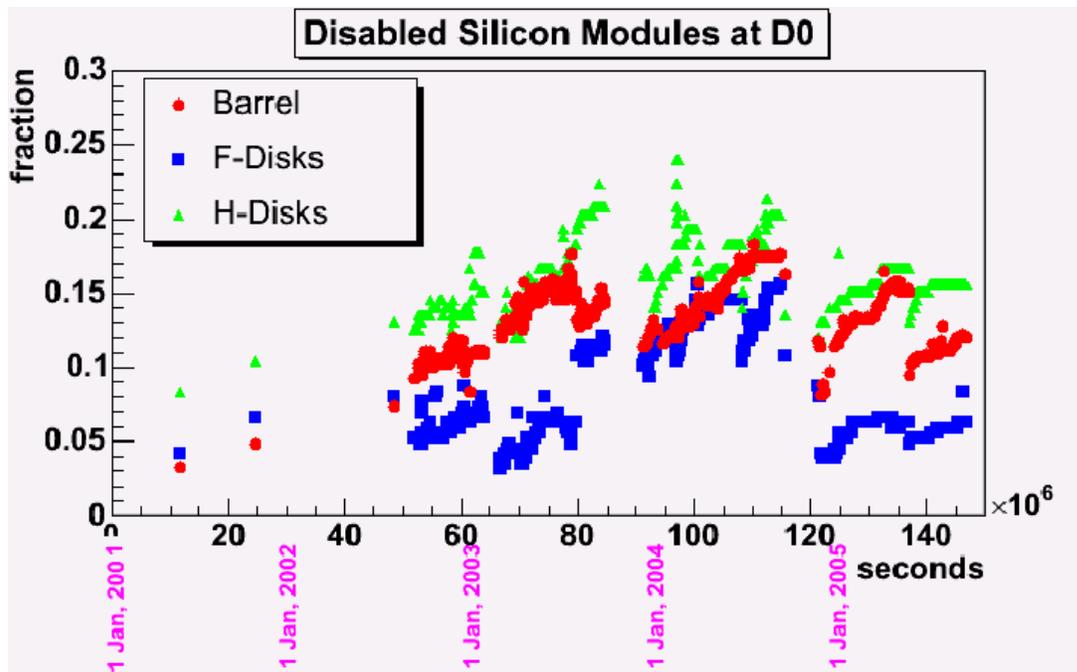}
\end{center}
\vspace*{-10mm}
\caption{\label{fig:longevity}
Fraction of disabled elements as a function of time.
}
\vspace*{-0.4cm}
\end{figure}

Figure~\ref{fig:longevity} shows the variation of the number of disabled elements with time.
During each data-taking shutdown several disabled elements were repaired.

\clearpage
\section{Single wafer alignment in data rerun} 

\begin{figure}[h!]
\begin{minipage}{0.23\columnwidth}
\begin{tabular}{l|c|c}
\small $P_t^{\rm track}$     &\small  high &\small  low \\ \hline
                   &      &     \\
\small $R^{\rm mean}_{\rm before}$     &\small 16.0 &\small 13.3 \\
                   &      &     \\
\small $R^{\rm mean}_{\rm after}$       &\small  3.6  &\small  0.1 \\
\end{tabular}
\end{minipage}
\begin{minipage}{0.75\columnwidth}
\begin{center}
\includegraphics[width=0.7\columnwidth,height=8cm]{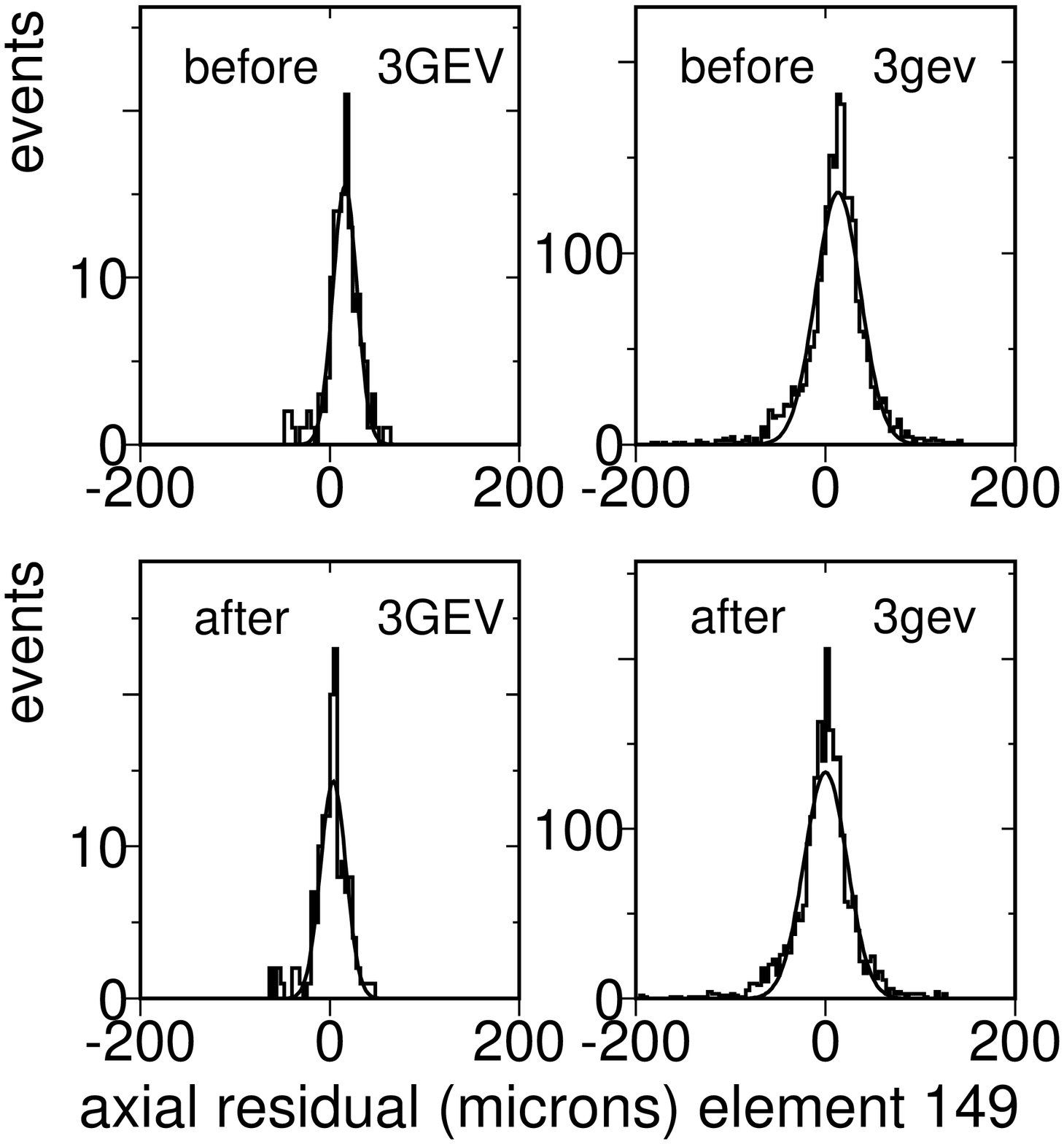}
\end{center}
\end{minipage}
\vspace*{-5mm}
\caption{\label{fig:single_wafer}
Residual mean values (in $\mu$m) of a single wafer (mean value of the fitted Gaussian) before and after alignment
for high and low $P_t$ tracks. GEV refers to $P_t>3$ GeV and gev to $P_t<3$ GeV.  
}
\end{figure}

After data-taking shutdowns some additional wafers become operational and require alignment.
Figure~\ref{fig:single_wafer} 
shows an example of residuals for a single wafer (mean value of the fitted Gaussian) before and after alignment.

\section{Alignability of wafers from different runs}
As some wafers are non-operational depending on the time period of data-taking,
the combination of data from different time periods improves the overall alignment.
Table~\ref{tab:alignability} summarizes the numbers of aligned wafers for different data-taking periods.

\begin{table}[h!]
\begin{center}
\caption{\label{tab:alignability}
Number of aligned wafers in different running periods.
}
\begin{tabular}{l|c|c}
 Period & dates  &  alignable wafers \\  \hline
 B1   &  Oct. 24, 2002     & 827   \\
 B2  &  Oct. 24, 2002     &  827   \\
 A  &  April 20, 2003      & 813   \\
 C    &  Aug. 10, 2003     & 794   \\
 A+B1+B2+C &  mixed &     843 
\end{tabular}
\end{center}
\vspace*{-2mm}
\end{table}

\section{Time stability of detector alignment}
\begin{figure}[h!]
\vspace*{-4mm}
\begin{minipage}{0.45\columnwidth}
\begin{tabular}{l|c}
\small Period & \small dates   \\  \hline
\small 1 (B1,B2)   & \small Oct. 24, 2002     \\
\small 2 (A)  & \small April 20, 2003    \\
\small 3 (C)  & \small Aug. 10, 2003     \\
\small 4 (D)  & \small Jan. 18, 2004     \\        
\small 5 (E)  & \small Aug. 17, 2004     \\
\small 6 (F)  & \small Dec. 18, 2004      
\end{tabular}
\end{minipage} \hfill
\begin{minipage}{0.45\columnwidth}
\includegraphics[width=\columnwidth,height=6.5cm]{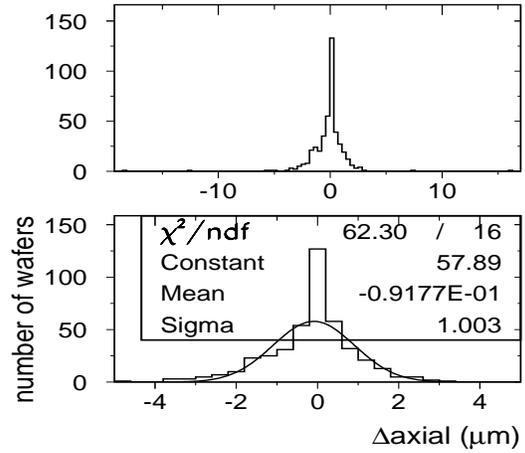}
\end{minipage}
\vspace*{-3mm}
\caption{\label{fig:stability}
Left: list of data sets and their date of data taking.
Right: axial differences between wafers of the aligned geometries for time period 1 and 5.
No significant variation between the aligned geometries is observed.
}
\vspace*{-8mm}
\end{figure}

\begin{figure}[h!]
\begin{center}
\includegraphics[width=0.45\columnwidth,height=7cm]{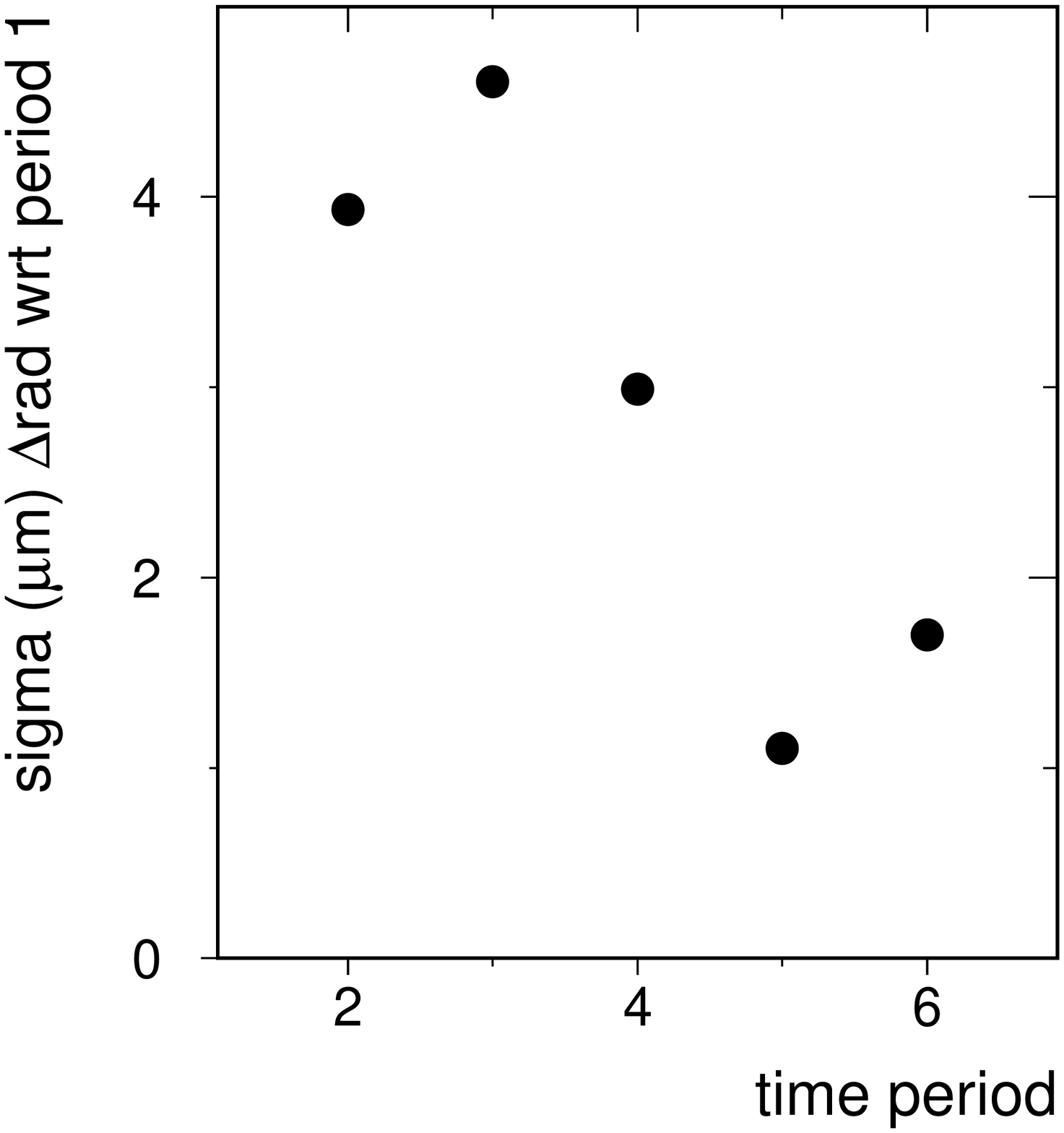} \hfill
\includegraphics[width=0.45\columnwidth,height=7cm]{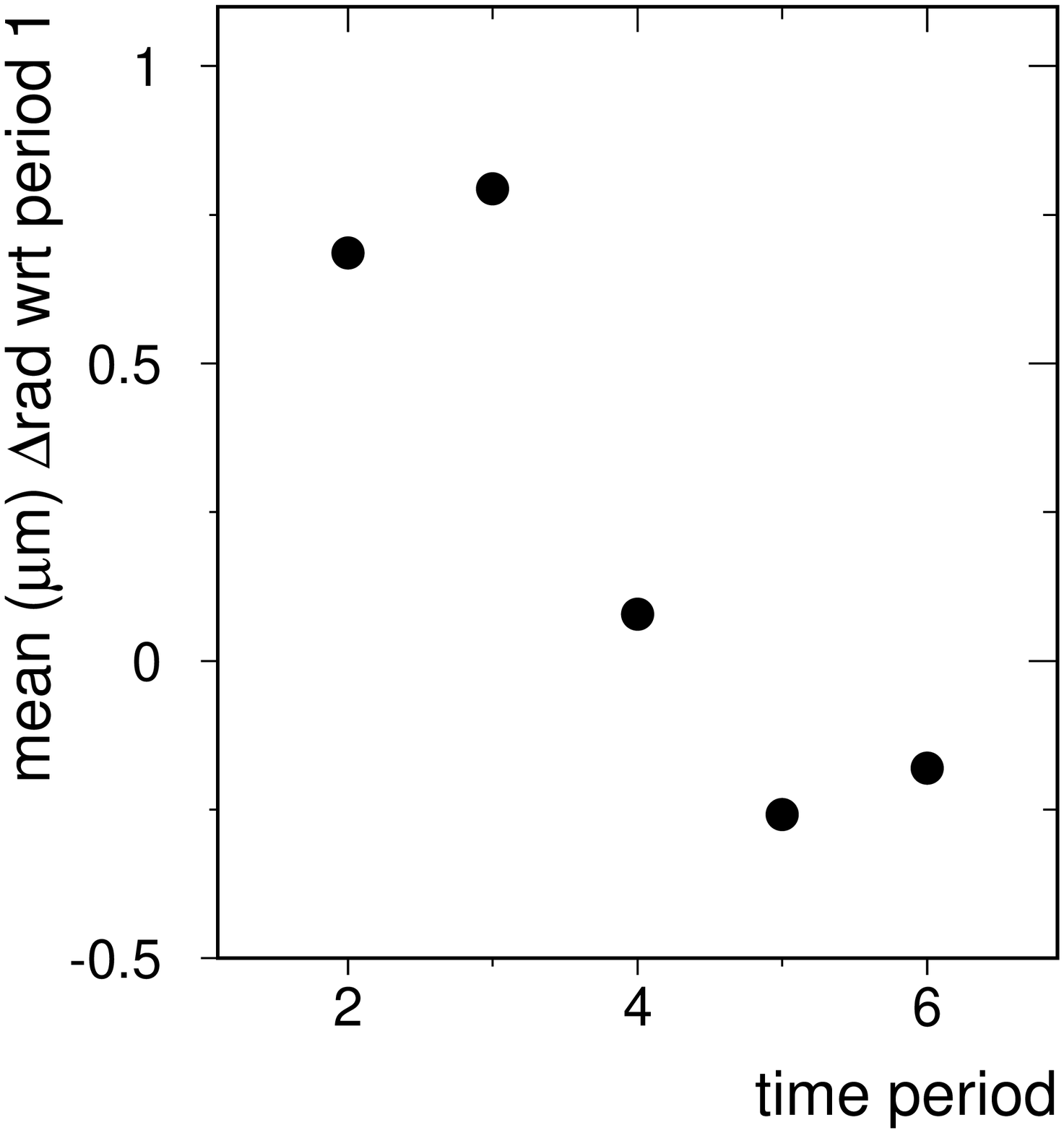}
\end{center}
\vspace*{-10mm}
\caption{\label{fig:stability2}
Radial differences between wafers of the aligned geometries for time periods 
2 to 6 w.r.t. period 1 (see Fig.~\ref{fig:stability}).
No significant variation between the aligned geometries is observed.
}
\end{figure}

The detector alignment has been performed for different
time periods. In order to determine the alignment precision
two aligned geometries from different time periods
are compared. An example is given in Fig.~\ref{fig:stability}
for period 1 and 5.
The time stability of the detector for various time
periods between April 2002 and December 2004 
has been studied and no significant variation
is observed (Fig.~\ref{fig:stability2}).

\clearpage
\section{Local alignment: metrology}
\begin{figure}[h!]
\begin{center}
\vspace*{-0.2cm}
\includegraphics[width=0.8\columnwidth]{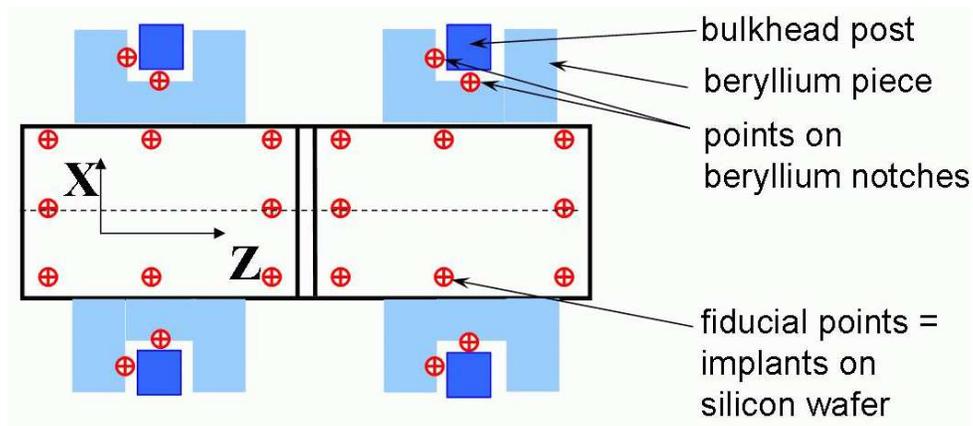}
\end{center}
\vspace*{-6.5mm}
\caption{\label{fig:fiducials}
Positions of fiducial points on the silicon wafers.
}
\end{figure}

\begin{figure}[h!]
\includegraphics[width=0.7\columnwidth]{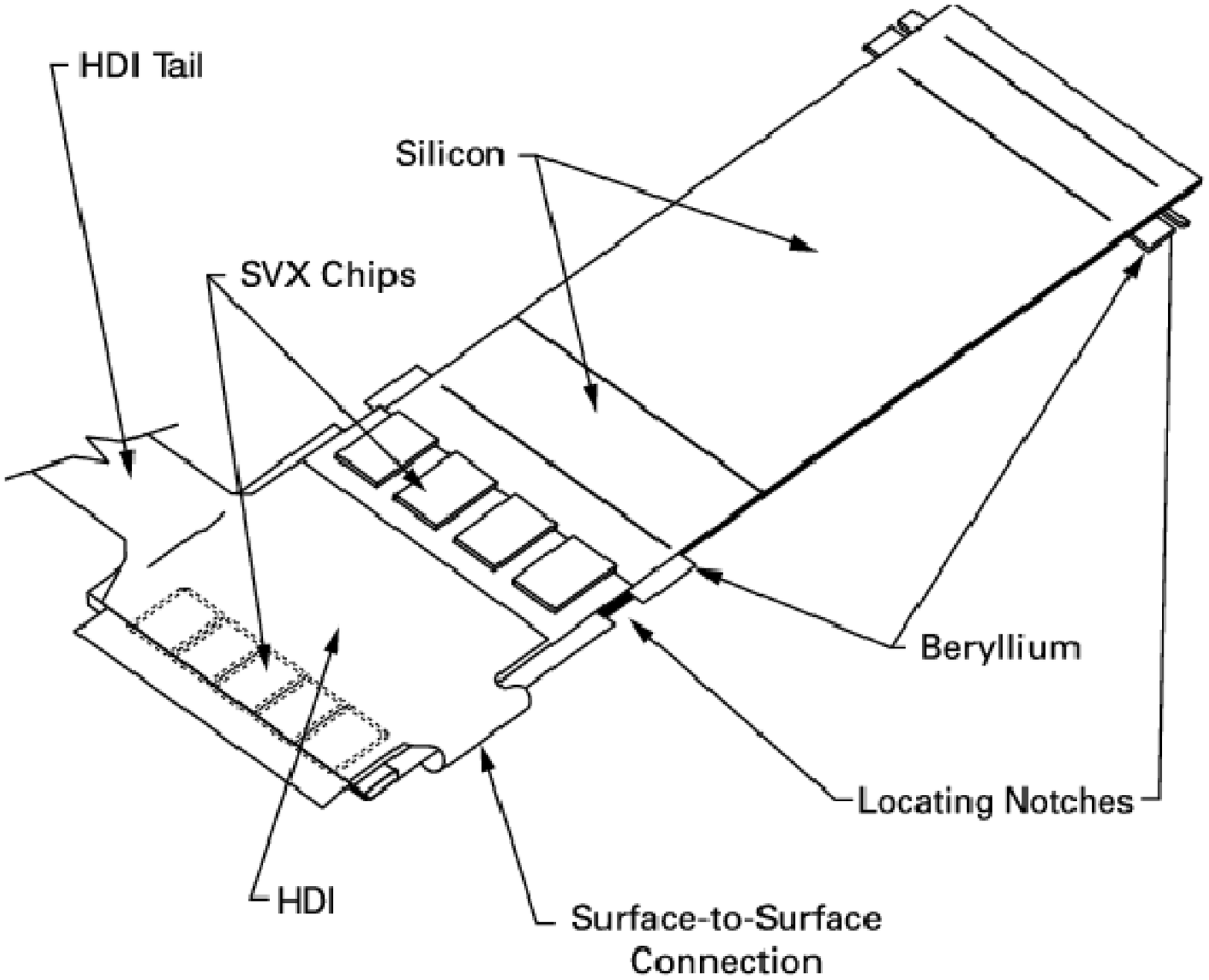} \hfill
\includegraphics[width=0.6\columnwidth,angle=90]{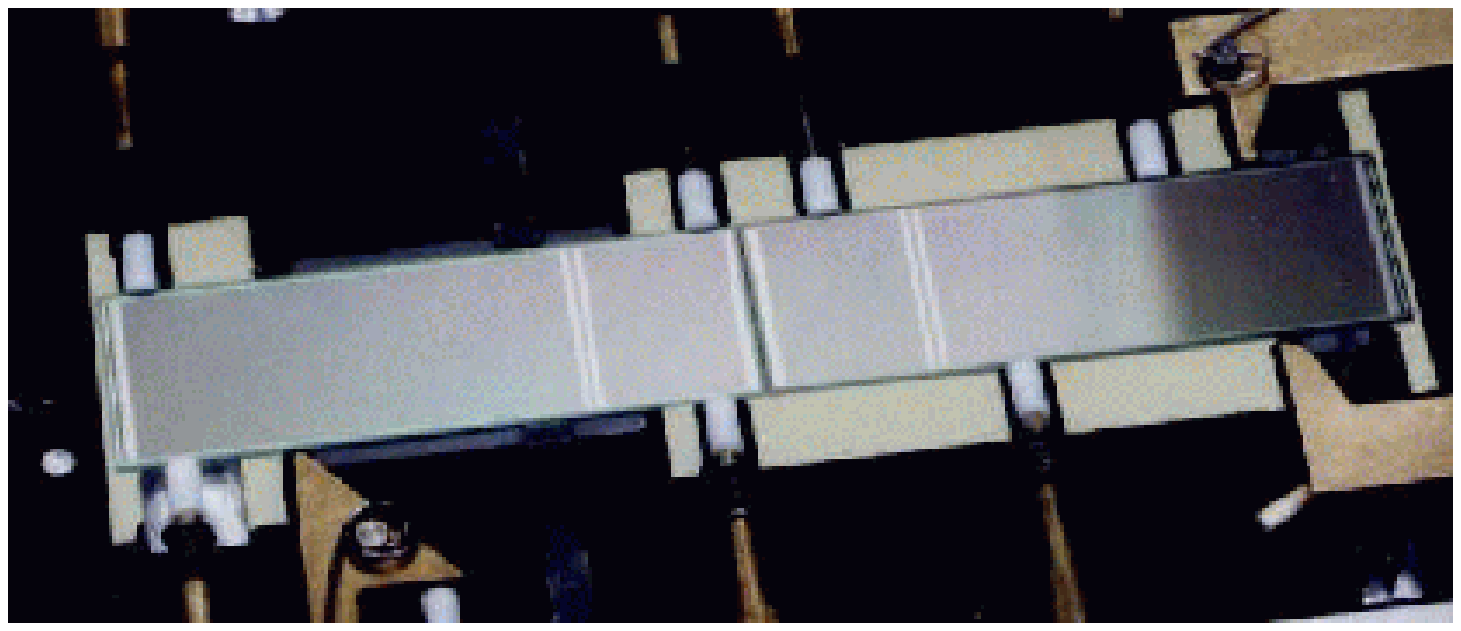}
\vspace*{-2mm}
\caption{\label{fig:separation}
Left: sensitive silicon area, HDI readout with SVX chips. 
Right: joining two silicon plates in support structure. 
}
\end{figure}

\begin{figure}[tp]
\begin{center}
\vspace*{-2mm}
\includegraphics[width=0.9\columnwidth,height=12cm]{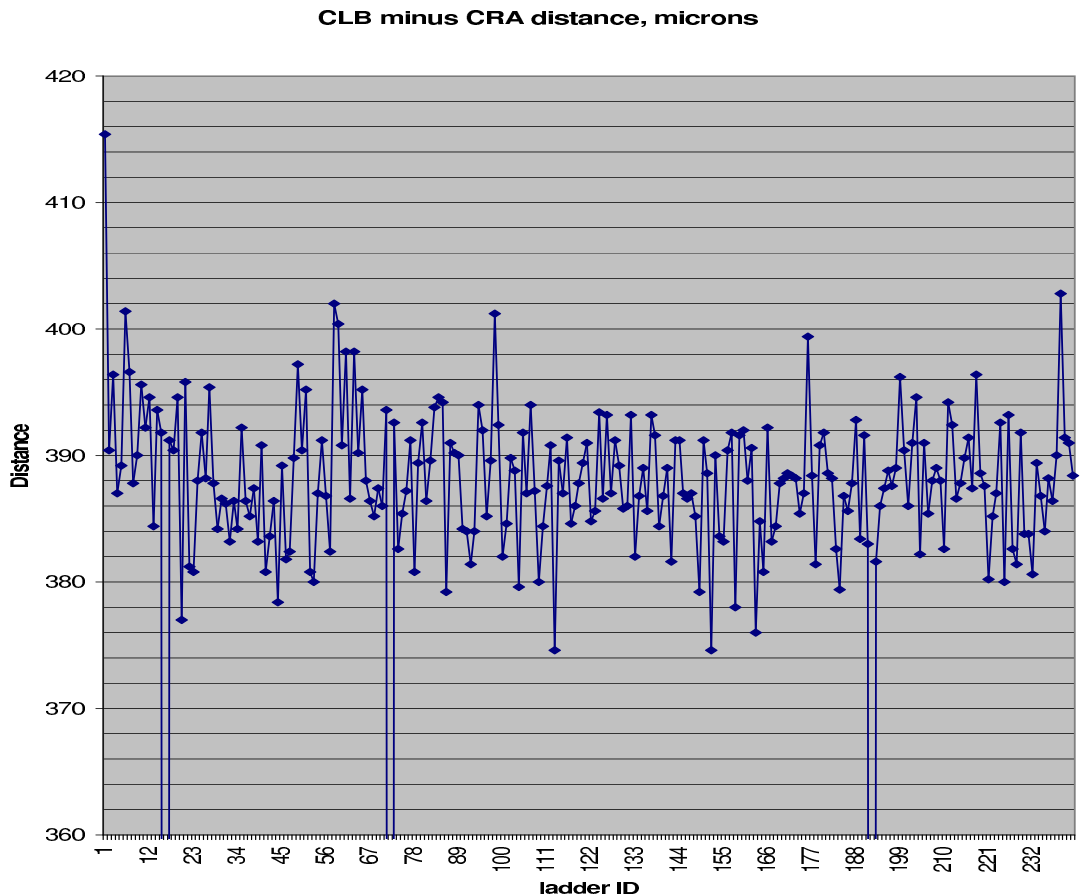}
\end{center}
\vspace*{-9mm}
\caption{\label{fig:metrology}
Metrology z-distance measurements (in $\mu$m) between two fiducial points on the joined 
silicon plates.
No significant variation from their design position (of about $390~\mu$m) is observed.
}
\end{figure}

In addition to the previously described global alignment, where wafers
are considered ideal planes with no structure, the local alignment has been
investigated.
Local alignment refers to the alignment on a given sensitive element.
As an example, the wafer geometry and the 
separation in z-direction is shown in Fig.~\ref{fig:separation}.
Figure~\ref{fig:fiducials} illustrates the positions 
of the fiducial points on the wafers. 
This is particulary interesting as some wafers are made of two independent silicon plates.
The precision in the distance $\Delta z$ of two fiducial points for detector elements
made of the two sensors is given in Fig.~\ref{fig:metrology} from metrology.
No indication of a shift in the survey between these plates is observed and variations 
are within $\pm10\mu$m.


\clearpage

\section{Influence on physics results}
\begin{figure}[h!]
\vspace*{-3mm}
\begin{center}
\includegraphics[width=0.325\columnwidth,height=8cm]{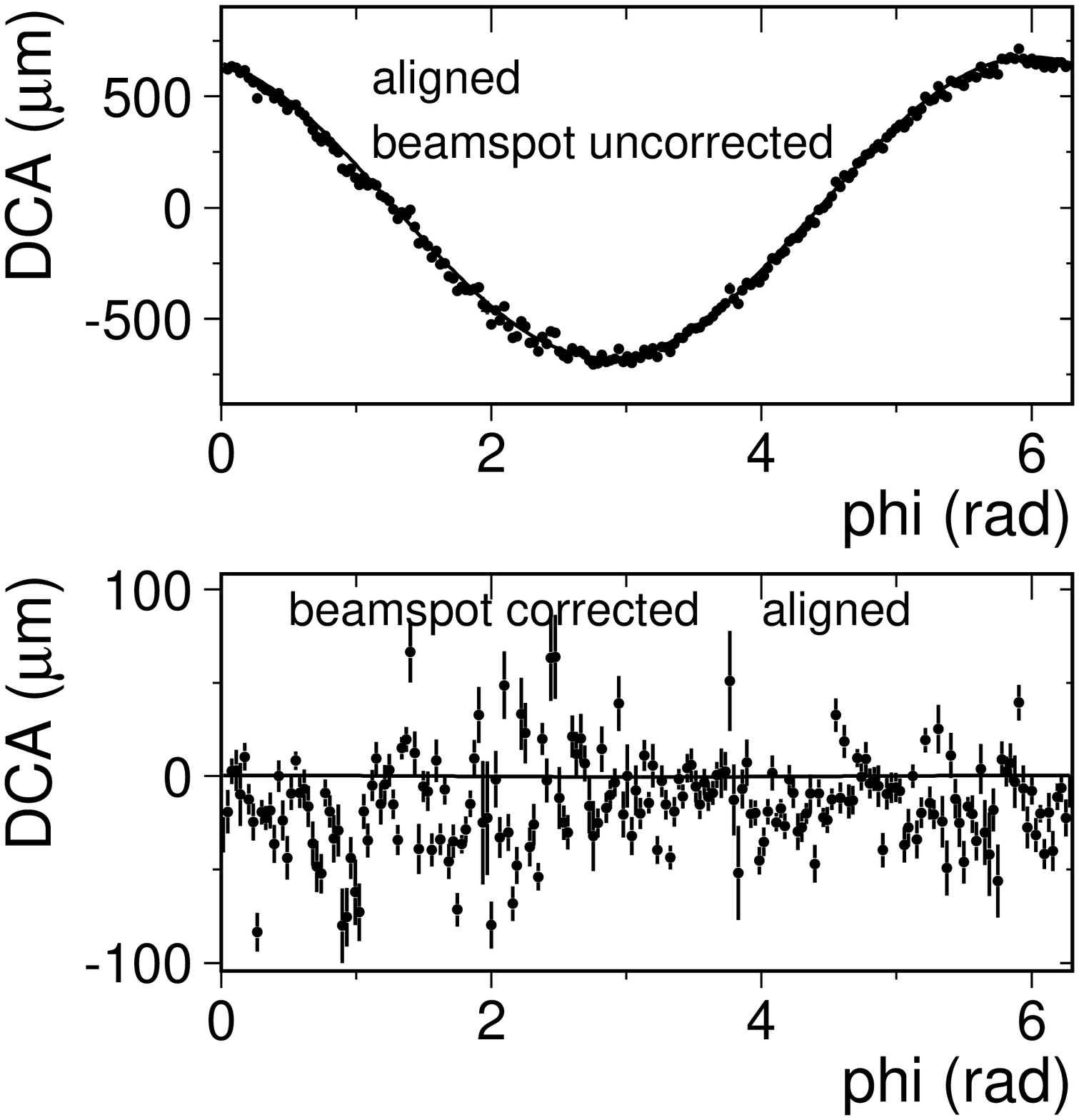} \hfill
\includegraphics[width=0.325\columnwidth,height=8cm]{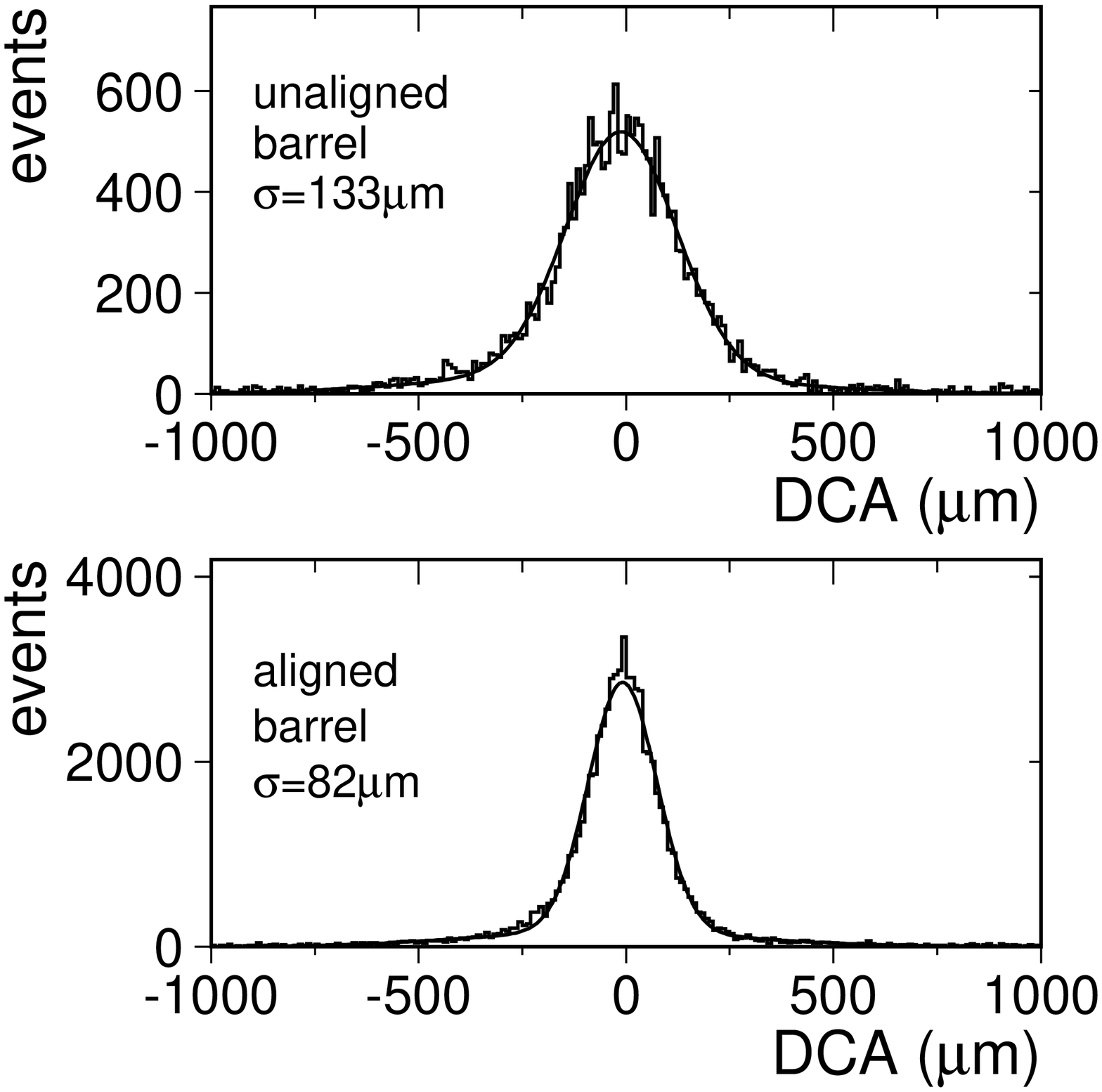}  \hfill
\includegraphics[width=0.325\columnwidth,height=8cm]{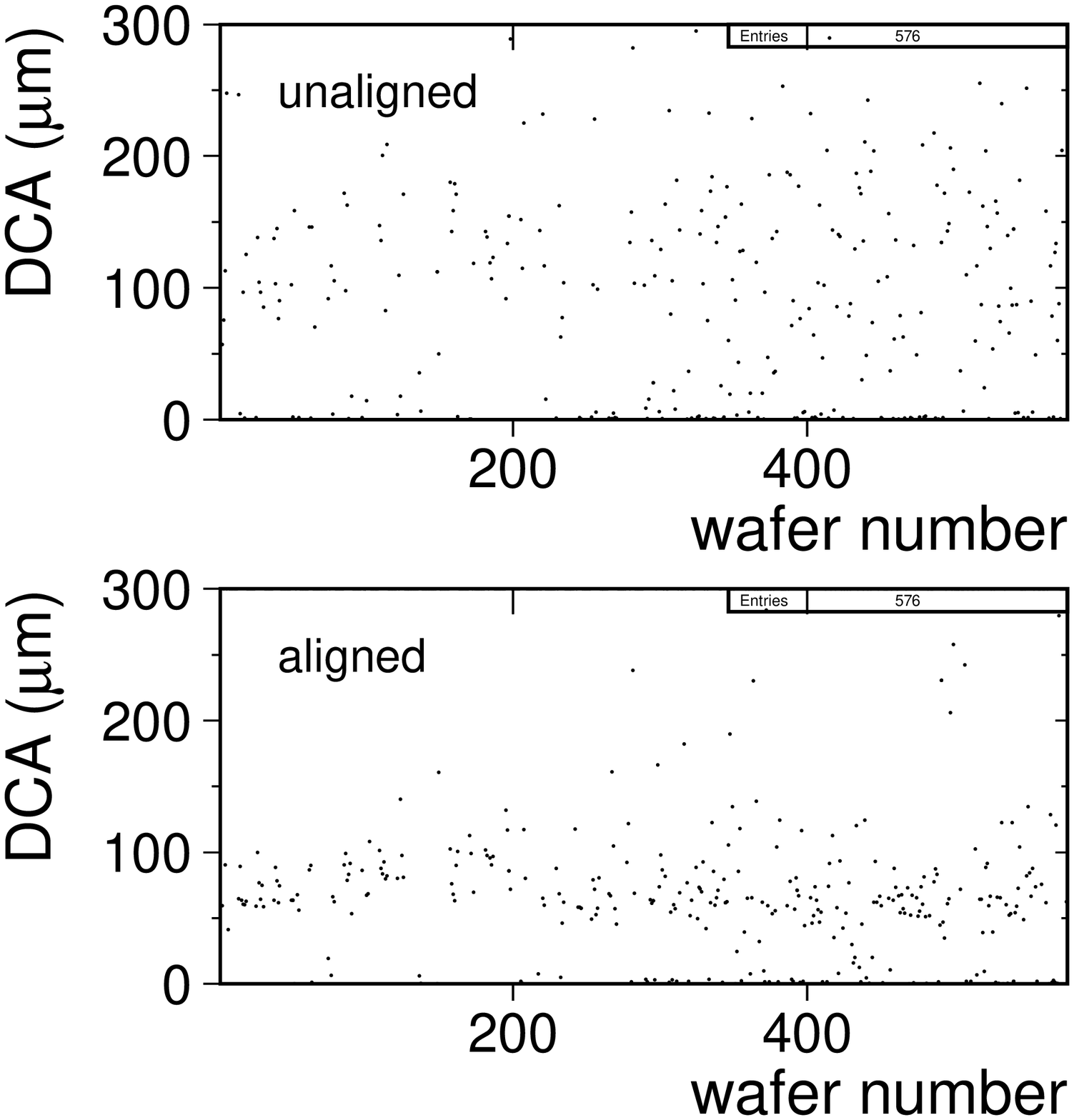}
\end{center}
\vspace*{-14mm}
\caption{\label{fig:beamspot}
Upper left: displaced beamspot position in the r-$\phi$ plane
as a function of $\phi_{\rm track}$.
Lower left: corrected beamspot position.
Center: DCA resolution for unaligned and aligned detector after beamspot correction.
Right: DCA resolution for each barrel wafer.
}
\end{figure}

In order to determine the resolution of the Distance of Closest Approach, DCA,
the position of the beamspot has been determined.
A displaced beamspot position in the r-$\phi$ plane corresponds to a sine dependance 
of the DCA $\delta$ as a function of the track direction $\phi_{\rm track}$. 
It is parametrized like
$\delta = \delta_0 - P_1\sin\phi_{\rm track} + P_2\cos\phi_{\rm track}$ and this 
function is fitted to the data as shown in Fig.~\ref{fig:beamspot}.
The figure shows also the DCA resolution.
The DCA resolution depends on the transverse momentum of the track and
Table~\ref{fig:dca} gives the DCA resolutions for different $P_t$ ranges.
The DCA resolution consists of the beam-spot size convoluted with the
Impact Parameter (IP) resolution.
The beam-spot size is approximately 30-40$\mu$m and depends on the machine optics.

\begin{figure}[h!]
\vspace*{-5.5mm}
\center
  \includegraphics[width=\columnwidth]{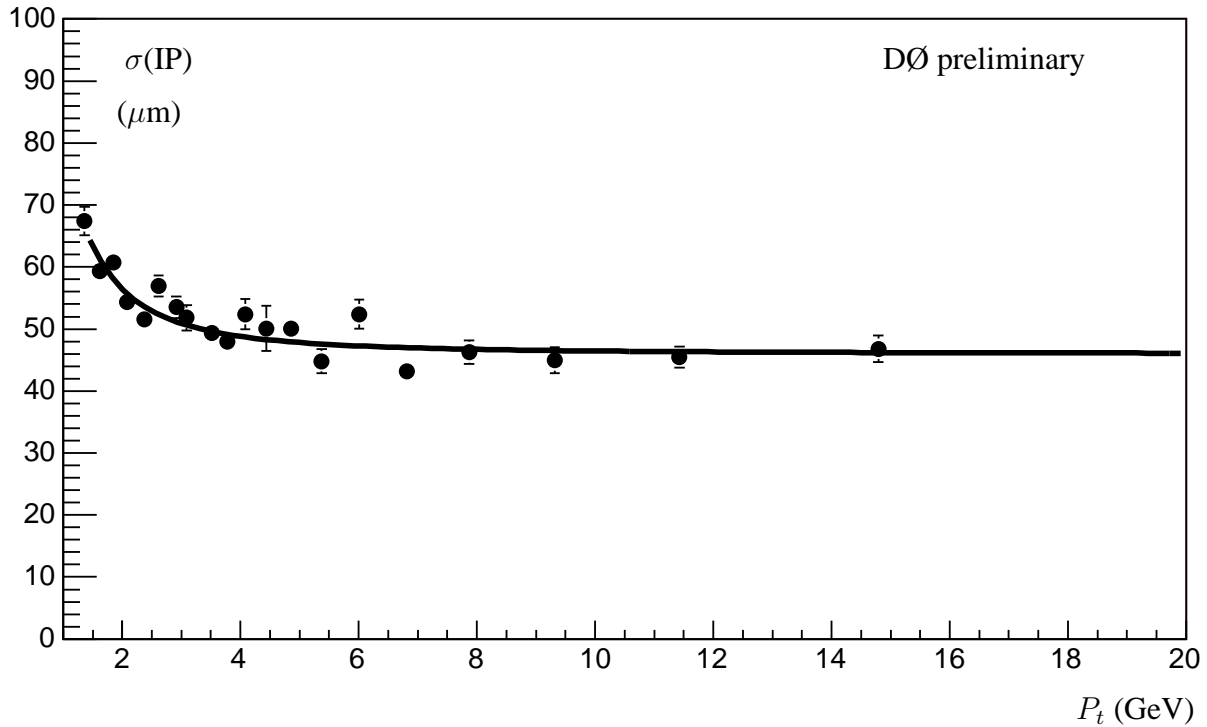}
\put(-120.0,245.0){D\O\ preliminary}
\put(-50.0,-1.0){ $P_t$ (GeV)}
\put(-410,245.0){ $\sigma$(IP)}
\put(-410,225.0){($\mu$m)}
\vspace*{-2mm}
\caption{\label{fig:stt}
Impact parameter resolution for the online Silicon Track Trigger (STT).
}
\vspace*{-2mm}
\end{figure}

The impact parameter resolution is crucial for the Silicon Track Trigger (STT).
The beamspot determination of the previous run is used. The resulting 
impact parameter resolution is shown in Fig.~\ref{fig:stt}.
%
%

\vspace*{-2mm}
\begin{table}[h!]
\begin{center}
\caption{\label{fig:dca}
DCA resolution for different $P_t$ ranges.
}
\begin{tabular}{l|c|c|c|c|c|c|c}
$P_t$ (GeV) &  0.2-0.5 &  0.5-1  &  1-3 &  3-5 &  5-10 & 10-20 & 20-50\hspace*{-0.5mm} \\ \hline
\hspace*{-0.5mm}DCA res. ($\mu$m)\hspace*{-0.5mm} & 203 & 118 & 77 & 60 & 55 & 53 & 50
\vspace*{-1mm}
\end{tabular}
\end{center}
\end{table}

\begin{figure}[thp]
\begin{minipage}{0.5\columnwidth}
\includegraphics[width=\columnwidth]{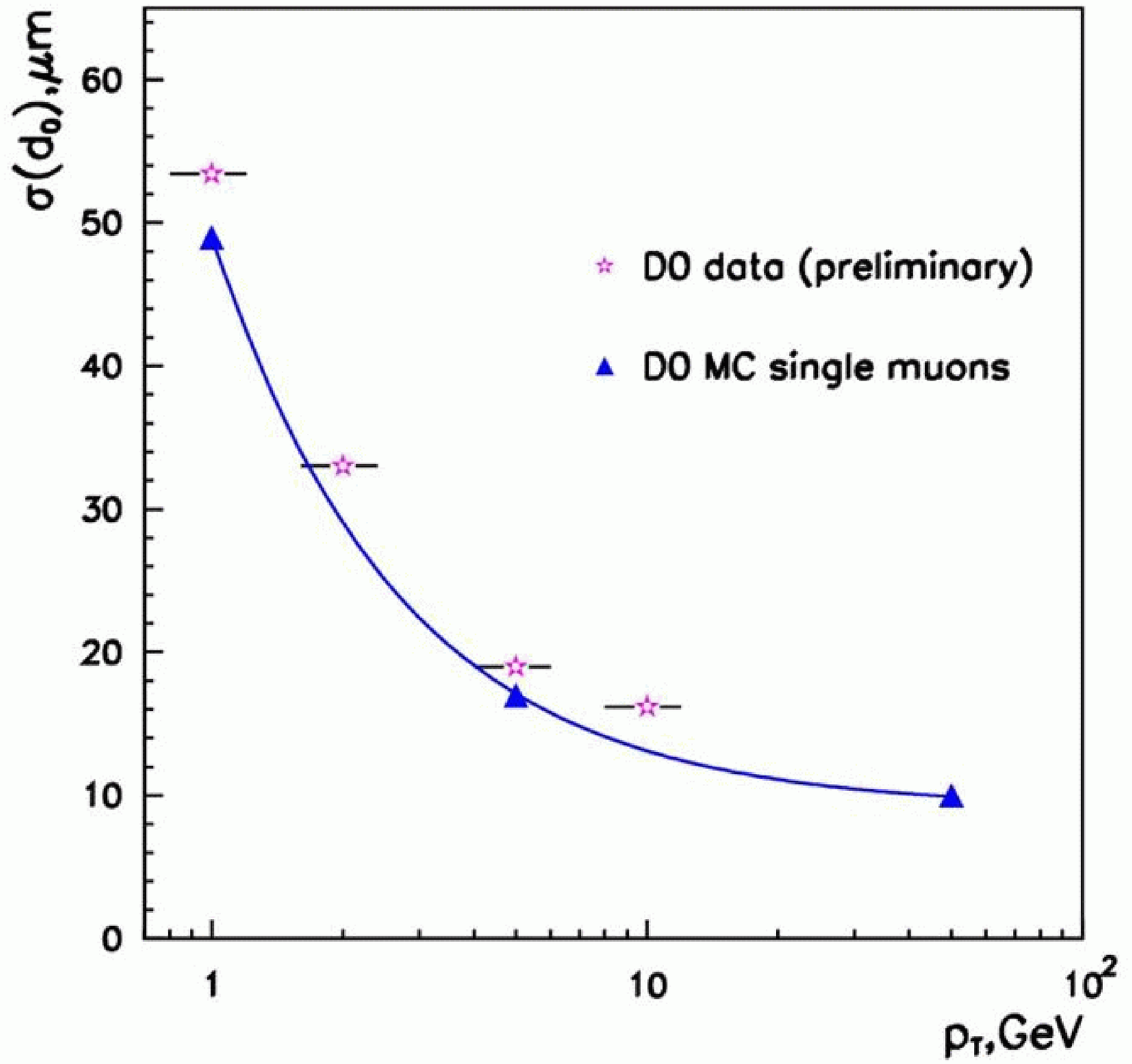}
\end{minipage}\hfill
 \begin{minipage}{0.49\columnwidth}
\includegraphics[width=\columnwidth]{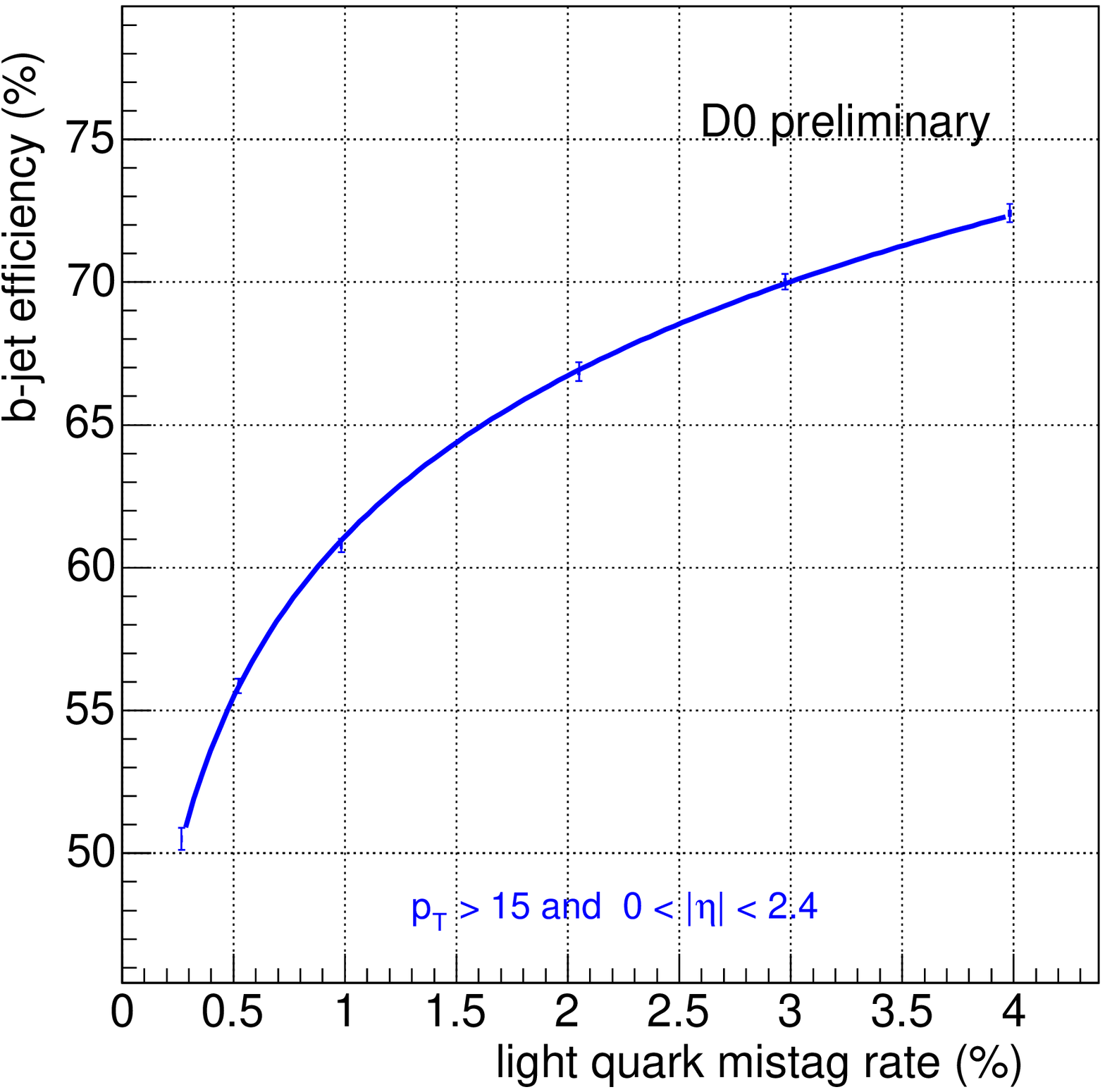}
\end{minipage}
\vspace*{-2mm}
\caption{\label{fig:btag}
Left: impact parameter resolution after alignment as a function of the 
transverse momentum.
The beamspot size has been taken into account.
Right: b-quark tagging performance using a neural network algorithm.
}
\vspace*{-1mm}
\end{figure}

The impact parameter measurement is also an important aspect for b-quark 
tagging. Its resolution after alignment in the offline analysis together 
with the Monte Carlo prediction is shown in Fig.~\ref{fig:btag}.
Multiple scattering is the dominant source of resolution degradation
at small $P_t$ values.
The figure shows also the b-quark tagging efficiency versus the light quark
mistag rate.

\begin{figure}[h!]
\begin{center}
\includegraphics[width=0.45\columnwidth,height=8cm]{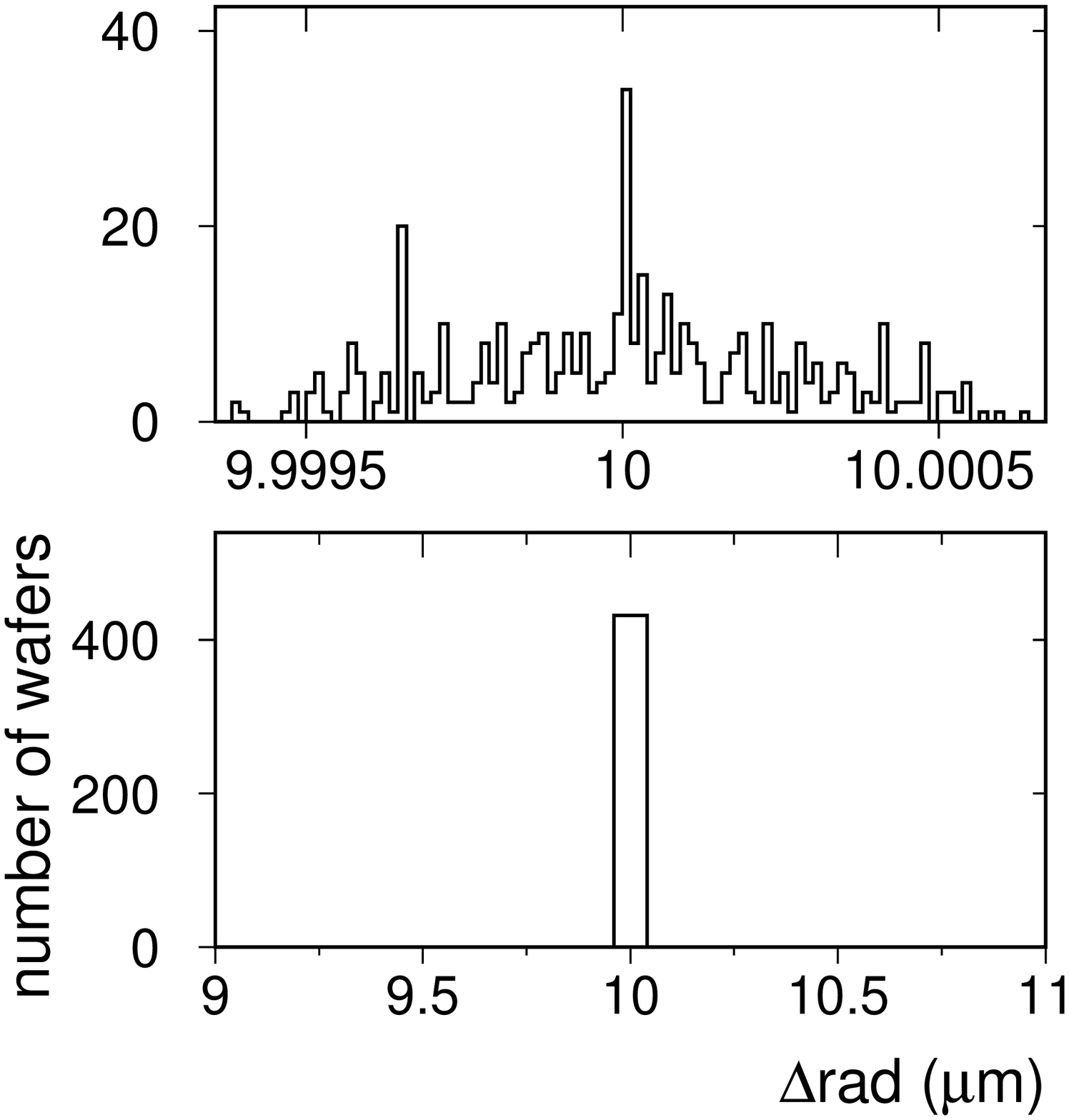} \hfill
\includegraphics[width=0.45\columnwidth,height=8cm]{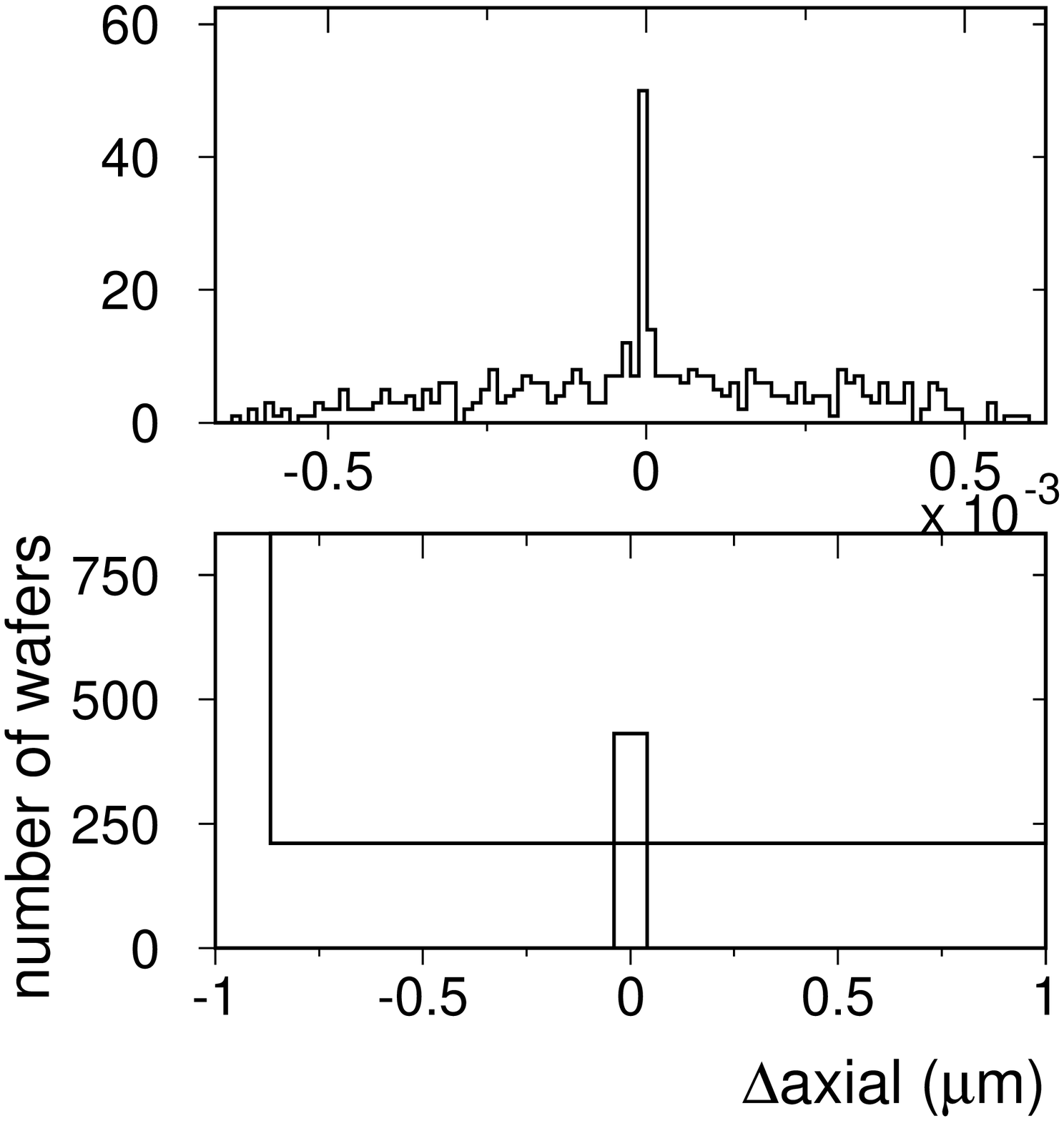}
\end{center}
\vspace*{-12mm}
\caption{\label{fig:outward}
For systematic error studies: comparison of geometries, $10\mu$m radial shift outwards.
}
\end{figure}

The alignment uncertainty contributes to the systematic errors in several
physics analyses. The effect of the alignment uncertainty has been studied, 
for example, by assuming a constant shift of 10$\mu$m in radial direction outward
(Fig.~\ref{fig:outward}) in order to estimate the impact on B-meson lifetime 
measurements.
Only a small contribution to the systematic uncertainty from alignment
in B-meson lifetime measurements is observed (Table~\ref{tab:lifetime}).

\clearpage

\begin{table}[h!]
\vspace*{-5mm}
\caption{\label{tab:lifetime}
Systematic uncertainties $\rm B_{s}\rightarrow J/\psi \phi$ lifetime measurement.
}
\begin{center}
\vspace*{-2mm}
\begin{tabular}{c|c}
                     & $c\tau(B_{\rm s}) $ ($\mu$m) \\ \hline
{ Alignment}         &       2                  \\ 
$\rm J/\psi$\ vertex &       3                  \\
Model for resolution &       3                  \\
Background           &       4                  \\ \hline

Total                &       6                   \\
\end{tabular}
\end{center}
\vspace*{-16mm}
\end{table}

\begin{figure}[h!]
\includegraphics[width=0.49\columnwidth,height=8cm]{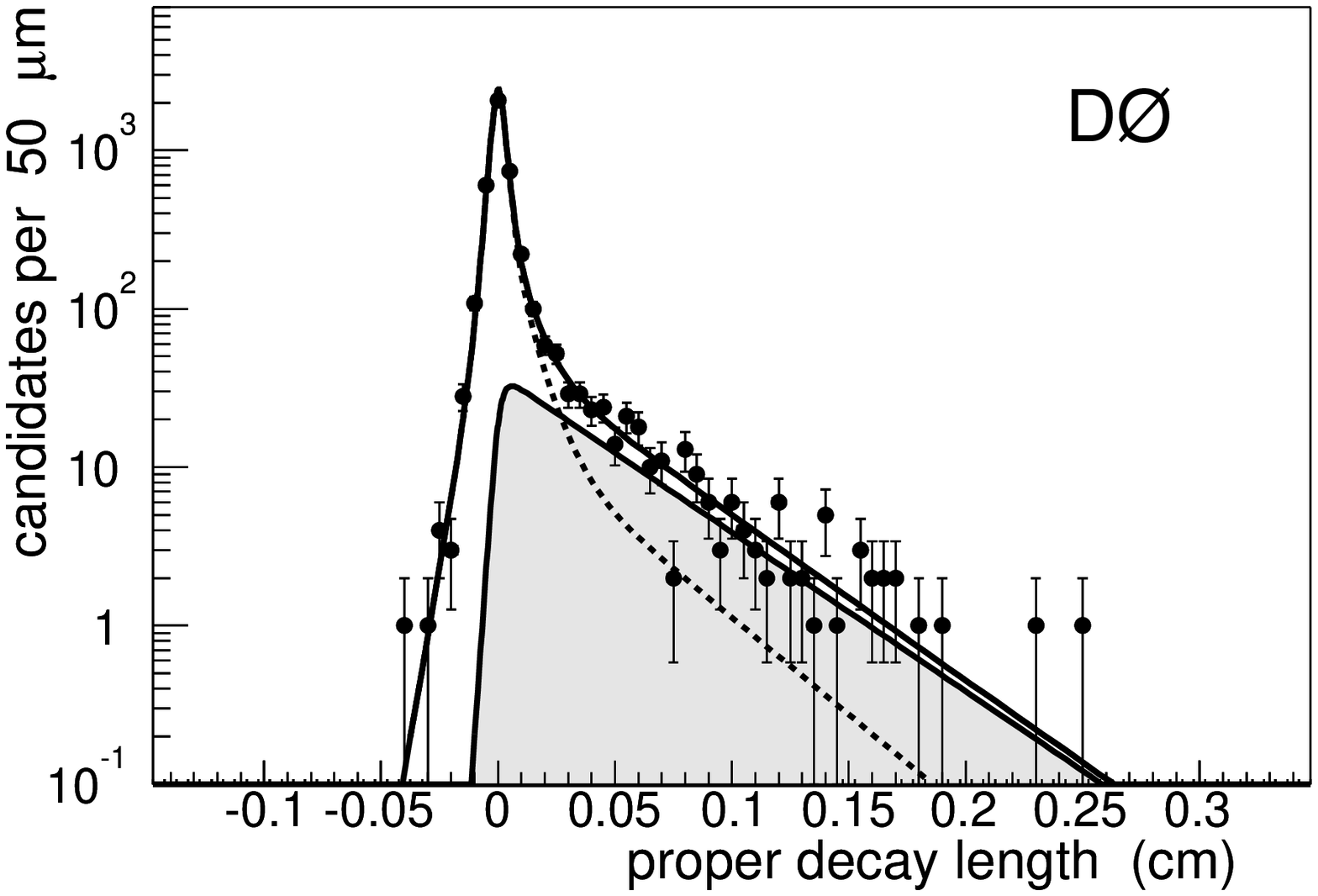}\hfill
\includegraphics[width=0.49\columnwidth,height=8cm]{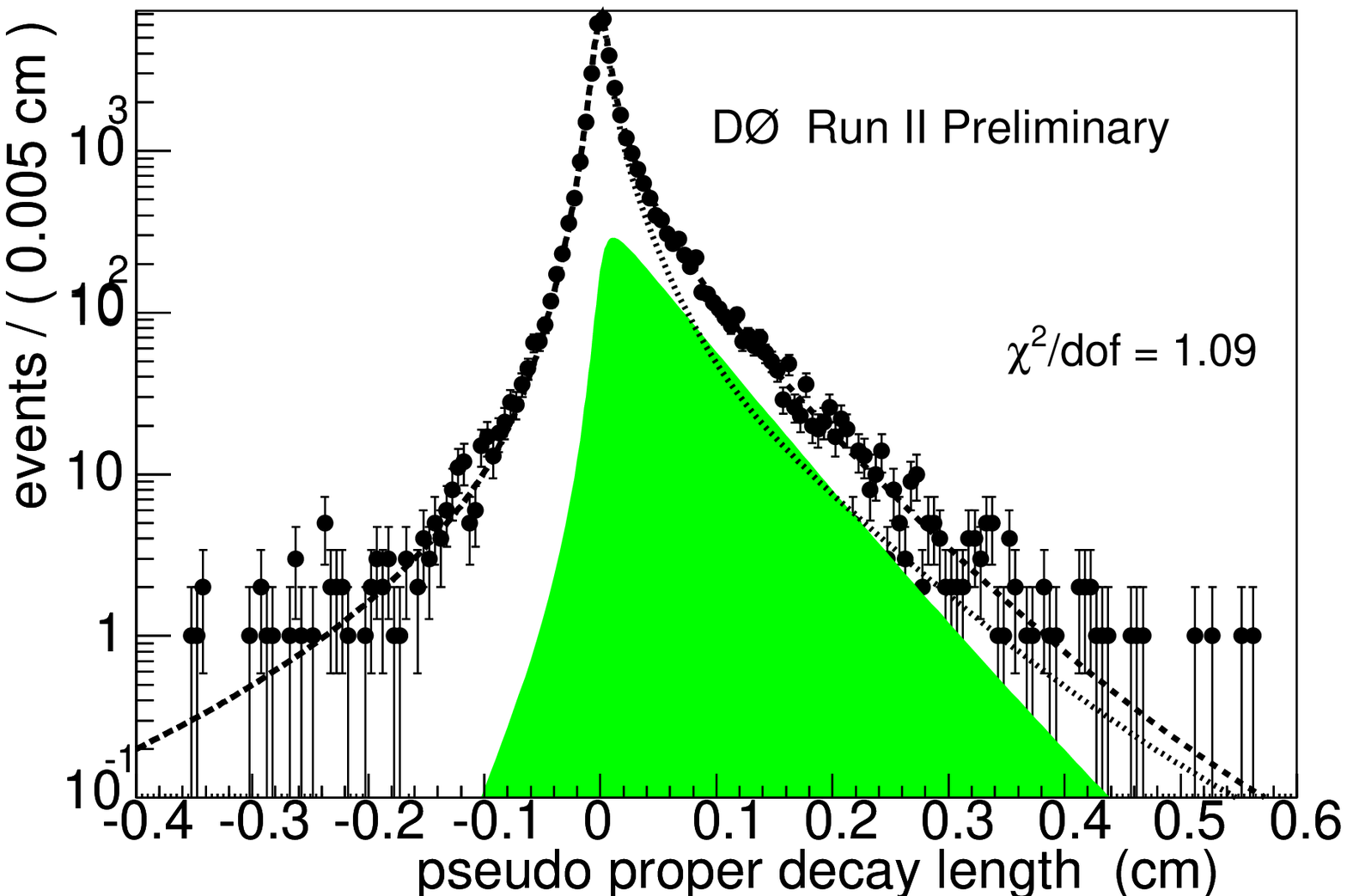}
\vspace*{-4mm}
\caption{\label{fig:lifetime1}
B-meson lifetime measurements.
Left: $\rm B_{s}\rightarrow J/\psi \phi$.
Right: $\rm B_{s}\rightarrow D_s\mu\nu X $.
}
\end{figure}

\begin{figure}[h!]
\vspace*{-2mm}
\begin{center}
\includegraphics[width=0.49\columnwidth,height=8cm]{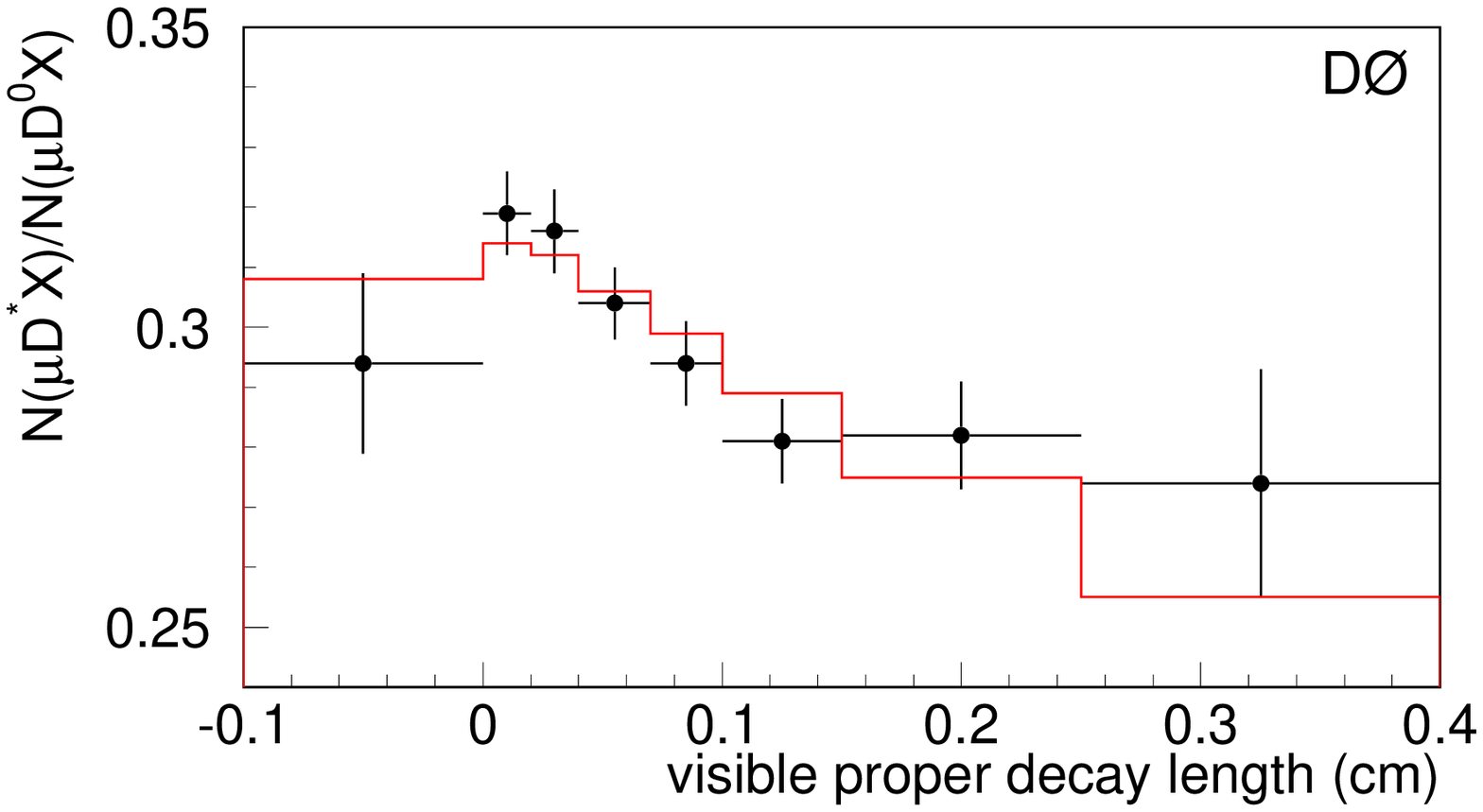} \hfill
\includegraphics[width=0.49\columnwidth,height=8cm]{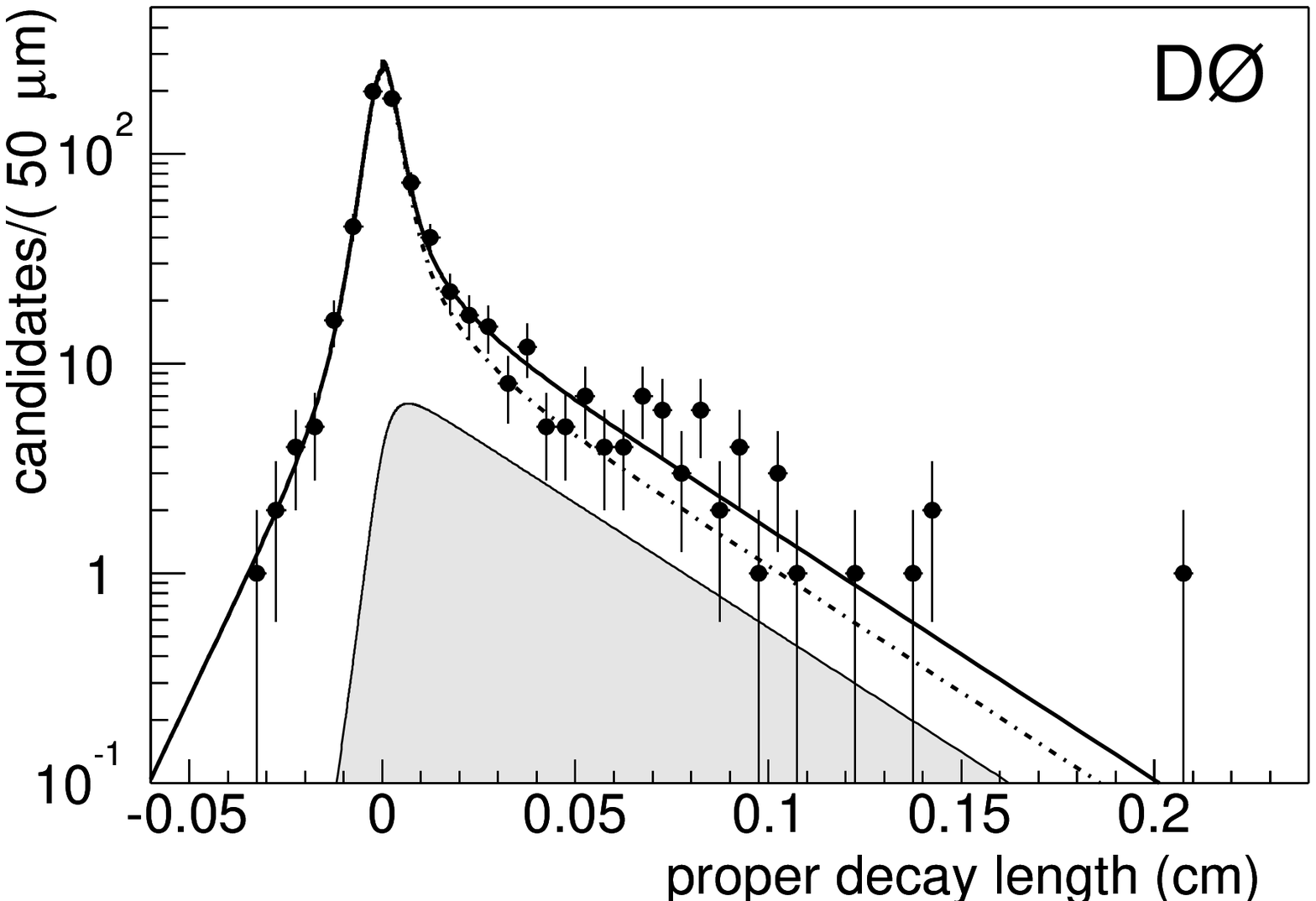}
\end{center}
\vspace*{-6mm}
\caption{\label{fig:lifetime2}
Lifetime measurements.
Left: $\rm B_{d}$.
Right: $\rm \Lambda_{b}\rightarrow J/\psi\Lambda$.
}
\end{figure}

\clearpage
As examples of physics analyses where the detector alignment is crucial,
results from lifetime measurements are shown in Figs.~\ref{fig:lifetime1} 
and~\ref{fig:lifetime2}. The signal (shaded region) is clearly visible
over the background (dotted line).

\clearpage
\section{Conclusions and outlook}
\begin{figure}[hcbtp]
\vspace*{5mm}
\begin{center}
\begin{minipage}{0.6\columnwidth}
\includegraphics[width=\columnwidth]{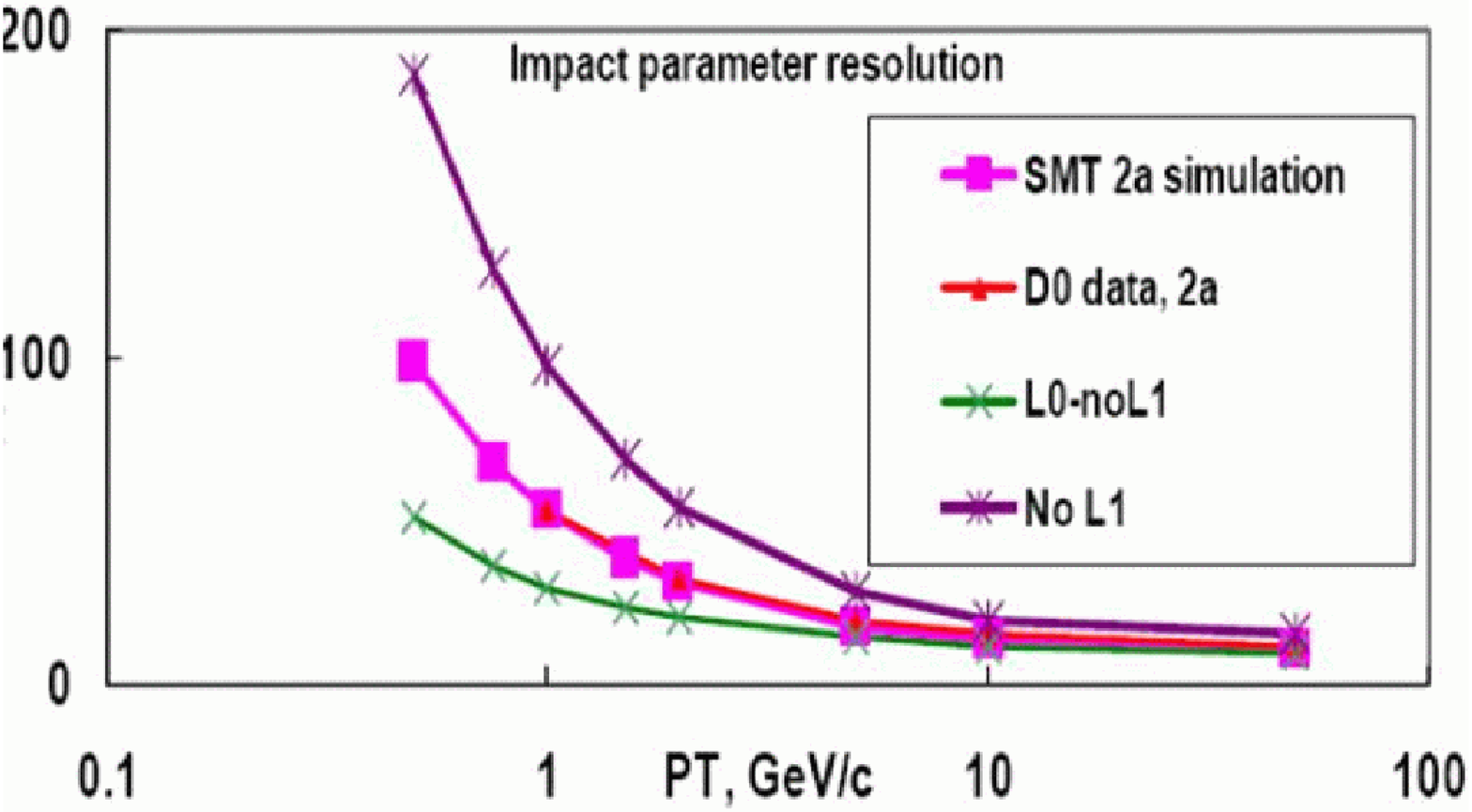}
\put(-250,128.0){\small $\sigma$(IP)}
\put(-250,114.0){\small ($\mu$m)}
\end{minipage}
\end{center}
\vspace*{-5mm}
\caption{\label{fig:layer0}
Expected layer-0 (L0) improvement for the impact parameter (IP) resolution. 
Simulation and data (2a) overlap. Also shown is the impact parameter 
resolution without layer-1.
}
\vspace*{-3mm}
\end{figure}

About 850 sensitive elements have been aligned.
The alignment precision is close to design value 
(e.g. residuals: data $20\mu$m, simulation $16\mu$m). 
Some alignment parameters have been optimized.
Systematic uncertainties of the alignment procedure are less than about $10\mu$m.
The monitoring of the detector stability showed no significant movement.
The alignment ensures excellent on-line and off-line b-tagging, and lifetime measurements,
and is therefore crucial for Higgs, top, and B-physics.
In spring 2006, a new inner layer~\cite{michele}, Layer-0,
at 1.6 cm from the interaction point will be installed inside the current vertex detector, 
which will significantly improve the impact parameter resolution, as 
illustrated in Fig.~\ref{fig:layer0}.

\section*{Acknowledgements} I would like to thank the organizers of the TIME'05 conference for
their kind hospitality, and 
Tim Brodbeck, 
Aran Garcia-Bellido, 
Mike Hildreth, 
Alex Melnitchouk,
Ulrich Straumann,
Mike Strauss and
Rick van Kooten 
for comments on the manuscript. 
Contributions from Guennadi Borrisov and Brian Davies are particularly acknowledged.

\end{document}